\documentclass{aa}  
\usepackage{graphicx}
\usepackage{txfonts}
\usepackage{natbib}
\bibpunct{(}{)}{;}{a}{}{,} 
\usepackage{ulem} 
\usepackage[colorlinks]{hyperref}

\def\aaps{A\&AS}
\def\aap{A\&A}
\def\apj{ApJ}
\def\apjs{ApJS}
\def\aj{AJ}

\def \hi {\ion{H}{i}}

\def\kms{km\,s$^{-1}$}
\def\kmss{km\,s$^{-1}~$}

\def\deg{\hbox{$^\circ$}}
\def\arcmin{\hbox{$^\prime$}}

\def\fdg{\hbox{$.\!\!^\circ$}}
\def\farcm{\hbox{$.\mkern-4mu^\prime$}}

\defcitealias{Kalberla2016b}{Paper~I}

\begin{document}

\title{\hi\, anisotropies associated with radio-polarimetric filaments. }

   \subtitle{Steep power spectra associated with cold gas}

\author{P.\ M.\ W.\ Kalberla  \inst{1} 
        \and 
        J. Kerp   \inst{1} 
        \and 
        U. Haud   \inst{2}
        \and 
        M. Haverkorn \inst{3}
}

\institute{Argelander-Institut f\"ur Astronomie, 
           Auf dem H\"ugel 71, 53121 Bonn, Germany \\
           \email{pkalberla@astro.uni-bonn.de} 
           \and 
           Tartu Observatory, 61602 T\~oravere, Tartumaa, Estonia 
           \and 
           Department of Astrophysics/IMAPP, Radboud University Nijmegen, 
           P.O. Box 9010, 6500 GL Nijmegen, The Netherlands \\
           }

   \authorrunning{P.\,M.\,W. Kalberla et al. } 

   \titlerunning{\hi~and radio-polarimetric filaments}

   \offprints{P.\,M.\,W. Kalberla}

   \date{Received 1 September 2016 / Accepted 18 July 2017 }

  \abstract 
  {LOFAR detected toward 3C 196 linear polarization structures which
    were found subsequently to be closely correlated with cold
    filamentary \hi\, structures. The derived direction-dependent
    \hi~power spectra revealed marked anisotropies for narrow ranges in
    velocity, sharing the orientation of the magnetic field as expected
    for magneto hydrodynamical (MHD) turbulence. }
  {Using the Galactic portion of the Effelsberg--Bonn \hi~Survey (EBHIS)
    we continue our study of such anisotropies in the \hi~distribution
    in direction of two WSRT fields, Horologium and Auriga; both are
    well known for their prominent radio-polarimetric depolarization
    canals.  At 349 MHz the observed pattern in total intensity is
    insignificant but polarized intensity and polarization angle show
    prominent ubiquitous structures with so far unknown origin.  }
  {Apodizing the \hi~survey data by applying a rotational symmetric 50\%
    Tukey window, we derive average and position angle dependent power
    spectra. We fit power laws and characterize anisotropies in the
    power distribution. We used a Gaussian analysis to determine relative
    abundances for the cold and warm neutral medium.  }
  {For the analyzed radio-polarimetric targets significant anisotropies
    are detected in the \hi~power spectra; their position angles are
    aligned to the prominent depolarization canals, initially detected
    by WSRT. \hi~anisotropies are associated with steep power
    spectra. Steep power spectra, associated with cold gas, are detected
    also in other fields. }
  {Radio-polarimetric depolarization canals are associated with
    filamentary \hi~structures that belong to the cold neutral medium
    (CNM). Anisotropies in the CNM are in this case linked to a
    steepening of the power-spectrum spectral index, indicating that
    phase transitions in a turbulent medium occur on all
    scales. Filamentary \hi~structures, driven by thermal instabilities,
    and radio-polarimetric filaments are associated with each other. The
    magneto-ionic medium that causes the radio-polarimetric filaments is
    probably wrapped around the \hi.  }

  \maketitle
%

\section{Introduction}
\label{Intro}

A significant fraction of the cold neutral medium (CNM) filamentary
structures is aligned with filamentary structures in polarized dust
emission, both elongated along the mean magnetic field
\citep{Clark2014,Planck2016,Kalberla2016}. The correlation is well
defined over large areas. Constructing a dust model by incorporating CNM
column density maps from the Galactic All Sky Survey
\citep[GASS,][]{Kalberla2015} as tracers of the dust intensity
structures and using a phenomenological description of the mean
Galactic magnetic field, \citet{Ghosh2017} were able to reproduce the
{\it Planck} dust observations at 353 GHz for the southern Galactic cap.

LOFAR radio-polarimetric observations in a field toward 3C~196
show striking filamentary structures in polarized intensity. This is one
of three primary fields of the LOFAR-Epoch of Reionization key science
project. It shows many degrees long filamentary structures, the most
striking of which is remarkably straight, at a Faraday depth of +0.5 rad
m$^{-2}$, oriented in equatorial coordinates in an approximately 
north-south direction and parallel to the Galactic plane
\citep{Jelic2015}. If this filament is assumed to lie within the Local
Bubble \citep{Lallement2014}, it shows an excess in thermal electron
density compared to its surroundings. Radio-polarimetric
depolarization canals appear to define boundaries around this
filamentary structure, and are most probably the result of beam
depolarization due to discontinuities in polarization angle orientation
\citep[][Sect. 5]{Jelic2015}. These filamentary structures are
surprisingly well correlated with the magnetic field orientation, probed
by the {\it Planck} satellite \citep{Zaroubi2015}.

Characteristic for a radio-polarimetric depolarization canal is that the
observed polarized intensity falls to zero; the polarization angle
changes by 90\deg across the canal which is observed to be only a
telescope beam wide. In case of the fields studied here and in
\citep[][Paper I]{Kalberla2016b} , many of the observed
radio-polarimetric depolarization canals are rather straight with well
defined preferential directions.

Radio-polarimetric depolarization canals are believed to have
heterogeneous causes. Canals can be created by a boundary between two
magneto-ionized regions with different properties and-or strong rotation
measure (RM) gradients, that cause a 90-degree polarization angle change
between the regions \citep{HaverkornHeitsch2004}. These canals are
characterized by a location-independence as a function of frequency and
trace lines of constant RM gradient. The canals around the straight
filaments in the LOFAR 3C~196 observations are likely caused by this
effect. 

Radio-polarimetric depolarization canals due to fortuitous properties of
the magneto-ionized medium completely nulling polarization along the
line of sight (differential Faraday rotation) have been called Faraday
ghosts \citep{Shukurov2003}. These should only occur in very homogeneous
media, change location with frequency, and trace lines of constant
RM. In a turbulent, ionized medium in which synchrotron emission and
Faraday rotation are mixed, the situation is more complex.
\citet{Fletcher2007} argue that straighter canals may arise from RM
discontinuities in the diffuse ISM, in particular shocks, and the
twisting canals have a different origin such as differential Faraday
rotation.

We recently explored filamentary structures and anisotropies in the
power distribution in \hi~gas on scales of arcminutes to a few degrees
in the field toward 3C~196 (\citetalias{Kalberla2016b}).  There,
we reported on strong anisotropies in the \hi~distribution that are best
described as anisotropies in the power distribution for narrow
ranges in velocity.  For a narrow range in position angle, oriented
perpendicular to the filamentary structures and the mean magnetic
fields, the spectral power (measured in the Fourier plane) is on average
more than an order of magnitude higher than parallel. These
observational results are consistent with predictions by
\citet{Goldreich1995}; the scale dependent anisotropy of the
  turbulence increases with spatial frequency but the spectral power
distribution orthogonal to the filamentary structures is left nearly
unaffected.

With respect to some aspects however, the observational findings do not
agree with the theoretical expectations. \citet{Kandel2016} extended
recently the velocity channel analysis (VCA), introduced by
\citet{Lazarian2000}, and found that for magneto hydrodynamical (MHD)
turbulence anisotropies should increase proportional to the thickness of
the velocity slice.  This proposal could not be confirmed by
observations \citepalias{Kalberla2016b}, anisotropies tend to be best
defined for narrow velocity intervals. The observations show in general
that the strongest anisotropies are associated with rather cold
filamentary \hi~structures. 

Absorption measurements against continuum background sources are
required to determine the temperature of the CNM
\citep{Dickey1990,Kalberla2009}. In practice the number of available
sufficiently strong continuum sources is however rather limited for a
complete census of the CNM \citep[e.g.,][]{Heiles2003}. But in case that
the optical depth of the CNM is not too high ($\tau ~\la 0.5$) it is
possible to use alternative methods to determine upper limits of the
kinetic temperature from line widths. The \hi~emission data are first
filtered by unsharp masking (USM)\footnote[1]{We generated USM maps by
  subtracting from the observed $T_{\rm B}$ distribution a smoothed
  brightness temperature distribution with an effective resolution of
  0\fdg5; spatial frequencies $ k < 0.033$ arcmin$^{-1}$ are attenuated
  this way.}, afterwards Doppler temperatures $T_{\rm D}$ are derived
from line widths, see Sects. 3, 5.8 and 5.9 of \citet{Kalberla2016} for
details.  Typical Doppler temperatures, upper limit to the kinetic or
excitation temperature of the \hi~gas, are $T_{\rm D} \sim 223 $ K. At
typical turbulent CNM Mach numbers of $M_{\rm t} \sim 3.7$, such Doppler
temperatures are characteristic for a thermal CNM gas temperature of $T
\sim 52$ K \citep{Heiles2005,Kalberla2016}. Consequently, radial
velocity channel maps (velocity slices) that are separated by more the 3
\kmss are uncorrelated and anisotropies are traceable for a very limited
number of subsequent velocity channels only. Increasing the velocity
slice thickness in such a situation does not necessarily improve the
signal-to-noise ratio (S/R) of the anisotropic power distribution but
may lead to the opposite result.

It is also striking to find filamentary CNM structures associated with
the magneto-ionic medium. MHD simulations of the interstellar medium
(ISM) by \citet[][their Fig. 8]{Choi2012} showed that thermal conduction
can play an important role in shaping structures formed by thermal
instabilities (TI). Anisotropic conduction in the presence of a regular
magnetic field can strongly affect the shapes and sizes of cold clouds,
possibly leading to thin filamentary \hi~structures as reported first by
\citet{Naomi2006} and later on larger scales by \citet{Clark2014} and
\citet{Kalberla2016}. However, even without magnetic fields the
formation of thin CNM sheets is considered to be feasible
\citep{Vazquez-Semadeni2006}.

Anisotropies in the power distribution of the \hi, analyzed in
\citetalias{Kalberla2016b}, are associated with CNM. \citet{Saury2014}
have shown that turbulent motions of the \hi~cannot provoke the phase
transition from warm neutral medium (WNM) to CNM.  An increase of the
WNM density by at least a factor two to four is needed to induce phase
transitions. Based on the morphology of the CNM clouds it was argued in
\citetalias{Kalberla2016b} that these anisotropies must have been caused
by shocks.  As a follow-up of this conjecture we consider the case
whether thermal instabilities might cause filamentary \hi~structures,
associated with ionized gas layers which are in the presence of a
magnetic field responsible for the linear polarization structures.

We study in detail two fields with prominent radio-polarimetric
depolarization canals aligned with polarized intensity filaments,
located in the constellations of Horologium and Auriga
\citep{Haverkorn2003a,Haverkorn2003b,Haverkorn2003c}. Observations and
data processing are presented in Sect. \ref{Data}. In
Sects. \ref{Horologium} and \ref{Auriga} respectively we derive power
spectra and power anisotropies for the Horologium and Auriga fields. For
several \hi~components we find dominant anisotropies and compare the
corresponding filamentary \hi~structures with images of the
radio-polarimetric Westerbork Synthesis Radio Telescope (WSRT)
data. Anisotropies and power spectral indices show significant
variations, depending on velocity. In Sect. \ref{Spectral_index} we
demonstrate that the spectral index is related to the temperature of the
\hi~gas and the column density ratio between CNM and WNM. We conclude
that phase transitions must be responsible for a steepening of the power
spectra and use in Sect. \ref{TI_Power} a heuristic description for the
changes in the turbulent power distribution caused by thermal
instabilities.  Section \ref{Discussion} discusses several possible
explanations for the observed filamentary structures. We conclude in
Sect. \ref{Summary} that regular magnetic fields may play a significant
role for phase transitions of compressed cold \hi~gas in a sheet-like
geometry.

\section{Data} 
\label{Data}

We compare data from WSRT continuum polarization observations with
\hi~data from the Effelsberg--Bonn \hi~Survey (EBHIS) and describe here
the observations and basics of the data reduction.

\subsection{Continuum polarization observations: Horologium and Auriga}

For the multi-frequency radio-polarimetric observations of the Galactic radio
background in Horologium and Auriga the WSRT was used. These fields were
observed in 8 frequency bands between 325 and 390~MHz simultaneously,
each with a band width of 5~MHz. To obtain a large field of view, and to
reduce off-axis instrumental polarization, the mosaicking technique was
used. In each case six 12 hr periods were observed, resulting in
baseline increments of 12 m with baselines between 36 m and 2700 m. The
resulting resolution is 1\arcmin~ but a Gaussian taper was applied later
to the (u, v)-data to increase the S/R. The derived
maps of linearly polarized intensities Stokes $Q$ and $U$ were used to
compute the polarized intensity and polarization angle, for details see
\citet{Haverkorn2003a,Haverkorn2003b,Haverkorn2003c}.

The Horologium field is a $\sim 5\deg\times7\deg$ field centered at ($l
\sim 137\deg, b \sim 7\deg$), and Auriga is $\sim 5\deg\times7\deg$ in
size, centered at ($l \sim 161\deg, b \sim 16\deg$).  So both fields are
located not far off the Galactic plane in the second Galactic quadrant,
in the Fan Region. The Fan Region is a large region in the sky, in the
range $\ell \sim [120\deg, 170\deg]$ and $b \sim [-5\deg, +20\deg]$ with
remarkably high polarized intensity and regular polarization angle,
interpreted as an especially regular magnetic field structure in that
direction \citep{Spoelstra1984}. The Fan Region has long thought to be a
local structure, although recent work suggests that Fan Region emission
may come from a range of distances out to the Perseus Arm
\citep{Wolleben2006, Hill2017}.  Over much of the Fan Region the
electric vectors are perpendicular to the Galactic plane, indicating
that the mean magnetic field is aligned with the plane.

Both the Horologium and the Auriga field show conspicuous linear
structures in polarized intensity at frequencies around 349~MHz. These
filaments are not present in total (synchrotron) intensity, indicating
that they are caused by Faraday rotation creating small-scale structure
in Stokes parameters Q and U, and not by enhanced filamentary
synchrotron emission. The Auriga field is dominated by linearly
polarized filaments up to 4$^{\circ}$ in length, dominated by two
directions one of which is parallel to the Galactic plane. The most
conspicuous feature in the Horologium field is a ring-like structure
(not discussed here), crossed by long linear polarized filaments, again
directed along the Galactic plane. Both fields show long and narrow
depolarized filaments, always one resolution element wide, called
depolarization canals \citep{Haverkorn2000}. 

\subsection{HI Survey data and processing}
\label{Data_processing}

For the \hi~distribution in direction to these WSRT fields we use the
first data release of the Galactic portion of the Effelsberg--Bonn
\hi~Survey \citep[EBHIS, ][]{Winkel2016a}.  This survey covers the
northern sky for declinations grater than $ -5\deg $ at a velocity
resolution of $\Delta v_{\rm LSR} = 1.44$ \kms. From the original EBHIS
data base, corrected for instrumental baselines, radio interference and
stray radiation, we extract FITS data cubes with an effective rotational
symmetric Gaussian beam-size of 10\farcm8 FWHM \citep{Winkel2016b}. The
brightness temperatures of individual velocity channel maps at this
resolution have rms uncertainties of 90\,mK.

To derive the spectral power distribution we use the same data
processing as described in detail in \citetalias{Kalberla2016b}.  We
first apodize the data with a rotational symmetric 50\% cosine taper
(Tukey) window \citep{Harris1978}. The tapered distribution is Fourier
transformed, the amplitudes are then squared and corrected for the beam
response.  We use polar coordinates $k, \Phi$ to determine dependencies
in the $u,v$ plane with a position angle $\Phi = {\rm
  atan2}(v,u)$. This, also the definition in \citetalias{Kalberla2016b},
differs from the standard north through east definition but we do not
expect conflicts since $\Phi$ is not used for comparison with published
position angles.

\hi~observations are here as usual processed as
position-position-velocity (PPV) data cubes. Individual channels
represent brightness temperatures (or column densities) at constant
velocities with a bandwidth of $\Delta v_{\rm LSR}$. This bandwidth can be
increased by integrating several channel maps.  Power spectra,
calculated from such channel maps, provide the so-called 2D power
distribution from which a 2D spectral index $\gamma$ can be
fit. Throughout this paper we provide without further notification
observed 2D power distributions and accordingly 2D spectral indices.

We derive the average power spectrum $P(k)$ by integrating the 2D power
distribution in annuli of constant spatial frequencies $ k = (k^2_u +
k^2_v)^{1/2} $, fitting the power distribution with a power law of the
form
\begin{equation}
P(k) = c \cdot k^\gamma + N(k),
\label{eq:Pav}
\end{equation}
here $c$ is an arbitrary scaling factor, $\gamma$ is the spectral
index and $N(k)$ the contribution due to instrumental
noise. Subsequently we correct for the contribution $N(k)$ to the power
spectrum by subtracting the matched noise template $N(k)$ as
  described in \citetalias{Kalberla2016b}. The noise stability of the
  EBHIS is excellent, it is not necessary to derive position or time
  dependent templates.

The derived normalized power spectrum $P(k) \propto k^\gamma $ is noise
limited for high spatial frequencies $k$; the limit depends on the 
  S/R of the observations but reflects also the limited spatial
sensitivity of the 100-m telescope. We interpret only data with a 
  S/R of three or better, comparing the noise corrected power $P(k)$
with the matched noise template $N(k)$. This limit, typically close to $
k < 0.07 $ arcmin$^{-1}$, is marked in all plots by a vertical line. The
lowest spatial frequency that we can use for our analysis depends on the
field of view and is $ k \sim 1.2 ~ 10^{-3}$ arcmin$^{-1}$.

To quantify anisotropies, we average data within sectors $\Phi \pm
\Delta\Phi$ to measure the position angle dependent power $P(\Phi,k)$;
we use $\Delta\Phi = 4\deg$. In case of significant anisotropies,
$P(\Phi,k)$ shows well defined maxima in the $u, v$ plane at similar
position angles $\Phi_{\perp}$ over a range of spatial frequencies
\citepalias[Fig. 7 of][]{Kalberla2016b}. Since the power spectrum is
defined in the $u,v$ plane, $\Phi_{\perp}$ is oriented perpendicular to
the position angle of filamentary structures in the image plane
\citepalias[][Figs. 3 to 5]{Kalberla2016b}. In presence of a magnetic
field the propagation of turbulence is affected by the field direction
and eddies are elongated along the magnetic field lines at
$\Phi_{\parallel}$\citep{Goldreich1995}.

As a measure of the local anisotropies we define the ratio between
maximum and minimum power, at position angles $\Phi_{\perp}$ and
$\Phi_{\parallel}$,
\begin{eqnarray}\nonumber
Q(k,v_{\rm LSR}) = P(\Phi_{\perp},k,v_{\rm LSR}) / P(\Phi_{\parallel},k,v_{\rm LSR}) \\
= P_{\perp}(k,v_{\rm LSR})/P_{\parallel}(k,v_{\rm LSR}),
\label{eq:Q}
\end{eqnarray}
where $\Phi_{\parallel} = \Phi_{\perp} + 90\deg$.

There are two different ways to derive characteristic power
anisotropies, the average $Q_{\rm aver}$, defined as the geometrical
mean anisotropy over a range in spatial frequencies and alternatively
$Q_{\rm peak}$, the peak anisotropy at a particular spatial
frequency. Both anisotropy measures should result in similar position
angles. The magnitude of $Q_{\rm aver}$ is arbitrary since it depends on
the spatial frequency range used. We use $Q_{\rm aver}$ for an automated
identification of velocity channels with significant anisotropies
according to Eq. \ref{eq:Q}. Based on this we search then for peak
anisotropies at particular interesting velocity channels. 

Our data processing methods, outlined above and in Paper I, are
discussed in more detail and compared with robust methods in Appendix
\ref{MBMpower}.

\section{Horologium}
\label{Horologium}

In this Section we derive parameters for the local \hi~distribution in
the Horologium region (RA = 48\deg, DEC = 66\deg (B1950.0), $l \sim
137\deg, b \sim 7\deg$) that may be compared with observations of the
linearly polarized component of the diffuse galactic radio background
\citep{Haverkorn2003b,Haverkorn2003c}. We analyze EBHIS data within a
diameter of 13\fdg7 (at 100\% taper). 

\begin{figure}[htbp]  
   \centering
   \includegraphics[width=6.5cm,angle=-90]{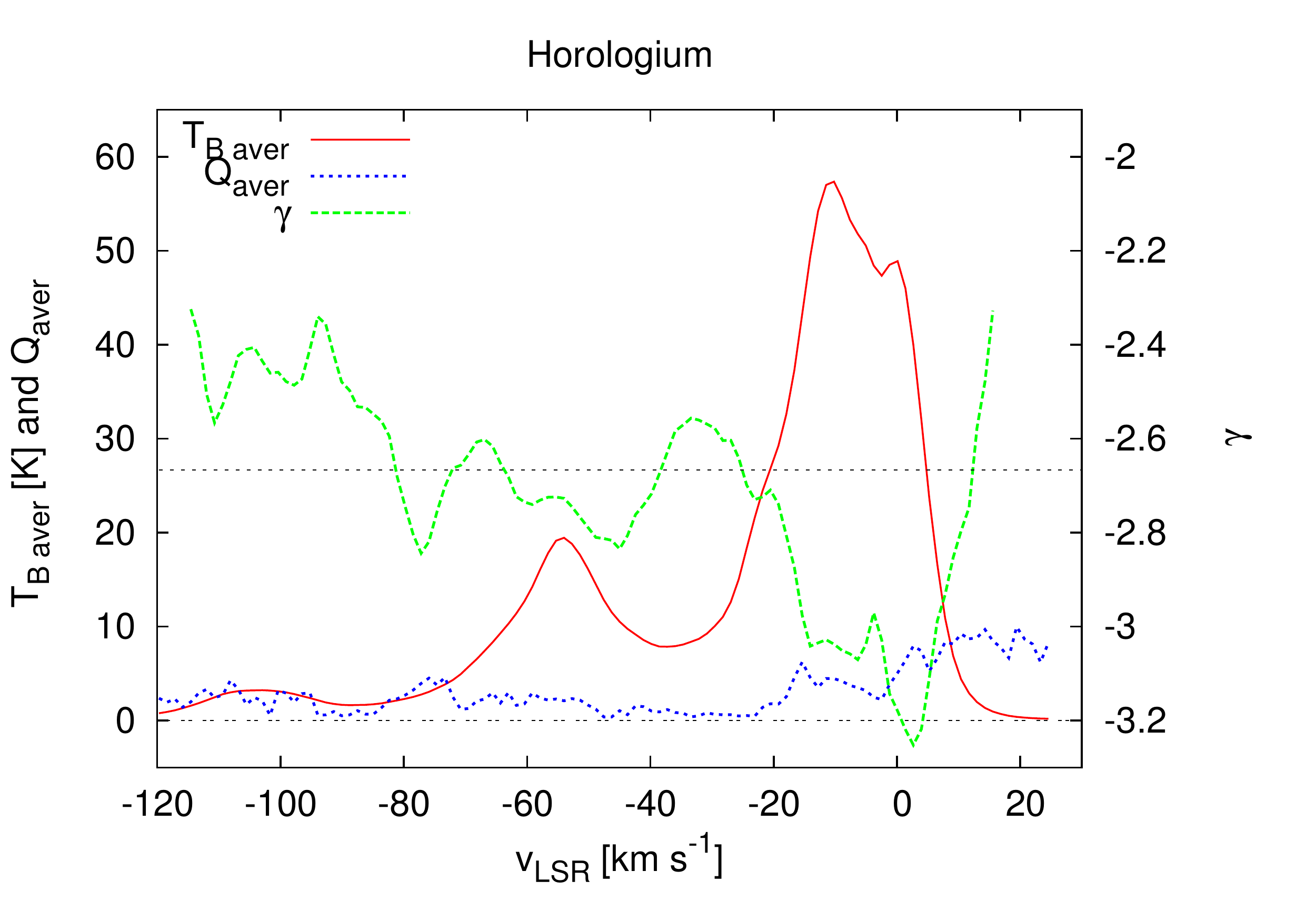}
   \caption{Comparison between average brightness temperature $T_{\rm
       B\, aver}$ (red), anisotropy factor $Q_{\rm aver}$ (blue) 
       for $0.002 < k < 0.015$ arcmin$^{-1}$, and spectral index
     $\gamma$ (green) for the Horologium field. The upper horizontal
     black dotted line indicates the Kolmogorov spectral index of $\gamma =
     -8/3$, the lower dash-dotted line $Q_{\rm aver} = 0$.  } 
   \label{Fig_HO_overview}
\end{figure}

\subsection{Average \hi~properties}
\label{Horo_aver} 

Figure \ref{Fig_HO_overview} gives an overview over global properties of
the \hi~gas in this region. The weighted mean brightness temperature
$T_{\rm B\, aver}$ (see \citetalias[][Eq. 4]{Kalberla2016b}), averaged
after apodization, shows for the local gas a broad emission line with
several components, peaking at a velocity of $v_{\rm LSR} = -10.2$ \kms.
 
From the rotation curve of the Milky Way we may map velocities to
distances \citep[e.g.,][]{Kalberla2008}.\footnote[2]{a simple tool is
  available at
  \url{https://www.astro.uni-bonn.de/hisurvey/euhou/LABprofile/}.  The
  \hi~components with $-18 \la v_{\rm LSR} \la 0 $ \kmss can accordingly
  be at distances up to 1.5 kpc.  In addition to the local gas we
  observe a component at a velocity of $v_{\rm LSR} = -54 $ \kms. This
  gas is according to the rotation curve located at a distance of 5.5
  kpc and belongs probably to the Perseus arm.}  The galactocentric
distance is $R \sim 13 $ kpc, its height above mid-plane $z \sim 0.7 $
kpc. In addition there is a faint high velocity component at $v_{\rm
  LSR} \sim -100 $ \kms. Using a dynamical distance estimate from
\citet{Kalberla2008}, this gas is located beyond the outer arm at a
distance of 19 kpc with a galactocentric radius of 26 kpc
\citep{Levine2006}. The anisotropies at high velocities are discussed in
the Appendix \ref{appendix}.

In Fig. \ref{Fig_HO_overview} we show the average (position
angle independent) spectral index $\gamma$ of the power distribution
according to Eq. \ref{eq:Pav}. For the local emission at $v_{\rm LSR} =
2.7 $ \kmss we find a well defined minimum at $\gamma = -3.25 \pm 0.03
$. The derivation of the anisotropy factor $Q_{\rm aver}$ is
explained in Sect. \ref{Horo_power}

\begin{figure}[thp]  
   \centering
   \includegraphics[width=6.5cm,angle=-90]{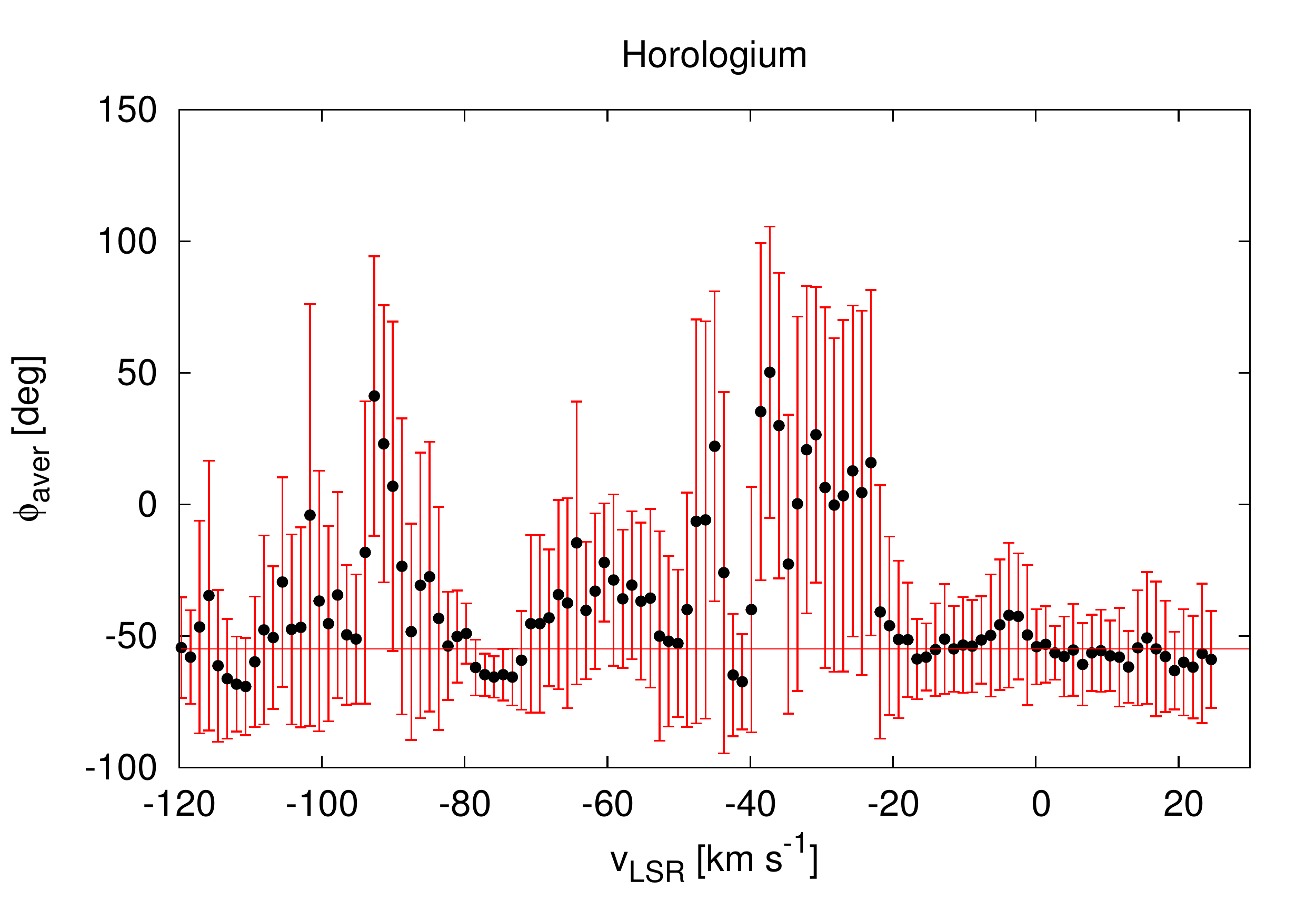}
   \caption{Average position angles $\Phi_{\rm aver}$ calculated for $
     0.002 < k < 0.015 $ arcmin$^{-1}$ and associated one $\sigma$ rms
     scatter for the Horologium field. The position angle $\Phi =
     -58.5\degr$, perpendicular to the position angle of the
     radio-polarimetric depolarization canals that are aligned parallel
     to the Galactic plane, is indicated with a horizontal line. }
   \label{Fig_HO_overview_angle}
\end{figure}

\begin{figure}[tbp]  
   \centering
   \includegraphics[width=6.5cm,angle=-90]{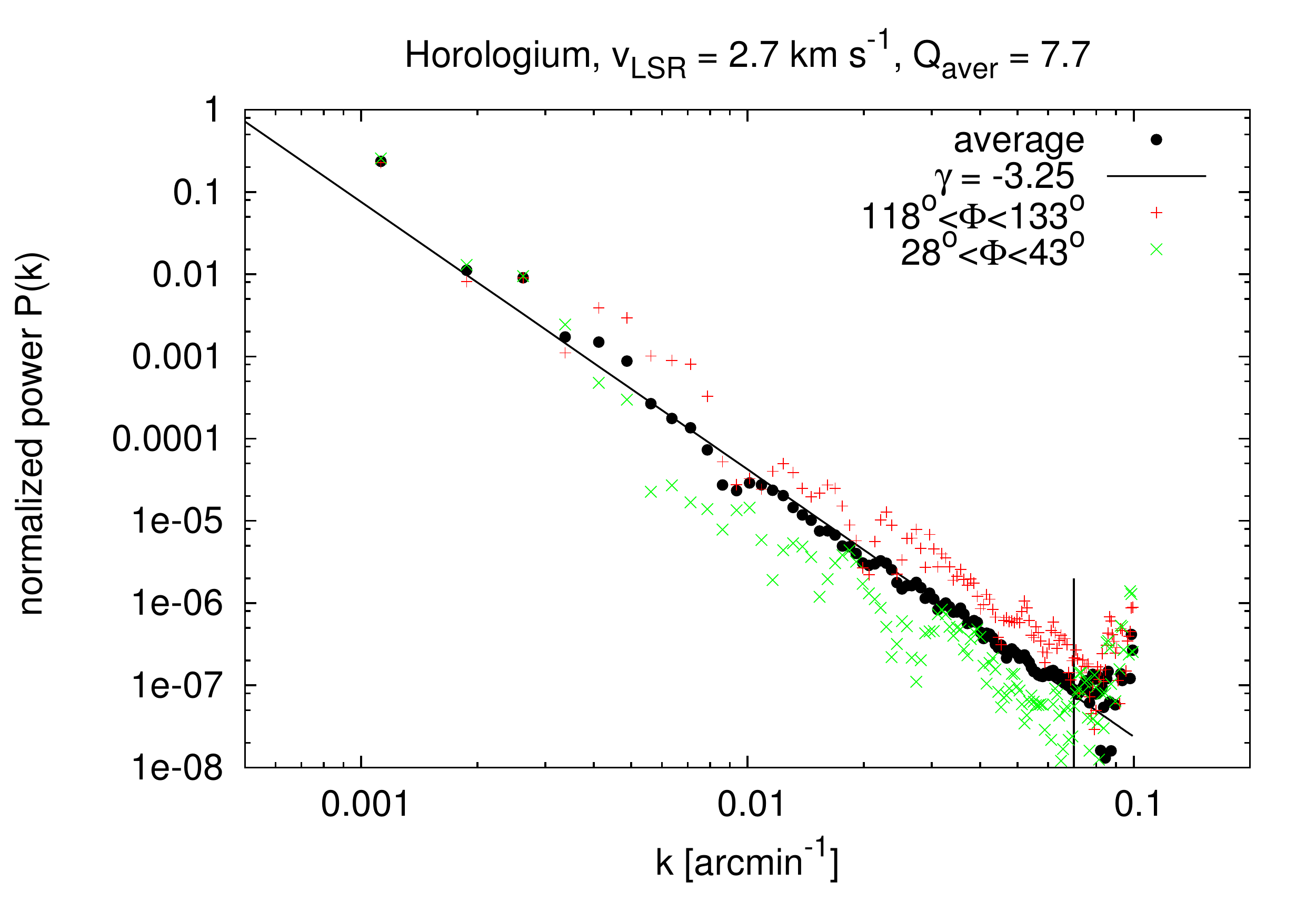}
   \caption{Average power spectrum observed for $v_{\rm LSR} = 2.7 $
     \kmss (black dots) and fit power law with $\gamma = -3.25 \pm 0.03
     $ for $ k < 0.07$ arcmin$^{-1}$ (vertical line). In addition the
     power spectrum for $118\degr < \Phi < 133\degr $ (red) and $28\degr
     < \Phi < 43\degr $ (green) is given. The average anisotropy factor
     for $0.002 < k < 0.015$ arcmin$^{-1}$ is $Q_{\rm aver} = 7.7$.
   }
   \label{Fig_spec_HO_22}
\end{figure}

\begin{figure}[tbp]  
   \centering
   \includegraphics[width=6.5cm,angle=-90]{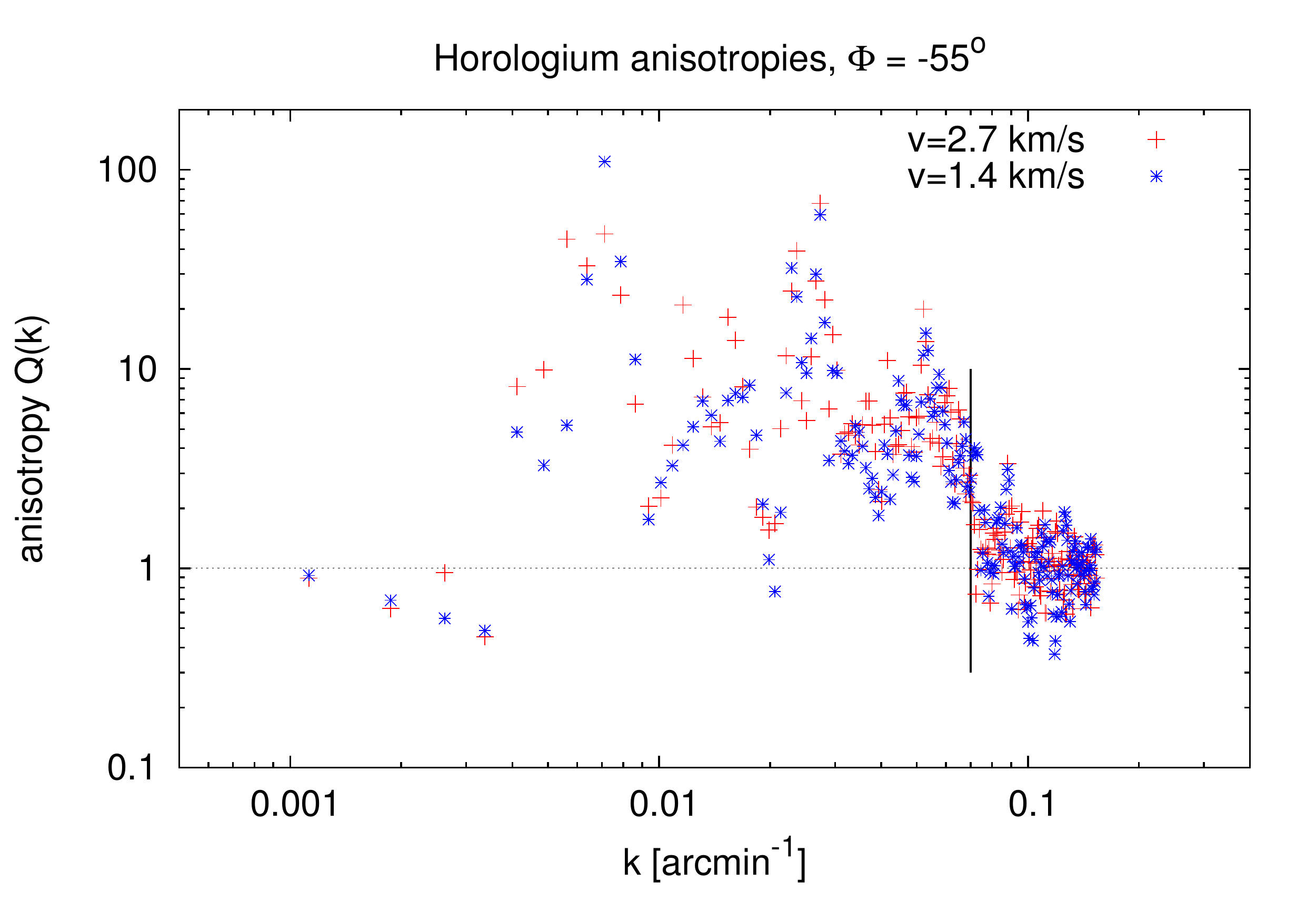}
   \caption{Anisotropies $Q(k)$ for the channels at $ v_{\rm LSR} =
     2.7 $ \kms (red) and $ v_{\rm LSR} = 1.4 $ \kms (blue).  }
   \label{Fig_Q_spec_HO_22}
\end{figure}

\begin{figure}[tbp]  
   \centering
   \includegraphics[width=7.5cm,angle=-90]{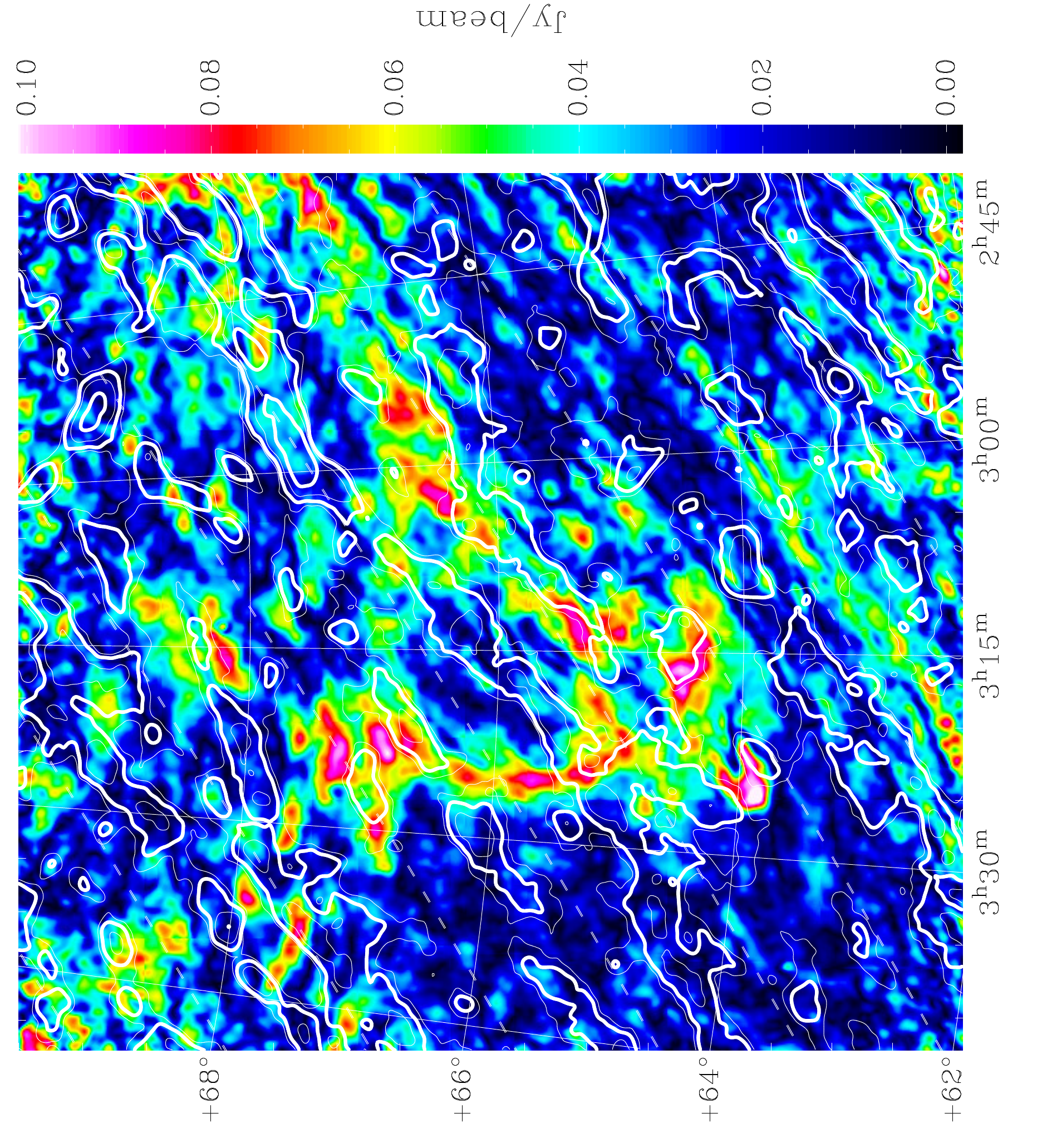}
   \caption{Polarized intensity map of Horologium in equatorial
     coordinates (B1950.0), observed at 349 MHz with the WSRT. The
     filamentary features from an \hi~USM map at $v_{\rm LSR} = 2.7 $
     \kmss are overlaid with contours of 0.1, 1, 2.5 and 5 K.  The
     dashed lines are parallel to the Galactic plane, in steps of
     $\Delta b = 1\deg$. }
   \label{Fig_Overlay_HO_2.7}
\end{figure}

\subsection{Position angle dependencies on $v_{\rm LSR}$}
\label{Horo_power} 

Local anisotropies and average position angles $\Phi_{\rm aver}(v_{\rm
  LSR})$, defined as averages over $\Phi_{\perp}(v_{\rm LSR})$, were
determined as described in Sect. 3.2 of \citetalias{Kalberla2016b}, see
there Fig. 7.\footnote[3]{angles are circular quantities, for the
  processing see e.g.,
  \url{https://en.wikipedia.org/wiki/Mean_of_circular_quantities}.  When
  calculating the dispersion, the nearest of both possible angular
  distance needs to be taken. } In spatial frequencies we used the range
$0.007 < k < 0.07$ arcmin$^{-1}$, corresponding to angular scales of
$140\arcmin \ga 1/k \ga 14\arcmin$.  In the velocity range $ -120 <
v_{\rm LSR} < -35 $ \kmss we found anisotropies predominantly at low
spatial frequencies, corresponding to large scale features. We repeated
therefore the determination within a range $0.002 < k < 0.015$
arcmin$^{-1}$. This choice resulted in better defined position angles
with lower dispersion at high velocities but left the results for the
local gas nearly unchanged.  Fig. \ref{Fig_HO_overview_angle} displays
the average position angles $\Phi_{\rm aver}$ and the associated
standard deviation.  For $ -20 < v_{\rm LSR} < 20 $ \kmss the position
angle is essentially constant, $\Phi_{\rm aver} \sim -55\degr$,
indicated in Fig. \ref{Fig_HO_overview_angle} with a horizontal
line. $\Phi = -55\degr$ is oriented perpendicular to the position angle
of the WSRT depolarization canals and the Galactic plane
\citep[][Fig. 3]{Haverkorn2003c}. This implies that filamentary
structures in the image plane over this velocity range are elongated
predominantly parallel to the Galactic plane, see
Figs. \ref{Fig_Overlay_HO_2.7} and \ref{Fig_Overlay_HO-16.6}.

From the position angles $\Phi_{\rm aver}(v_{\rm LSR})$ we derive
average anisotropies $Q_{\rm aver}$. The result is plotted in
Fig. \ref{Fig_HO_overview}. The Horologium field shows moderate
anisotropies with $Q_{\rm aver} \sim 7.7$ at $v_{\rm LSR} = 2.7$ \kms.
The anisotropies appear to increase for $v_{\rm LSR} \ga 0 $ \kms,
possibly related to a CNM component with a steep spectral index.

\subsection{Anisotropies at $v_{\rm LSR} = 2.7 $ {\rm km~s$^{-1}$} }

\label{Horo_2.7} 

The \hi~distribution at $v_{\rm LSR} = 2.7 $ \kmss is of particular
interest since the uncertainties in the average position angle are low
and the spectral index $\gamma = -3.25 \pm 0.03 $ is quite steep.
Anisotropic power spectra for this channel map are plotted in
Fig. \ref{Fig_spec_HO_22}. We find frequently that anisotropies can
change significantly from one velocity channel to another, although some
of the structures may be preserved. We therefore compare the
anisotropies of two neighbor channels (in comparing, please take the
logarithmic scale into account). Fig. \ref{Fig_Q_spec_HO_22} shows for
comparison the anisotropies $Q$ at $v_{\rm LSR} = 2.7 $ \kmss and
$v_{\rm LSR} = 1.4 $ \kms.

The velocity channel at $v_{\rm LSR} = 2.7 $ \kmss has the best defined
position angle with the lowest dispersion and we expect accordingly that
the \hi~distribution should show filamentary structures.  Small scale
structures are best visualized from USM maps, as introduced by
\citet{Kalberla2016} and demonstrated in \citetalias{Kalberla2016b}. In
Fig. \ref{Fig_Overlay_HO_2.7} we compare the USM filamentary
\hi~structures at $v_{\rm LSR} = 2.7 $ \kms, using isophotes, with the
color coded map of the polarized intensity observed at 349 MHz.  The
filamentary \hi~structures are preferentially oriented parallel to the
Galactic plane, in excellent agreement
with the orientation obtained from power anisotropies. The ring-like
structure close to the center of the WSRT map is missing in the \hi~data
and not discussed here.

\subsection{Anisotropies at $v_{\rm LSR} = -16.6 $ {\rm km~s$^{-1}$ } }
\label{Horo-16.6} 

\begin{figure}[tbp]  
   \centering
   \includegraphics[width=6.5cm,angle=-90]{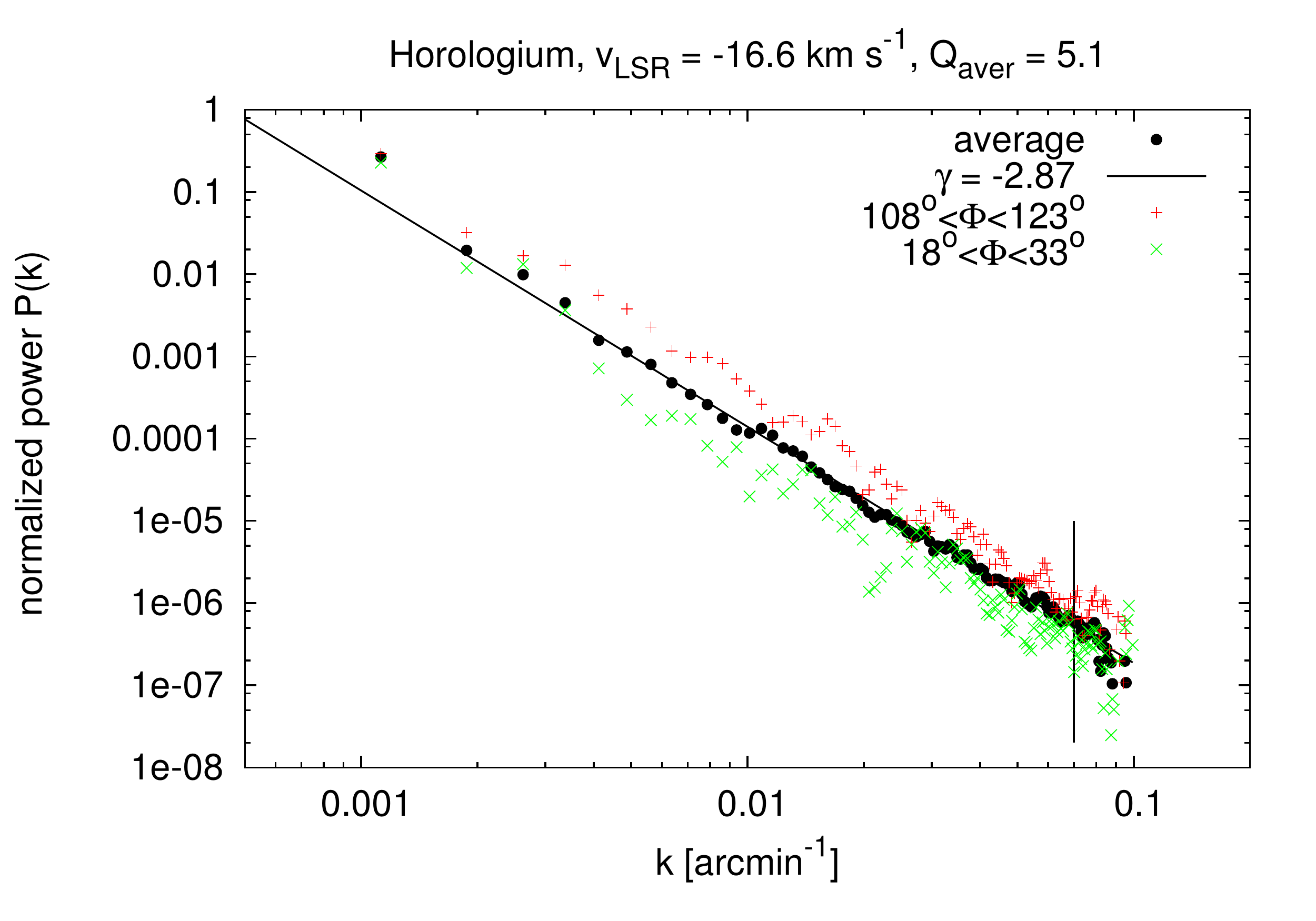}
   \caption{Average power spectrum observed for $v_{\rm LSR} = -16.6 $
     \kmss (black dots) and fit power law with $\gamma = -2.87 \pm 0.02$
     arcmin$^{-1}$  for $ k < 0.07$ (vertical line). In addition the
     power spectrum for $108\degr < \Phi < 123\degr $ (red) and $18\degr
     < \Phi < 33\degr $ (green) is given. The average anisotropy factor
     for $0.002 < k < 0.015$ arcmin$^{-1}$ is $Q_{\rm aver} =
     5.1$.  
  }
   \label{Fig_spec_HO_HV_81}
\end{figure}

\begin{figure}[tbp]  
   \centering
   \includegraphics[width=7.5cm,angle=-90]{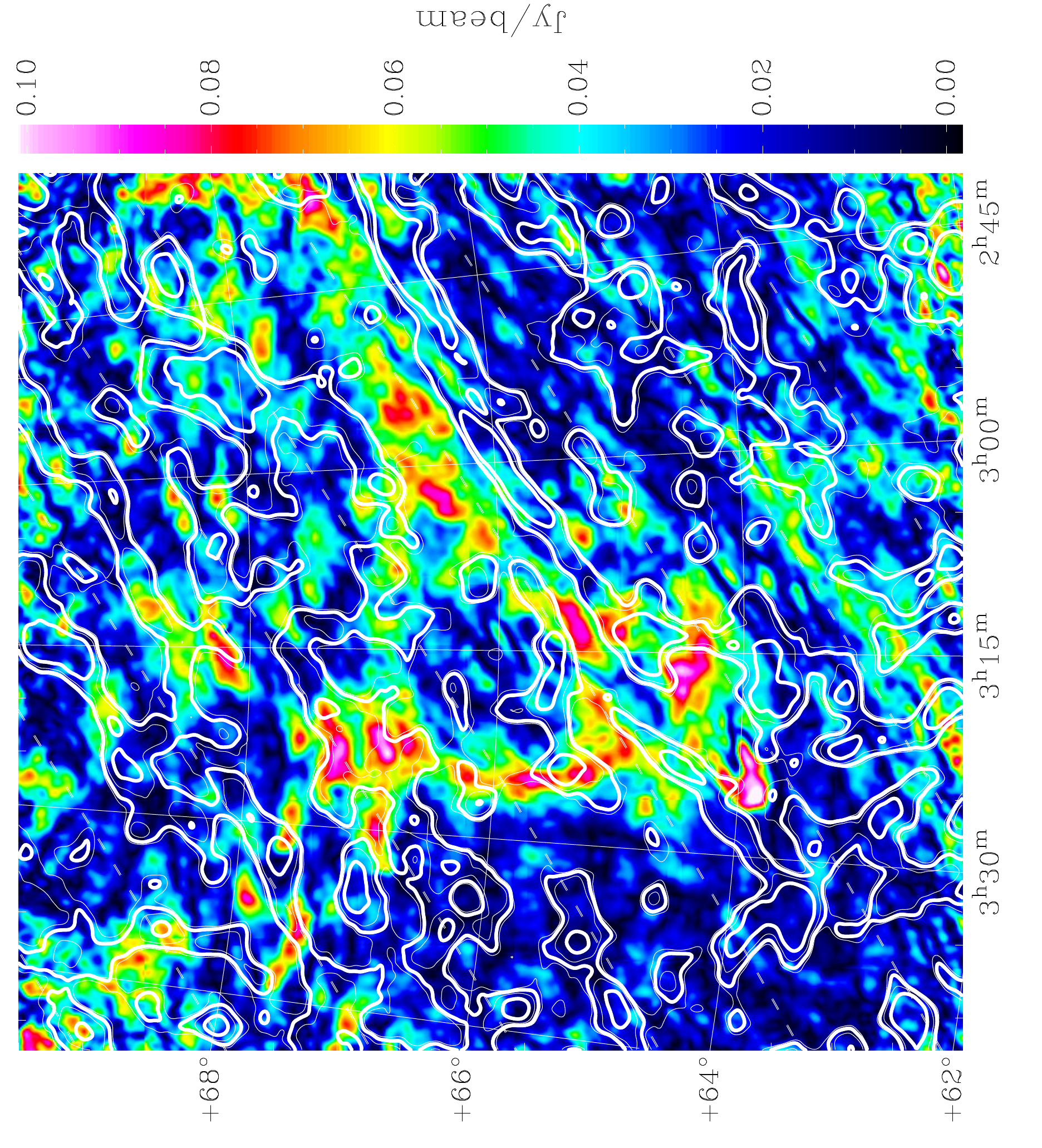}
   \caption{Polarized intensity map of Horologium in equatorial
     coordinates (B1950.0), observed at 349 MHz with the WSRT. The
     filamentary features from an \hi~USM map at $v_{\rm LSR} = -16.6 $
     \kmss are overlaid with contours of 0.1, 1, 2.5 and 5 K. The dashed
     lines are parallel to the Galactic plane, in steps of $\Delta b =
     1\deg$.  }
   \label{Fig_Overlay_HO-16.6}
\end{figure}

All channel maps at velocities $ -23.1 < v_{\rm LSR} < 5.25 $ \kmss show
filamentary structures that are aligned approximately parallel to the
Galactic plane. At velocities $v_{\rm LSR} = -16.6 $ \kmss we found
particular interesting features. We display in
Fig. \ref{Fig_spec_HO_HV_81} the average power spectrum, the fit
resulting in $\gamma = -2.87 \pm 0.02 $ arcmin$^{-1}$, and the
associated anisotropic power spectra. Figure \ref{Fig_Overlay_HO-16.6}
compares the USM \hi~distribution at this velocity (contours) with the
polarized intensity observed at 349 MHz with the WSRT (color coded
intensities). Near the field center (at RA = 3$^{\rm h}$ , DEC = 66\deg)
we observe a filamentary \hi~structure extending over a length of 4\deg
along one of the radio-polarimetric depolarization canals and parallel
to the Galactic plane.

\section{Auriga}
\label{Auriga}

This section deals with the analysis of the Auriga region
\citep{Haverkorn2003a,Haverkorn2003b} at RA = 92\fdg5, DEC = 52\fdg5,
(B1950.0), $l \sim 161\deg, b \sim 16\deg$. We used again the EBHIS to
generate a fits data cube with a diameter of 13\fdg8 (at 100\% taper). 

\begin{figure}[thbp]  
   \centering
   \includegraphics[width=6.5cm,angle=-90]{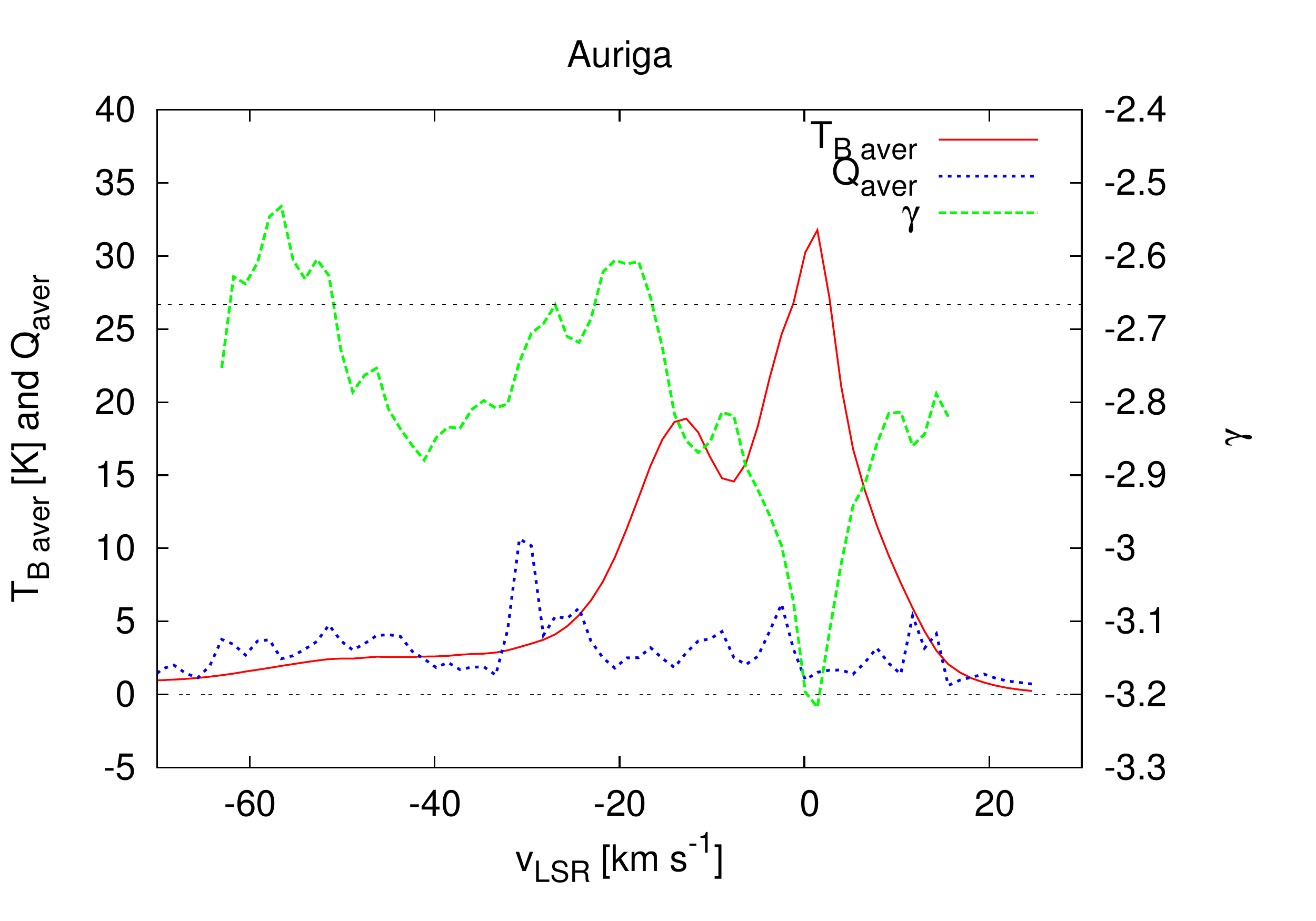}
   \caption{Comparison between the average apodized brightness
     temperature profile $T_{\rm B\, aver}$ (red) for the Auriga field,
     derived average anisotropy factor $Q_{\rm aver}$ for $0.007 < k <
     0.07$ arcmin$^{-1}$ (blue), and spectral index $\gamma$
     (green). The upper horizontal
     black dotted line indicates the Kolmogorov spectral index of $\gamma =
     -8/3$, the lower dash-dotted line $Q_{\rm aver} = 0$. }
   \label{Fig_AU_overview}
\end{figure}

\subsection{Average \hi~properties}
\label{Auriga_aver} 

Figure \ref{Fig_AU_overview} displays the global properties of the
\hi~gas in this region. The weighted mean brightness temperature $T_{\rm
  B\, aver}$, averaged after apodization, shows two major components,
peaking at velocities of $v_{\rm LSR} = 1.4 $ and $v_{\rm LSR} = -12.8$
\kmss respectively, with a long extended wing to more negative
velocities.  For comparison we give the average (position angle
independent) spectral index $\gamma$ of the power distribution for
individual velocity channels according to Eq. \ref{eq:Pav}. For the main
emission line we find at $v_{\rm LSR} = 1.4 $ \kms, $\gamma = -3.22 \pm
0.03 $ while $\gamma \sim -2.85 \pm 0.03 $ at the secondary peak.

\begin{figure}[thp]  
   \centering
   \includegraphics[width=6.5cm,angle=-90]{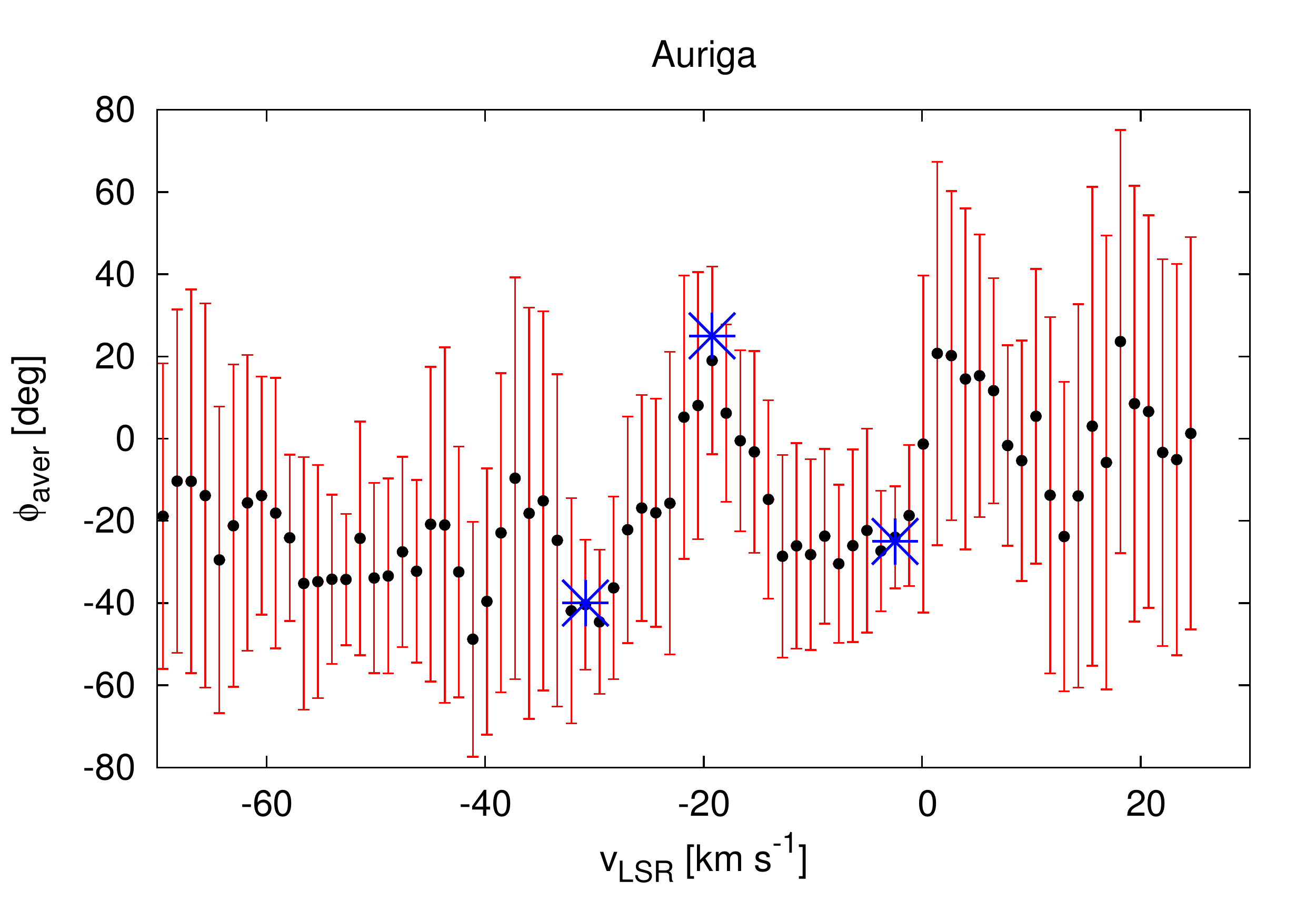}
   \caption{Average position angles $\Phi_{\rm aver}$ calculated for $
     0.007 < k < 0.07 $ arcmin$^{-1}$ and associated one $\sigma$ 
       rms scatter for the Auriga field. The blue asterisks
       denote peak anisotropies of the three CNM filamentary structures
       that we study in detail.}
   \label{Fig_AU_overview_angle}
\end{figure}

\subsection{Position angle dependencies on $v_{\rm LSR}$}
\label{Auriga_power}

To determine position angle dependencies of the power distribution on
the radial velocity, we calculate for each channel the average
anisotropy $Q_{\rm aver}$ in the range $0.007 < k < 0.07$ arcmin$^{-1}$
as well as the associated position angle $\Phi_{\rm aver}$ and its
standard deviation.

$\Phi_{\rm aver}$ is shown in Fig. \ref{Fig_AU_overview_angle}, the
corresponding $Q_{\rm aver}$ is displayed in
Fig. \ref{Fig_AU_overview}. The $\Phi_{\rm aver}$ distribution differs
significantly from Fig. \ref{Fig_HO_overview_angle}. For Horologium we
found little changes of $\Phi_{\rm aver}$ over the main emission
line. In Auriga we have distinct different $\Phi_{\rm aver}$ for
individual \hi~components.  $\Phi_{\rm aver}$ is ill-defined with large
uncertainties for positive velocities. The best defined position angle
with the lowest uncertainty is $\Phi_{\rm aver} = -25\degr \pm 12\degr $
at $v_{\rm LSR} = -2.5 $ \kmss and is marked in
Fig. \ref{Fig_AU_overview_angle}.  The second feature that we will
discuss is defined by a pronounced deviation from the vicinity with
$\Phi_{\rm aver} = 19\degr \pm 22\degr $ at $v_{\rm LSR} = -19.1 $
\kms. The peak anisotropy at this velocity channel is at $\Phi_{\rm
  peak} = 25\degr$, also marked. From the Galactic
rotation curve this component may be at a distance of 3.5 kpc. The third
feature at $v_{\rm LSR} = -30.8 $ \kmss shows a marked anisotropy with
$\Phi_{\rm aver} = -40\degr \pm 16\degr $ although the emission is low,
$T_{\rm B\, aver} = 3.2$ K.  According to the rotation curve this gas
may be at a distance of 7.5 kpc.

\begin{figure}[hbp]  
   \centering
   \includegraphics[width=6.5cm,angle=-90]{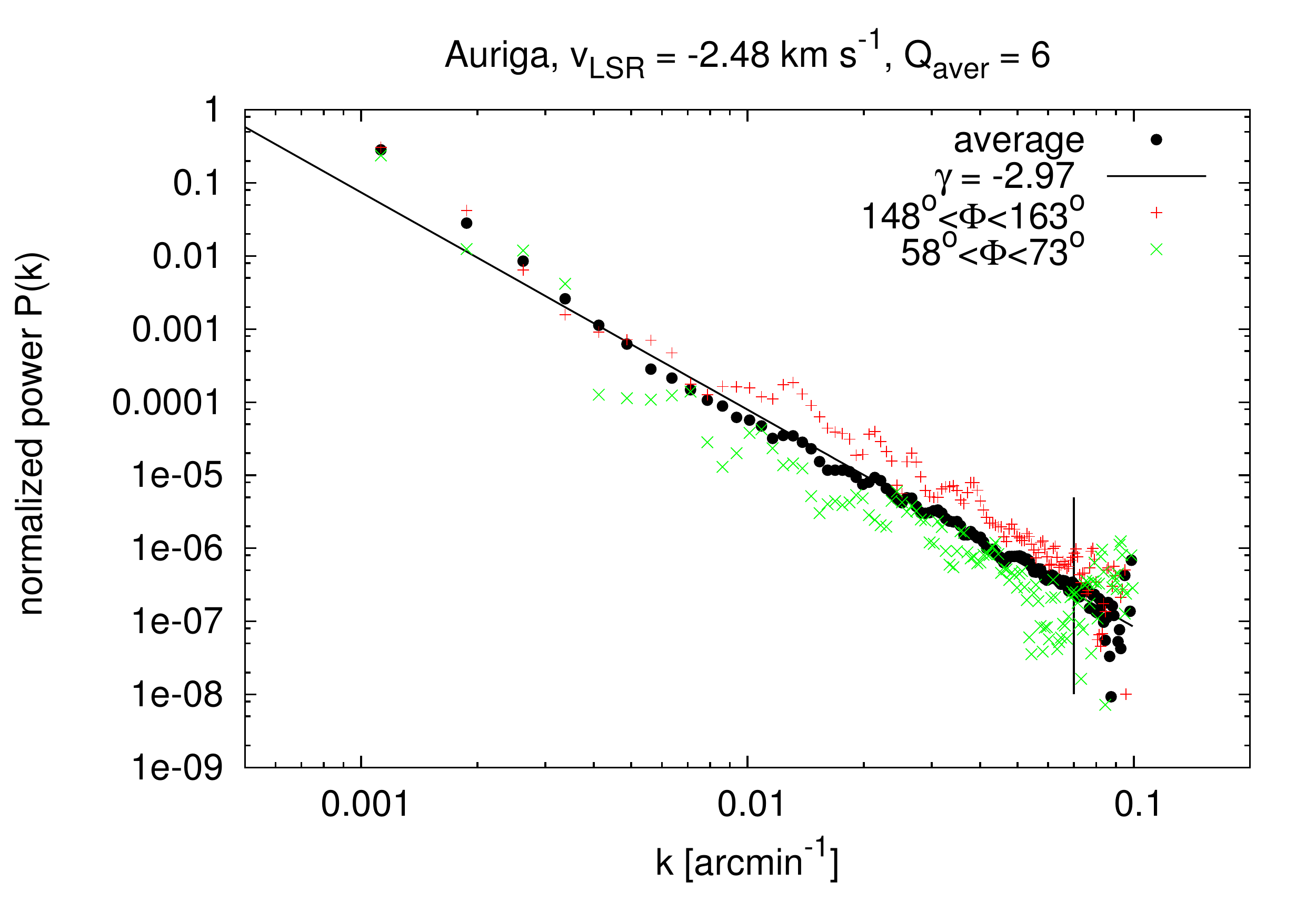}
   \caption{Average power spectrum observed for $v_{\rm LSR} = -2.5$
     \kmss (black dots) and fit power law with $\gamma = -2.97 \pm 0.03
     $ for $ k < 0.07$ arcmin$^{-1}$ (vertical line). In addition the
     power spectrum for $148\degr < \Phi < 163\degr $ (red) and $58\degr
     < \Phi < 73\degr $ (green) is given. The average anisotropy factor
     for $0.007 < k < 0.07$ arcmin$^{-1}$ is $Q_{\rm aver} = 6$. }
   \label{Fig_spec_au_18}
\end{figure}

\subsection{Anisotropies at $v_{\rm LSR} = -2.5 $ {\rm km~s$^{-1}$} }
\label{Auriga-2.5}

Figure \ref{Fig_spec_au_18} shows the \hi~power spectra at $v_{\rm LSR}
= -2.5 $ \kmss with the power law fit to the average (isotropic) power
spectrum. We fit $\gamma = -2.97 \pm 0.03 $ for $ 0.003 < k
< 0.07$ arcmin$^{-1}$. The position dependent power spectra at the
position angle with the most significant anisotropies 
appear in logarithmic presentation well displaced from the isotropic
distribution. The spectral shape for $ k < 0.003 $ arcmin$^{-1}$ is
unusual and not understood but will be discussed in
Sect. \ref{TI_deviations}. Instrumental problems of the EBHIS in this
range are unexpected and should be three orders of magnitude below the
observed power \citepalias[][Fig. 6]{Kalberla2016b}. Furthermore the
most important instrumental errors due to radio frequency interference,
baseline defects or stray radiation problems should be strongly
anisotropic but this is not observed.

\begin{figure}[htp]  
   \centering
   \includegraphics[width=6.5cm,angle=-90]{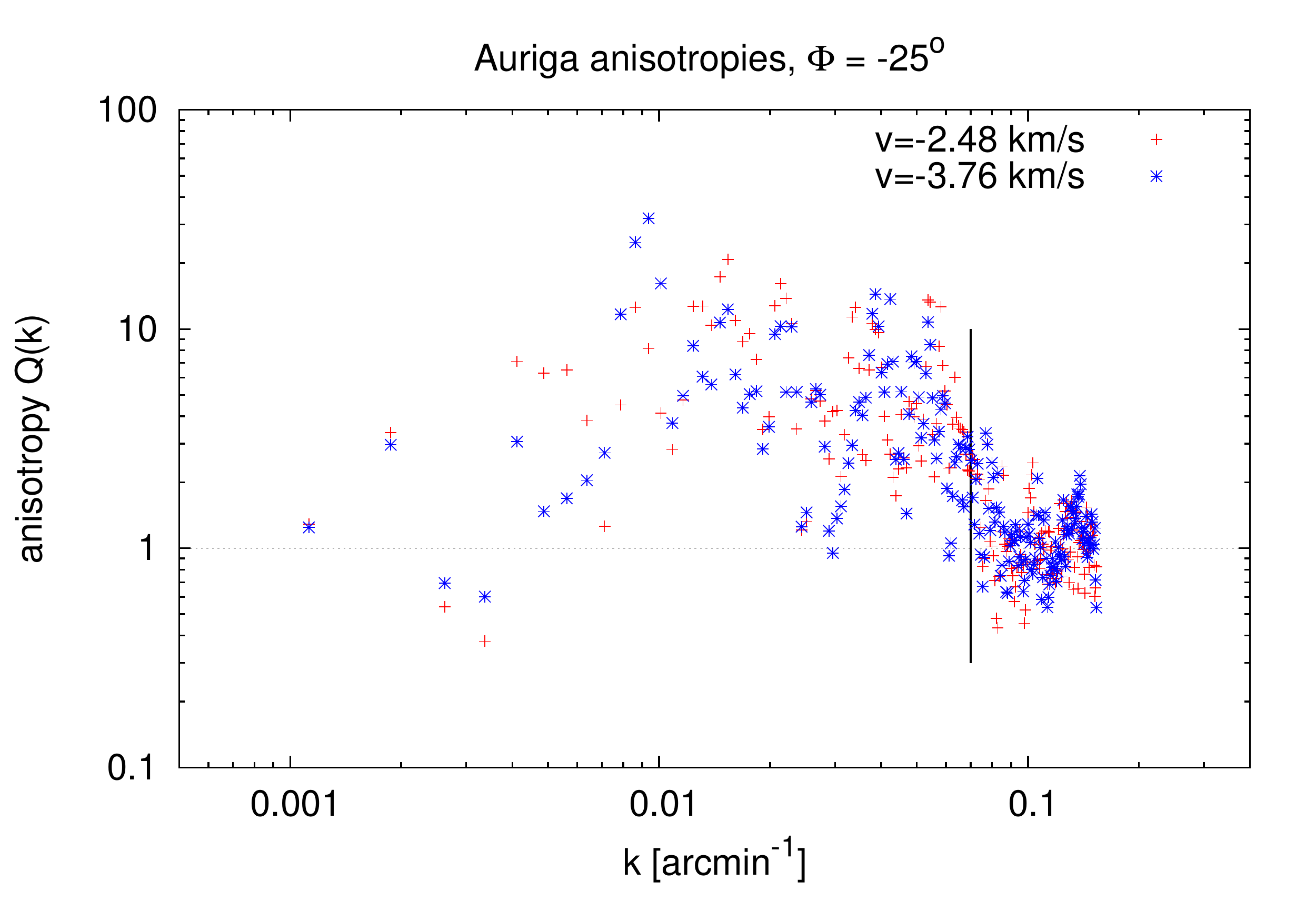}
   \caption{Anisotropies $Q(k)$ for the channels at $ v_{\rm LSR} =
     -2.5 $ \kms (red) and $ v_{\rm LSR} = -3.76 $ \kms (blue).  }
   \label{Fig_au_17_18}
\end{figure}

\begin{figure}[htp]  
   \centering
   \includegraphics[width=8.5cm]{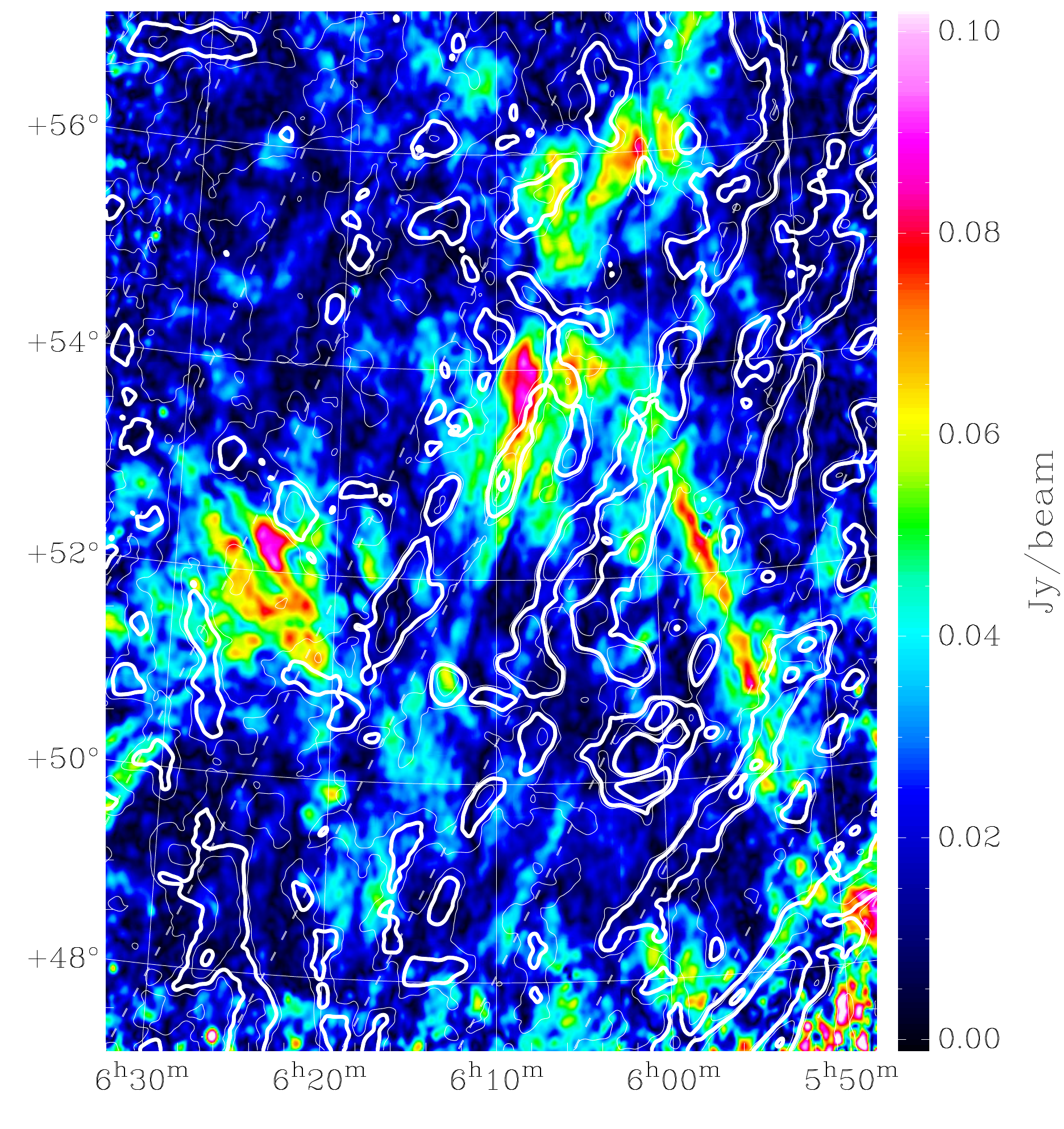}
   \caption{Polarized intensity map of Auriga in equatorial coordinates
     (B1950.0), observed at 349 MHz with the WSRT. The filamentary
     features from an \hi~USM map at $v_{\rm LSR} = -2.5 $ \kmss are
     overlaid with contours of 0.1, 1, 2.5 and 5 K.  The dashed lines
     are parallel to the Galactic plane, in steps of $\Delta b =
     1\deg$. }
   \label{Fig_Overlay_HO_2.5}
\end{figure}

Figure \ref{Fig_au_17_18} displays the derived power anisotropies
$Q(k)$ for $v_{\rm LSR} = -2.5 $ \kmss at a position angle $\Phi =
-25\deg$. We find significant fluctuations of $Q(k)$ that are common for
our analysis.  The scatter in $Q$ is significant, by far larger than
instrumental uncertainties that can be evaluated from the data points
located to the right of the vertical line for $k \ga 0.07$
arcmin$^{-1}$. For comparison we plot $Q(k)$ for the
neighbor channel at $v_{\rm LSR} = -3.76 $ \kmss which shows the
best agreement to the $v_{\rm LSR} = -2.5 $ \kmss channel also at a
similar well defined position angle, see 
Fig. \ref{Fig_AU_overview_angle}.  Peak anisotropies differ in spatial
frequencies and are in the range $20 \la Q(k) \la 30$.

We conclude that the velocity channel at $v_{\rm LSR} = -2.5 $ \kmss has
well defined anisotropies and expect accordingly that the
\hi~distribution should show well defined filamentary structures. In
Fig. \ref{Fig_Overlay_HO_2.5} we compare USM structures, using contours,
with the color coded map of the polarized intensity observed at 349 MHz
with the WSRT. The filamentary \hi~structures are preferentially
oriented parallel to the Galactic plane at a position angle of
$\Phi_{\rm gal} = 65\fdg4$, in excellent agreement with the angle
$\Phi_{\parallel} = 65\deg$ obtained from power anisotropies. The WSRT
maps show that also one arm of the prominent X-shaped structure is
oriented in this direction and partly well aligned with filamentary 
\hi~structures.

\begin{figure}[htbp]  
   \centering
   \includegraphics[width=6.5cm,angle=-90]{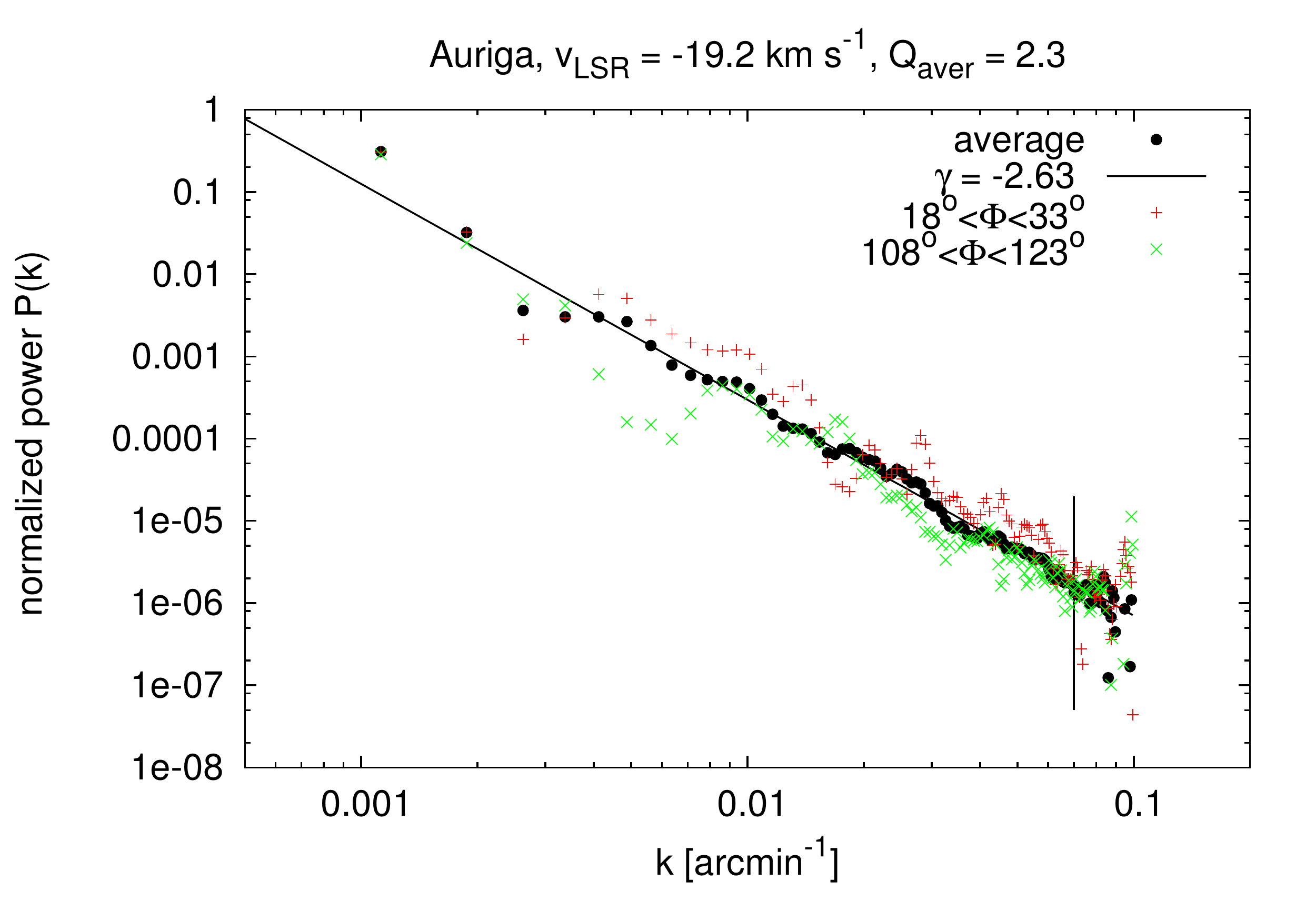}
   \caption{Average power spectrum observed for $v_{\rm LSR} = -19.2$
     \kmss (black dots) and fit power law with $\gamma = -2.63 \pm 0.03
     $ for $ k < 0.07$ arcmin$^{-1}$ (vertical line). In addition the
     power spectrum for $18\degr < \Phi < 33\degr $ (red) and $108\degr
     < \Phi < 123\degr $ (green) is given. The average anisotropy factor
     for $0.007 < k < 0.07$ arcmin$^{-1}$  is $Q_{\rm aver} = 2.3$. }
   \label{Fig_spec_au_5}
\end{figure}

\subsection{Anisotropies at $v_{\rm LSR} = -19.2 $ {\rm km~s$^{-1}$} }
\label{Auriga-19.2} 

The second feature, annotated in Fig. \ref{Fig_AU_overview_angle}, is at
a velocity of $v_{\rm LSR} = -19.2 $ \kms. Figure \ref{Fig_spec_au_5}
displays the average power spectrum, fit with a power law $\gamma =
-2.63 \pm 0.03 $ for $ k < 0.07$ arcmin$^{-1}$ (vertical line). Figure
\ref{Fig_AU_overview} shows that this \hi~gas belongs to the second
prominent $T_{\rm B\, aver}$ component at $v_{\rm LSR} = -12.8 $ \kms
and we safely may assume that this \hi~gas is not associated with the
first component at $v_{\rm LSR} = -2.5 $ \kms.

\begin{figure}[htbp]  
   \centering
   \includegraphics[width=6.5cm,angle=-90]{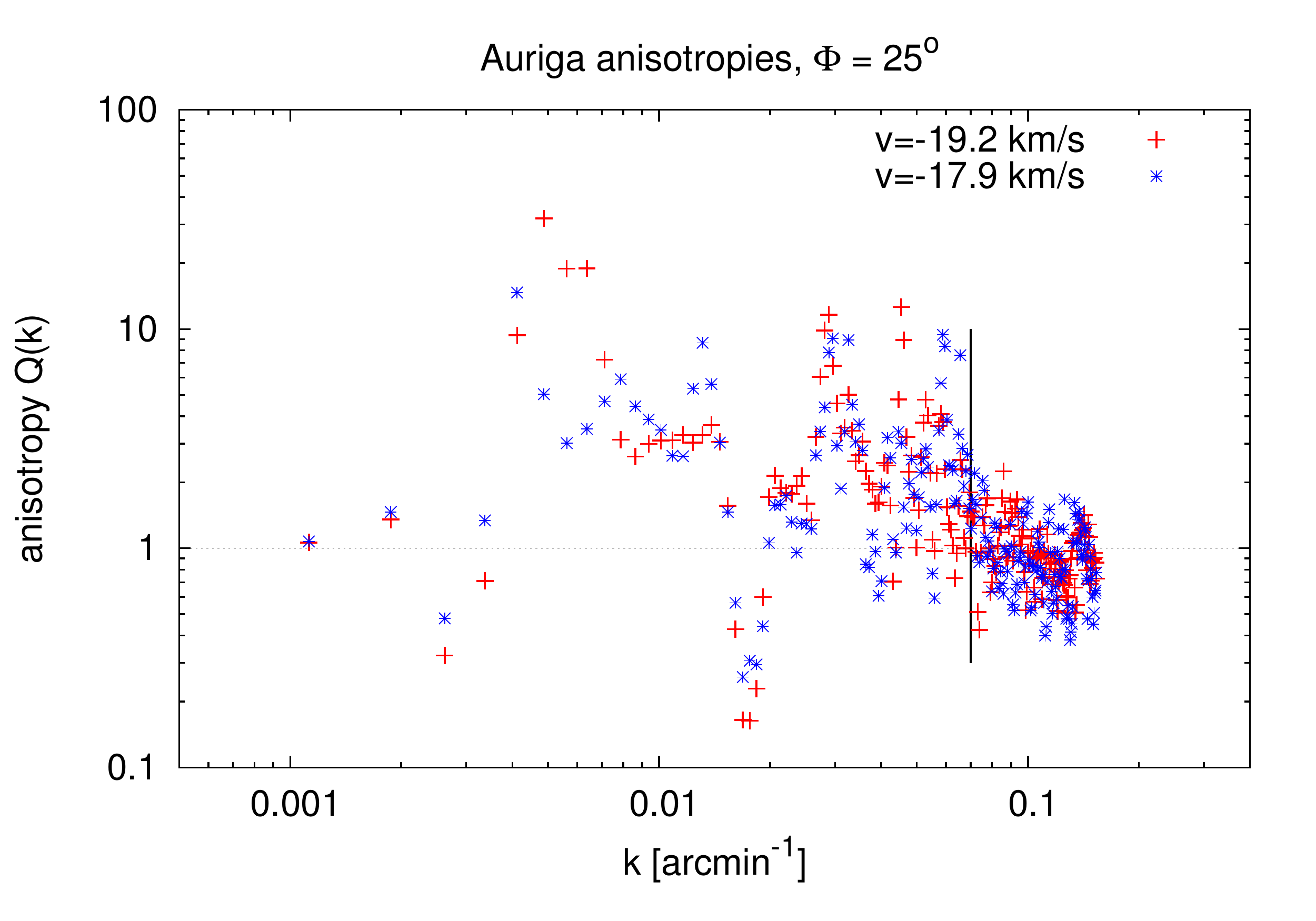}
   \caption{Anisotropies $Q(k)$ for the channel at $ v_{\rm LSR} = -19.2
     $ \kms (red) and at $ v_{\rm LSR} = -17.9 $ \kms (blue).  }
   \label{Fig_Q_au_05_06}
\end{figure}

\begin{figure}[htbp]  
   \centering
   \includegraphics[width=8.5cm]{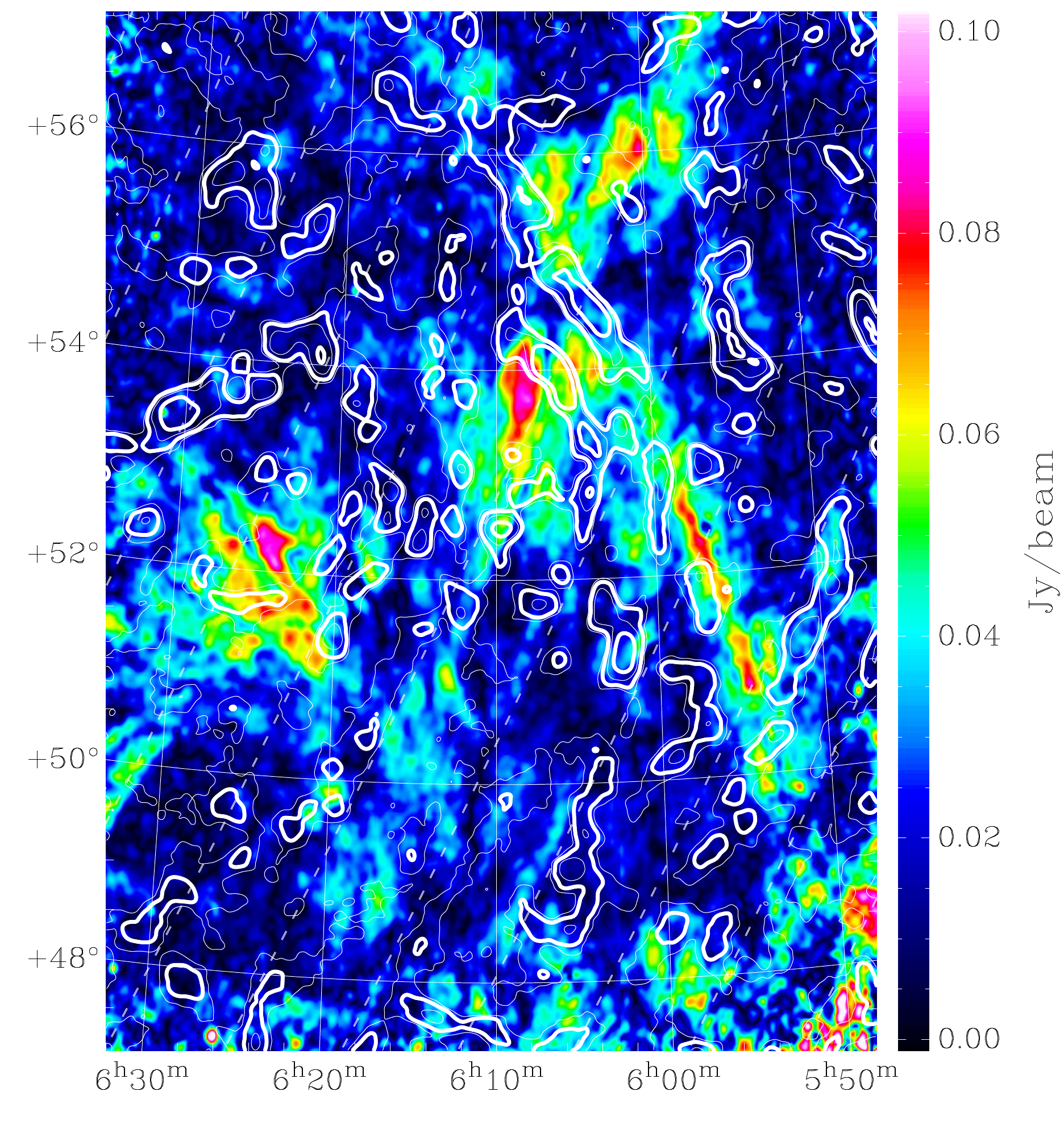}
   \caption{Polarized intensity map of Auriga in equatorial coordinates
     (B1950.0), observed at 349 MHz with the WSRT. The filamentary
     features from an \hi~USM map at $v_{\rm LSR} = -19.2 $ \kmss are
     overlaid with contours of 0.1, 1, 2.5 and 5 K.  The dashed lines
     are parallel to the Galactic plane, in steps of $\Delta b = 1\deg$.
      }
   \label{Fig_Overlay_HO-19.2}
\end{figure}

Calculating average anisotropies $Q_{\rm aver}$, we derived the position
angle $\Phi_{\rm aver} = 19 \degr \pm 22\degr$, see
Fig. \ref{Fig_AU_overview_angle}. The peak in the power anisotropy
however was found after a more detailed analysis at $\Phi_{\rm peak} =
25 \degr $.  Figure \ref{Fig_spec_au_5} displays the power spectra in
this direction and perpendicular. We plot in
Fig. \ref{Fig_Q_au_05_06} the power anisotropies $Q$ at $ v_{\rm LSR} =
-19.2 $ \kms together with the anisotropies of the neighbor
channel at $ v_{\rm LSR} = -17.9 $ \kms.

Most of the anisotropies appear to be weak, $Q(k) \la 10$ but
Fig. \ref{Fig_Q_au_05_06} shows several narrow spikes with large
anisotropies. At $k = 0.0049$ arcmin$^{-1}$ we determine $Q = 32$ for $
v_{\rm LSR} = -19.2 $ \kms. Interestingly, there are significant changes
in $Q$ for both neighbor channels. We conclude that this anisotropy must
be associated with a rather cold \hi~feature. The formal Doppler
temperature corresponding to the velocity resolution of $\Delta v_{\rm
  LSR} = 1.44$ \kms is $T_{\rm D} \sim 50$ K.  Fig. \ref{Fig_AU_overview2} in
Sect. \ref{T_D} shows a geometric mean Doppler temperature of $T_{\rm D}
\sim 150$ K, implying that the gas at this spatial frequency must be
super-sonic.

An other remarkable structure in Fig. \ref{Fig_Q_au_05_06} is the
anisotropy $Q \sim 0.16$ at $k \sim 0.017$ arcmin$^{-1}$. There is a
sharp cross-over of the anisotropies at this spatial frequency, implying
a local anisotropy up to $Q \sim 6$ in perpendicular direction.

Figure \ref{Fig_Overlay_HO-19.2} compares USM \hi~structures, using
contours, with the color coded map of the polarized intensity observed
at 349 MHz with the WSRT. The strongest filamentary \hi~structures
are in this case well aligned with the other arm of the X-shape
structure in polarized intensity. However we find also a few weaker 
  filamentary \hi~structures that are roughly aligned parallel to the
Galactic plane.

\begin{figure}[htp]  
   \centering
   \includegraphics[width=6.5cm,angle=-90]{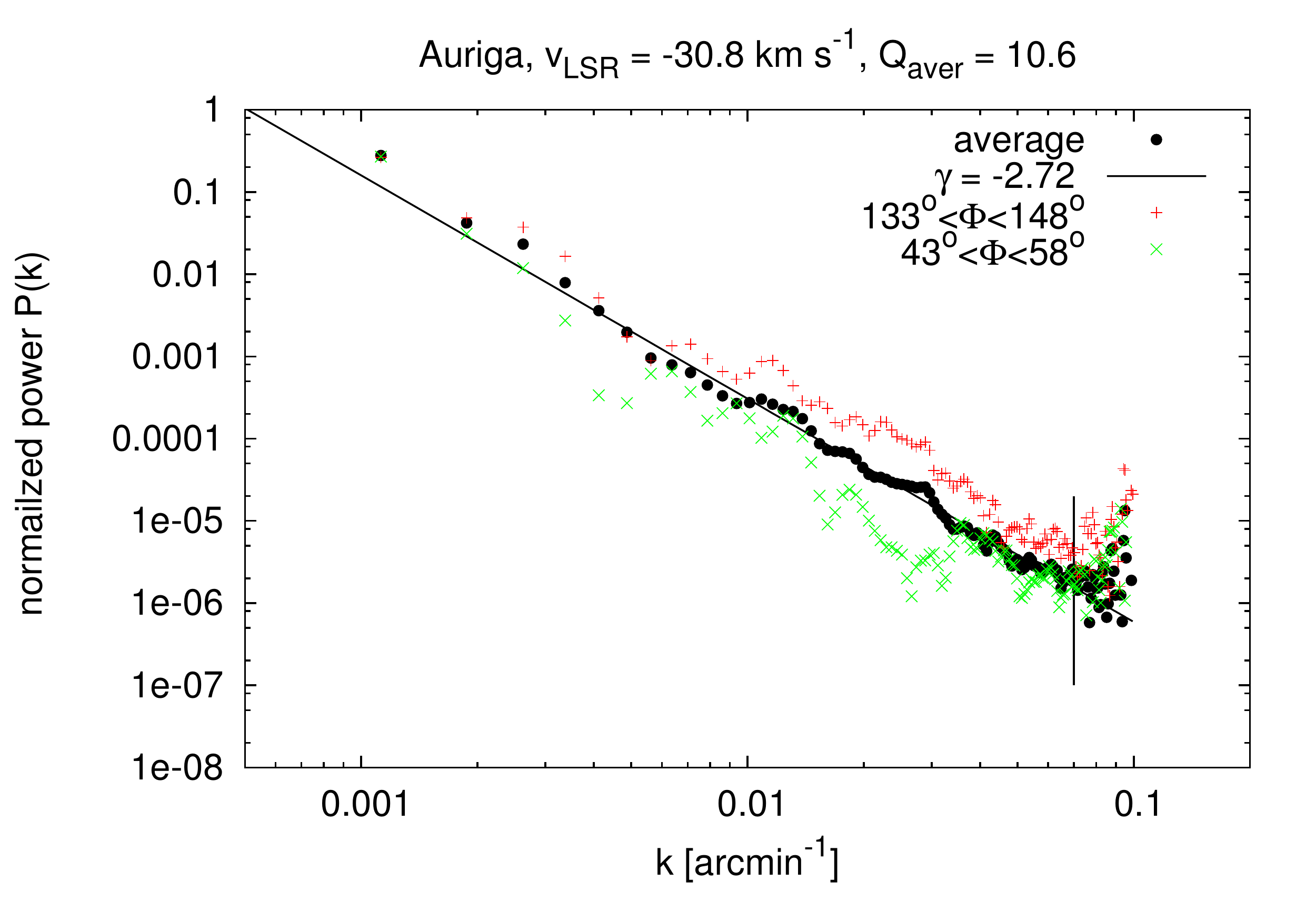}
   \caption{Average power spectrum observed for $v_{\rm LSR} = -30.8$
     \kmss (black dots) and fit power law with $\gamma = -2.72 \pm 0.03
     $ for $ k < 0.07$ arcmin$^{-1}$ (vertical line). In addition the
     power spectrum for $133\degr < \Phi < 148\degr $ (red) and $43\degr
     < \Phi < 58\degr $ (green) is given. The average anisotropy factor
     for $0.007 < k < 0.07$ arcmin$^{-1}$ is $Q_{\rm aver} = 10.6$.
   }
   \label{Fig_spec_au_HV_55}
\end{figure}

\subsection{Anisotropies at $v_{\rm LSR} = -30.8 $ {\rm km~s$^{-1}$} }
\label{Auriga-30.8}

\begin{figure}[htp]  
   \centering
   \includegraphics[width=6.5cm,angle=-90]{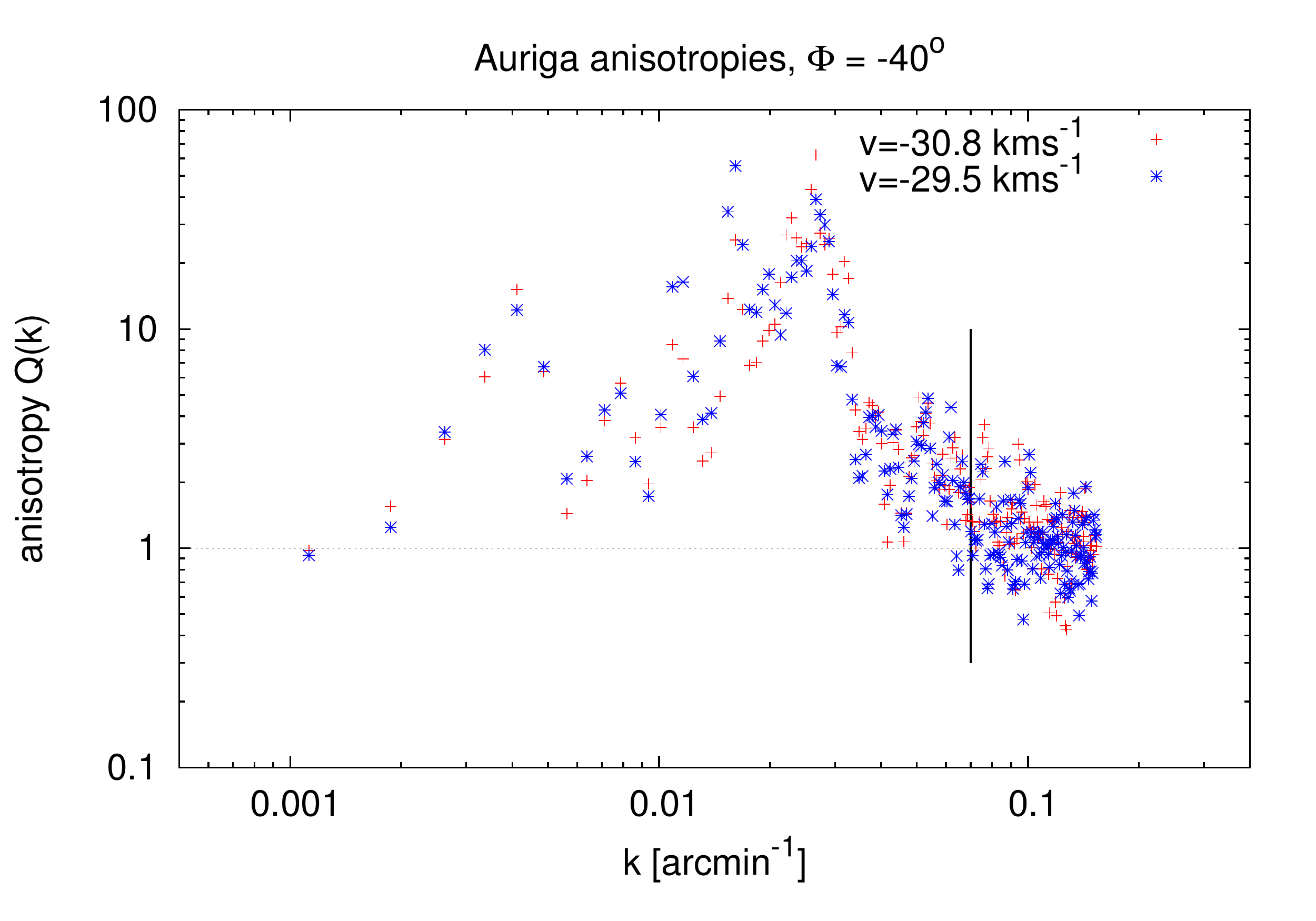}
   \caption{Anisotropies $Q(k)$ for the channel at $ v_{\rm LSR} = -30.8
     $ \kmss (red) and $v_{\rm LSR} = -29.5 $ \kmss (blue).  }
   \label{Fig_Q_spec_au_HV_55}
\end{figure}

\begin{figure}[htp]  
   \centering
   \includegraphics[width=8.5cm]{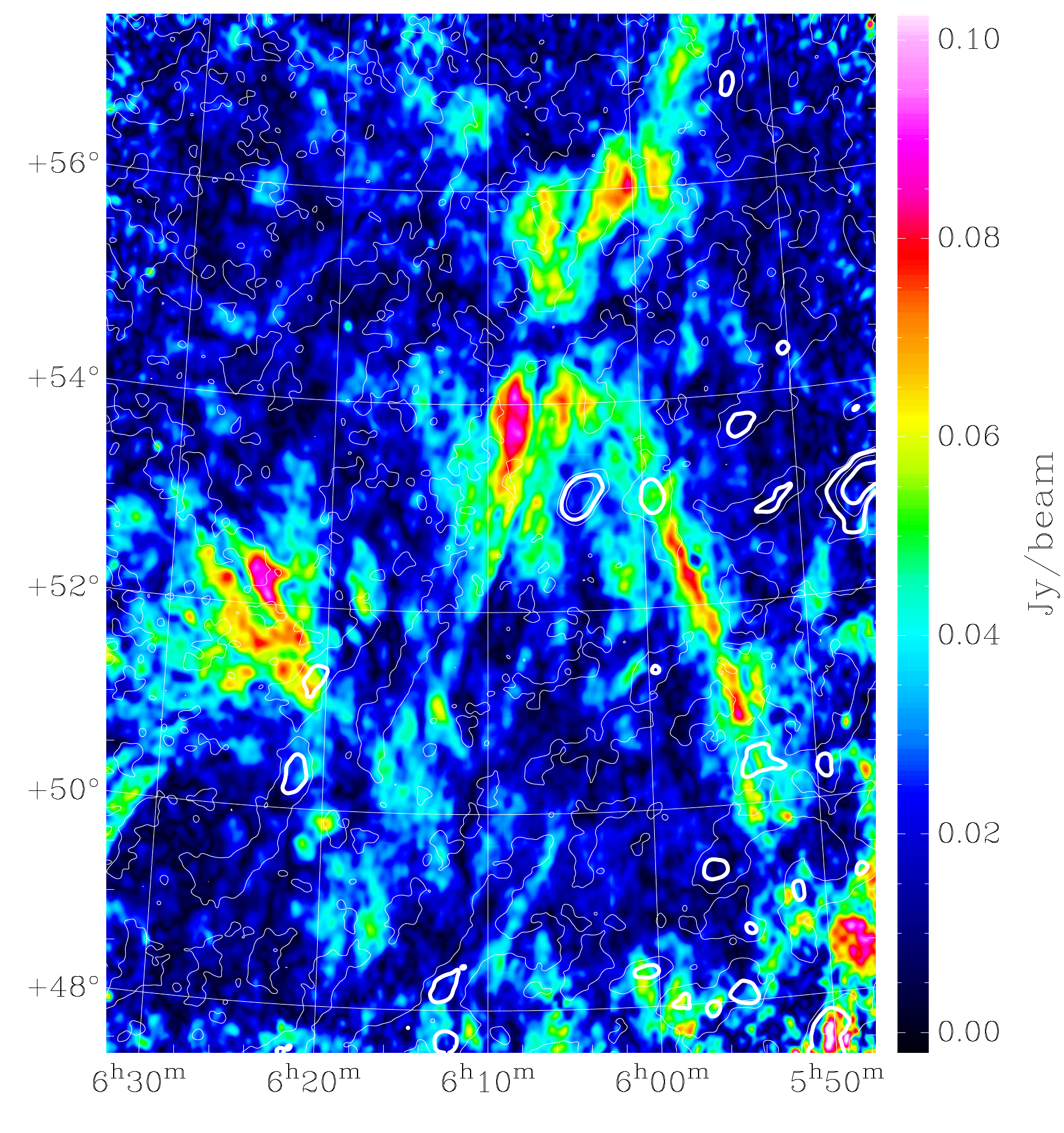}
   \caption{Polarized intensity map of Auriga in equatorial coordinates
     (B1950.0), observed at 349 MHz with the WSRT. The filamentary
     features from an \hi~USM map at $v_{\rm LSR} = -30.8 $ \kmss are
     overlaid with contours of 0.1, 1, 2.5 and 5 K.  }
   \label{Fig_Overlay_HO-30.8}
\end{figure}

A third weak \hi~feature does not stand out in the position angle
distribution in Fig. \ref{Fig_AU_overview_angle} but shows up with a
significant anisotropy $Q_{\rm aver} = 10.6$ at $v_{\rm LSR} = -30.8 $
\kmss in Fig. \ref{Fig_AU_overview}. This structure is located in the
wing of the main \hi~emission component. This component has a mean
Doppler temperature of $T_{\rm D} \sim 700$ K, implying that it is
somewhat warmer than the typical CNM, close to the upper limit of the
typical CNM temperature range \citep{Wolfire2003,Kalberla2016}. Its
position angle is $\Phi_{\rm aver} = -40 \degr \pm 16\degr$.

Figure \ref{Fig_spec_au_HV_55} displays the derived power spectra.  The
spectra are shallow, we fit $\gamma = -2.72 \pm 0.03$ for the isotropic
case.  Figure \ref{Fig_Q_spec_au_HV_55} shows the anisotropies for two
neighbor channels at $v_{\rm LSR} = -30.8 $ and $v_{\rm LSR} = -29.5 $
\kms. Strong anisotropies up to $Q \sim 60$ are found around $k \sim
0.02 $ arcmin$^{-1}$ but else both channels share similar features.

In Fig. \ref{Fig_Overlay_HO-30.8} we compare the corresponding
\hi~structures at this velocity, using contours, with the color coded
map of the polarized intensity observed at 349 MHz. Except for a few
structures, the agreement between both data-sets is less obvious. The
EBHIS data indicate that the \hi~distribution in the Auriga field is
dominated by two independent \hi~layers which can also be traced in
polarized intensity at 349 MHz. To distinguish a third weaker \hi~layer
is quite problematic.

\section{Spectral index dependencies }
\label{Spectral_index}

The power spectrum of a two-phase \hi~gas layer can be described as the
sum of three individual power spectra for WNM, CNM, and a spectrum that
describes the correlation between WNM and CNM components \citep[][their
Sect.  4.3]{Lazarian2000}. Unfortunately none of these power spectra is
directly observable. The \hi~gas is a mixture of WNM and CNM gas and
emission lines are usually dominated by the WNM. The accurate CNM column
density fraction has to be determined from absorption data. Continuum
sources for such an analysis are sparse and, as demonstrated by
\citet[][Fig. 7]{Heiles2003}, the CNM column density fraction has a very
broad distribution. Accordingly it is not possible to derive from
observations meaningful power spectra, characteristic either for WNM or
CNM. We found however significant fluctuations of the spectral index and
consider here the question whether the spectral index might depend on
the WNM or CNM composition. Similar to \citet{Lazarian2000} we consider
the idealized case of a two-phase medium without distinguishing whether
or not the WNM belongs to a stable phase or lies in the thermally
unstable region for temperatures of 500 to 5000 K \citep{Heiles2003}.

\subsection{Spectral indices from 2D images versus 3D turbulence}
\label{2d-3d}

Turbulence is a 3D phenomenon and under certain conditions 2D power
spectra can be converted to 3D power spectra in density and
velocity. One of the most important preliminaries for an easy
translation of 2D to 3D turbulent properties is that the extension of
the observed optical thin \hi~gas layer along the line of sight is
larger than the observed perpendicular extension. For our fields we
have no reliable information about distances and extensions, but usually
the \hi~gas layer is considered to be extended
\citep{Dickey1990,Kalberla2009}.

For details about the 2D to 3D conversion we refer to
\citet{Lazarian2000} or \citet{Miville-Deschenes2003} but we will give
later a few applications. This is an observational paper and we consider
in general 2D power spectra, regardless whether an unambiguous
conversion to 3D is possible or not.

\subsection{Spectral index versus Doppler temperature }
\label{T_D}

\begin{figure}[thbp]  
   \centering
   \includegraphics[width=6.5cm,angle=-90]{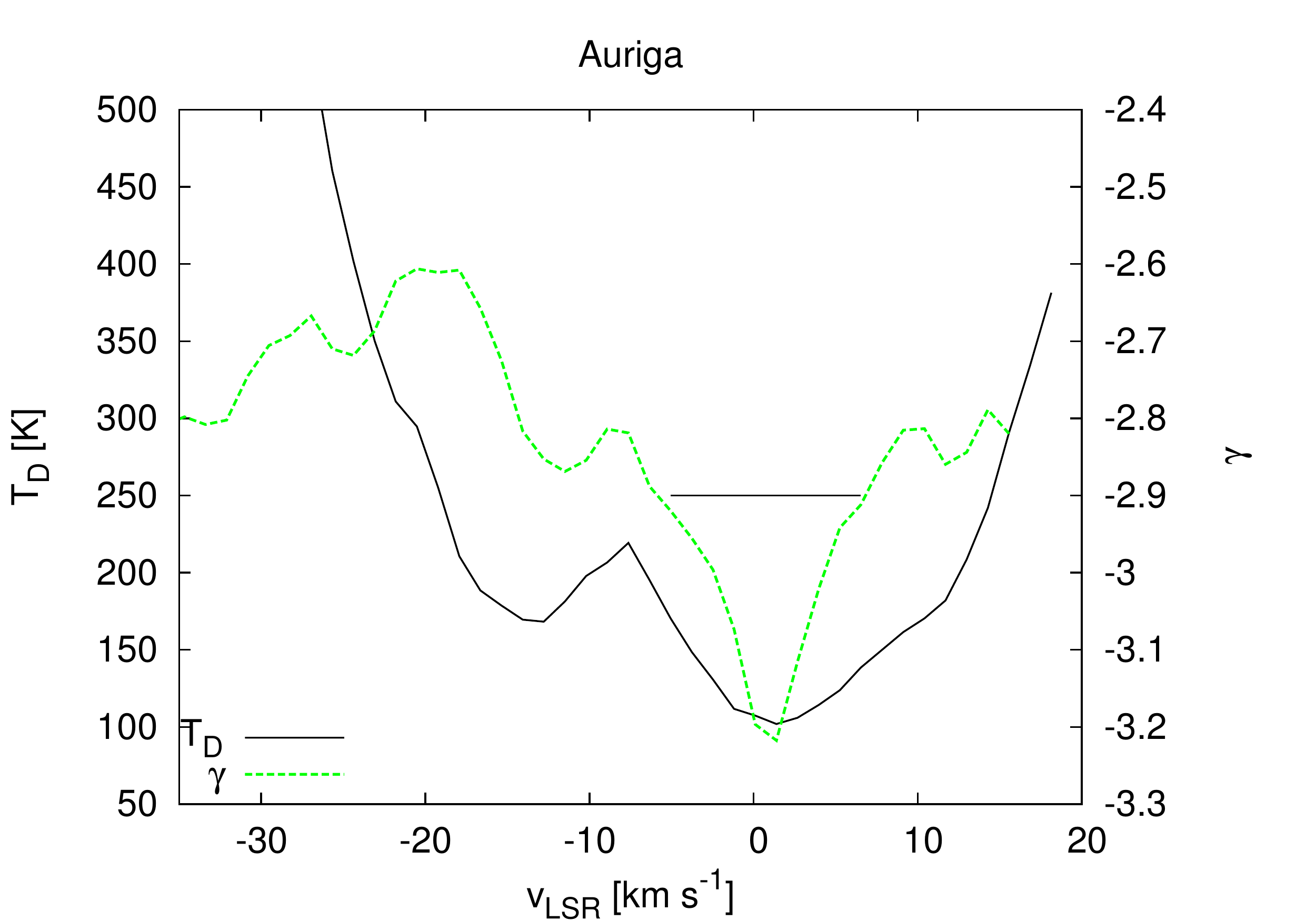}
   \includegraphics[width=6.5cm,angle=-90]{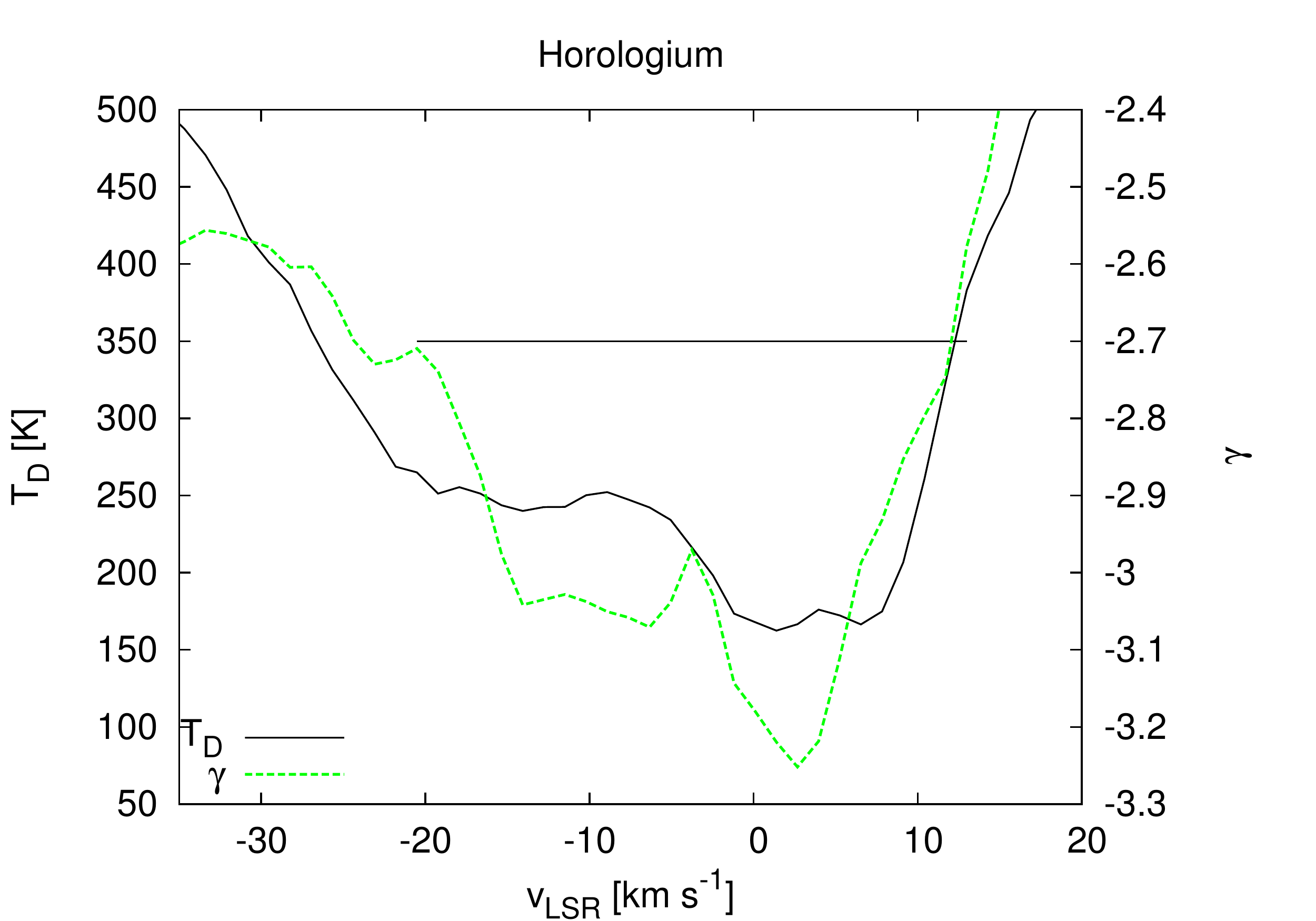}
   \caption{Comparison between the geometric mean Doppler temperature
     $T_{\rm D}$ (black) and the spectral index $\gamma$ (green dashed)
     for Horologium (bottom) and Auriga (top). The horizontal lines
     indicate the velocity ranges used for the determination of the
     spectral steepening discussed in Sect. \ref{TI_Power}. }
   \label{Fig_AU_overview2}
\end{figure}

We use USM maps to derive Doppler temperatures for filamentary CNM
structures and characterize the temperature distribution of the CNM by
its geometric mean Doppler temperatures $T_{\rm D}$, see 
Fig.  \ref{Fig_AU_overview2}. The minima for the derived values, $T_{\rm
  D} \ga 100$ K for Auriga (top) and $T_{\rm D} \ga 160$ K for
Horologium (bottom), are low compared to the median value $T_{\rm D} =
223$ K for filamentary structures at high and intermediate latitudes,
determined by \citet{Kalberla2016}. For both targets Doppler
temperatures tend to be the lowest at those velocities where the
spectral index $\gamma$ is the steepest.

The derivation of $T_{\rm D}$ at low Galactic latitudes is affected by
confusion. USM channel maps show there in general a wealth of
filamentary structures. A reliable determination of Doppler temperatures
is however only possible for \hi~structures that are isolated in 3D. We
only can observe projections in the position-velocity space. Multiple
filamentary features along the line of sight may blend, causing an
unknown overestimation of the derived geometrical mean $T_{\rm
  D}$. Hence differences in geometric mean Doppler temperatures between
Auriga ($T_{\rm D} \ga 100$ K) and Horologium ($T_{\rm D} \ga 160$ K)
are probably not significant since Horologium is at a lower Galactic
latitude with larger confusion.

Doppler temperatures are upper limits to the thermal (excitation or
spin) temperatures of the filamentary CNM structures. Strong background
sources are needed to derive this.  The continuum source 3C\,147 is
located outside the Auriga field but nearby at RA = 84\fdg7, DEC =
49\fdg8 (B1950.0), $l = 161.7\deg, b= 10.3\deg$ and strong enough to
allow a determination of accurate spin temperatures $T_{\rm
  spin}$. \citet{Kalberla1985} derived for small scale \hi~clumps
temperatures between 37 and 74 K for the observed absorption
components. For $v_{\rm LSR} = 0.2 $ \kms, $T_{\rm spin} = 37 \pm 24 $ K
was obtained, allowing to assign a typical turbulent CNM Mach number of
$M_{\rm T} \sim 2.7$ at $v_{\rm LSR} = 0.2 $ \kms. For such CNM optical
depth and self absorption may affect observed emission features at a few
positions, however our data do not allow to correct such effects
reliably. We find that power spectra for individual velocity channels
tend to be steepest at velocities with low $T_{\rm D}$ and $T_{\rm
  spin}$, at the same time coincident with the velocity at the peak of
the average \hi~emission (Fig. \ref{Fig_AU_overview}).

Continuum point sources, suitable for a determination of spin
temperatures in the Horologium field, are unfortunately not
available. The strongest point source has there only a flux
density of 400 mJy at 1.4 GHz.

\subsection{Spectral index and the $T_{\rm WNM}/T_{\rm   WNM+CNM} $ ratio}
\label{TI_gamma}

We find clear indications that the distributions of spectral indices
derived from single velocity channels, have well defined narrow minima
for velocities with strong \hi~emission lines
(Figs. \ref{Fig_HO_overview}, \ref{Fig_AU_overview}  and in
\citetalias{Kalberla2016b} Figs. 12 and 18). The associated CNM, derived
from USM data, shows also pronounced filamentary structures with low
Doppler temperatures that depend strongly on the observed radial
velocity (Fig. \ref{Fig_AU_overview2}).

Since in general most of the observed \hi~emission data are dominated by
the WNM, we determine in the following the velocity dependence of the
WNM fraction, defined as the average brightness temperature ratio in the
observed field, $T_{\rm B\, WNM}(v_{\rm LSR})/T_{\rm B\, WNM+CNM}(v_{\rm
  LSR})$. We use a Gaussian analysis for an estimate $T_{\rm G\,
  WNM}(v_{\rm LSR})/T_{\rm G\, WNM+CNM}(v_{\rm LSR}) $ of this
ratio. For the decomposition we use mostly the same approach which was
described by \citet{Haud2000} and applied earlier to the
Leiden/Argentine/Bonn \citep[LAB,][]{Kalberla2005} and the Galactic All
Sky Survey \citep{Kalberla2015}, see also \citet{Haud2013}. A similar
analysis was applied to the EBHIS and used by \citet{Kalberla2016}.

\begin{figure}[tbp]  
   \centering
   \includegraphics[width=6.5cm,angle=-90]{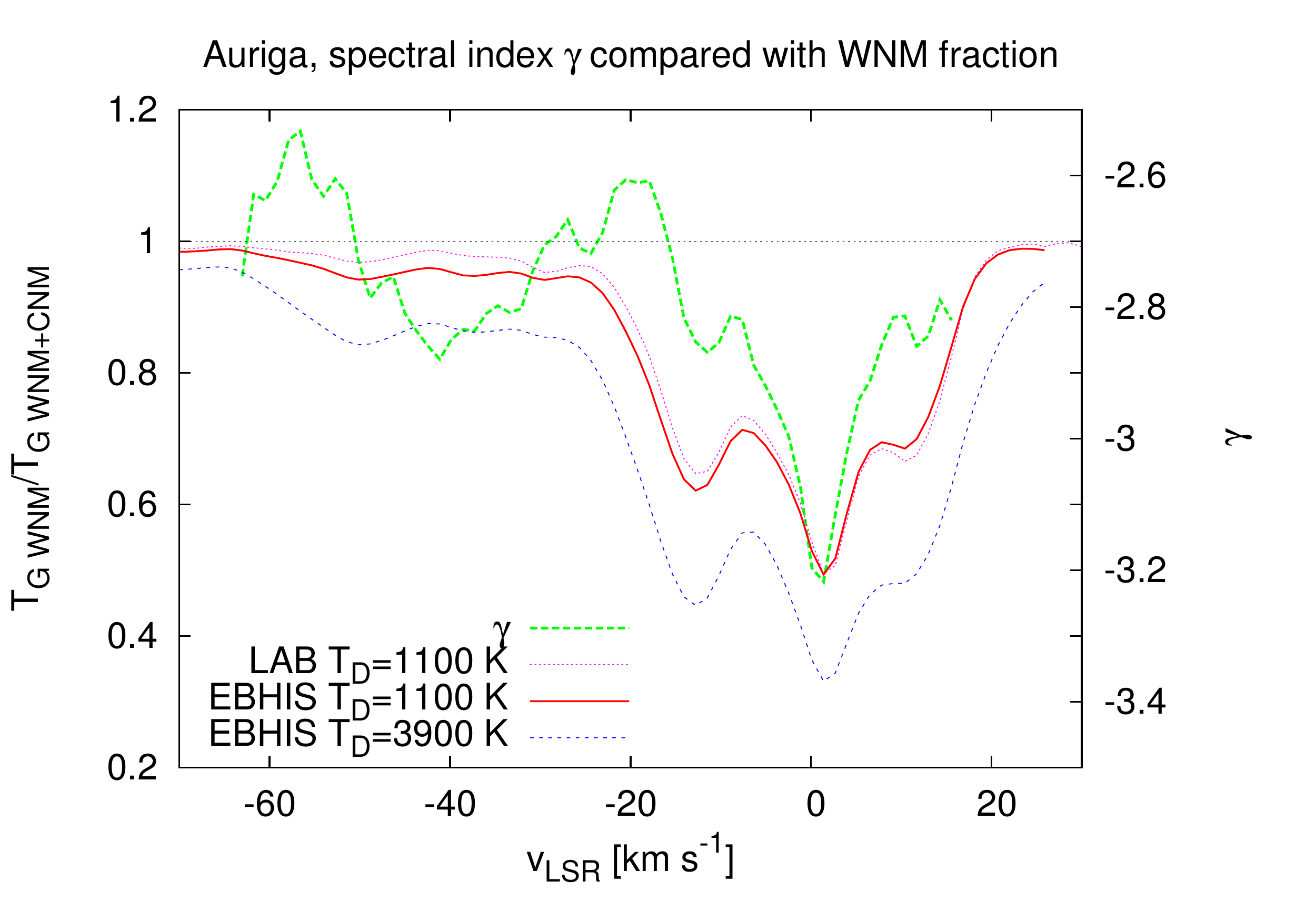}
   \includegraphics[width=6.5cm,angle=-90]{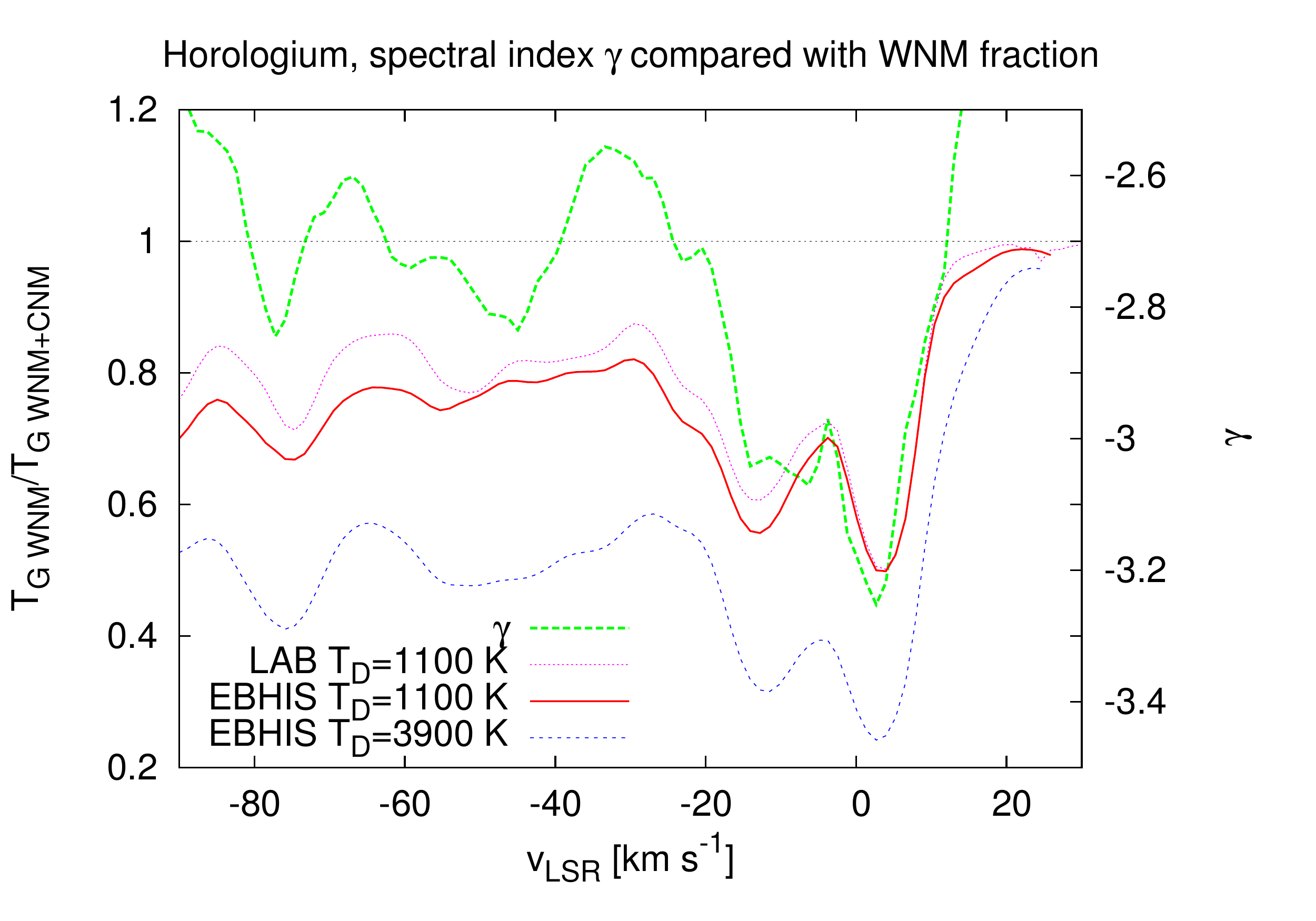}
   \caption{Bottom: Horologium, top: Auriga. Spectral index
     $\gamma(v_{\rm LSR})$ (green) compared with the WNM fraction
     $T_{\rm WNM}(v_{\rm LSR})/T_{\rm WNM+CNM}(v_{\rm LSR}) $ derived
     from LAB (pink) and EBHIS Gaussians (red). An upper limit for the
     Doppler temperature of the CNM of $T_{\rm D} < 1100 $ K,
     corresponding to a turbulent CNM  Mach number of $M_{\rm T} =3.7$ was
     applied. For comparison we display the WNM fraction $T_{\rm
       WNM}(v_{\rm LSR})/T_{\rm WNM+CNM}(v_{\rm LSR}) $ for a Doppler
     temperature of $T_{\rm D} < 3900 $ K, corresponding to a turbulent
     CNM  Mach number of $M_{\rm T} =7.7$ (blue).  }
   \label{Fig_Gauss}
\end{figure}

Each \hi~profile on a HEALPix nside = 1024 grid \citep{Gorski2005}
within the apodized region was decomposed into Gaussian components. For
each velocity channel we calculated the average brightness temperature
contribution from the CNM and the WNM. To distinguish between the two
\hi~phases we select Gaussians representing the CNM by applying a limit
of $T_{\rm D} < 1100 $ K to the Doppler temperatures of the Gaussian
components. All remaining components were assigned to the WNM.
According to \citet[][Sect. 5.1]{Kalberla2016} this limit corresponds
for a typical turbulent CNM Mach number $M_{\rm T} =3.7$
\citep{Heiles2003} and a typical thermal temperature $T = 52 $ K to a
maximum temperature of $ T = 258$ K for a stable CNM phase
\citep{Wolfire2003}. Increasing the $T_{\rm D}$ threshold leads to a
general decrease of the WNM fraction $T_{\rm G\, WNM}(v_{\rm
  LSR})/T_{\rm G\, WNM+CNM}(v_{\rm LSR}) $ at all velocities, as shown
in Fig. \ref{Fig_Gauss} with the example $T_{\rm D} < 3900 $
K. Alternatively we obtain a general increase of the WNM fraction for a
lower $T_{\rm D}$ threshold. The minima of $T_{\rm G\, WNM}(v_{\rm
  LSR})/T_{\rm G\, WNM+CNM}(v_{\rm LSR}) $ do not depend on the $T_{\rm
  D}$ threshold.

Figure \ref{Fig_Gauss} displays a comparison of the spectral index
$\gamma(v_{\rm LSR})$ with WNM fraction $T_{\rm G\, WNM}(v_{\rm
  LSR})/T_{\rm G\, WNM+CNM}(v_{\rm LSR}) $ derived from LAB and EBHIS
Gaussians. We find that the minimum of the spectral index is
mirrored by a well defined minimum of the WNM fraction. This
minimum is associated with rather low Doppler temperatures and a maximum
for $ T_{\rm G\, CNM}(v_{\rm LSR})$ (not shown). 

The separation of WNM and CNM components by a Gaussian analysis is based
on the model assumption that the observed \hi~spectrum can be
decomposed in a meaningful way into Gaussian components. There may be
systematic biases, for example separate narrow emission lines at similar radial
velocities can be blended in a way that the Gaussian decomposition
results in a single broader component. This bias, leading usually to an
underestimation of CNM lines, is more serious for weak emission lines
and observations from telescopes with a broader beam. We therefore
compare in Fig. \ref{Fig_Gauss} the results of two independent
decompositions, using LAB and EBHIS data. Discrepancies in the WNM
fraction are most probably caused by line blending but such problems
appear unimportant in regions with well defined CNM components
\citep[see also][]{Haud2010,Haud2013}.

The WNM fraction depends critically on the turbulent Mach number of the
\hi~gas. In addition to the CNM Mach number $M_{\rm T} =3.7$
\citep{Heiles2003,Kalberla2016}, we plot the WNM fraction for a very
high CNM Mach number $M_{\rm T} =7.7$ as determined by
\citet{Chepurnov2010}. In this case the WNM fraction decreases
significantly but location and shape of the minima of the WNM fraction
are still in good agreement with the minima of $\gamma$. The case
$M_{\rm T} =7.7$ is for comparison only, in the following we will use
the $T_{\rm G\, WNM}(v_{\rm LSR})/T_{\rm G\, WNM+CNM}(v_{\rm LSR}) $ for
a CNM Mach number $M_{\rm T} =3.7$ (thick line in
Fig. \ref{Fig_Gauss}) for discussion. The investigations by
\citet{Heiles2003} have shown that Mach numbers as high as 7.7 can exist
but these are not frequent.

\subsection{Velocity field analysis: VCA and velocity centroids}

For an understanding of the observed systematic changes in the spectral
power distribution we consider first theoretical investigations.
\citet{Lazarian2000,Lazarian2004} have introduced an analytic relation
for the change of the spectral index of velocity channel maps with the
thickness of the velocity slice. They have shown that the integration
over the full velocity dispersion provides the statistics that depends
only on the density field. Hence, to determine the  3D \hi~density
spectrum one needs to consider power spectra for thick velocity slices
while properties of the 3D turbulent velocity spectrum can be
derived from thin velocity slices.

A slice is defined to be thin if the velocity width of the investigated
\hi~distribution is small compared to the FWHM velocity width of the
gas, alternatively a thick slice needs to be larger than the
\hi~velocity width. Typical numbers for the width of the WNM lines are
considered to be $\Delta v_{\rm LSR} = 17$ \kmss and $\Delta v_{\rm LSR}
= 2.6$ \kmss for the CNM respectively \citet{Lazarian2000}. For
comparison, the instrumental velocity resolution of the EBHIS is $\Delta
v = 1.44$ \kmss and such a velocity channel represents a thin slice.

Alternatively to VCA, velocity centroids have been used to characterize
the  3D  velocity field of a turbulent medium
\citep{Lazarian2003,Miville-Deschenes2003}.  In the image plane the
(normalized and restricted) centroid velocity $VC(x,y)$ is for the
velocity interval $v_1 < v_{\rm LSR} < v_2 $ defined as
\begin{equation}
  VC(x,y) = \frac {\int_{v1}^{v2} T_{\rm B}(x,y,v_{\rm LSR}) v_{\rm LSR} \Delta v_{\rm LSR}}{\int_{v1}^{v2} T_{\rm B}(x,y,v_{\rm LSR}) \Delta v_{\rm LSR}}.
\label{eq:VC}
\end{equation}
The application of a velocity window $v_1 < v_{\rm LSR} < v_2 $ is
necessary to restrict the centroid to a particular emission line feature
since we usually observe multiple emission lines along the line of sight
with different anisotropies, blending partly in velocity. To avoid
biases from unrelated components, the velocity window should be
restricted to velocities with significant line emission. Also it should
be symmetric with $T_{\rm B}(v_1) \sim T_{\rm B}(v_2) $ to avoid biases
from the wings of the line.

\begin{figure}[h]  
   \centering
   \includegraphics[width=6.5cm,angle=-90]{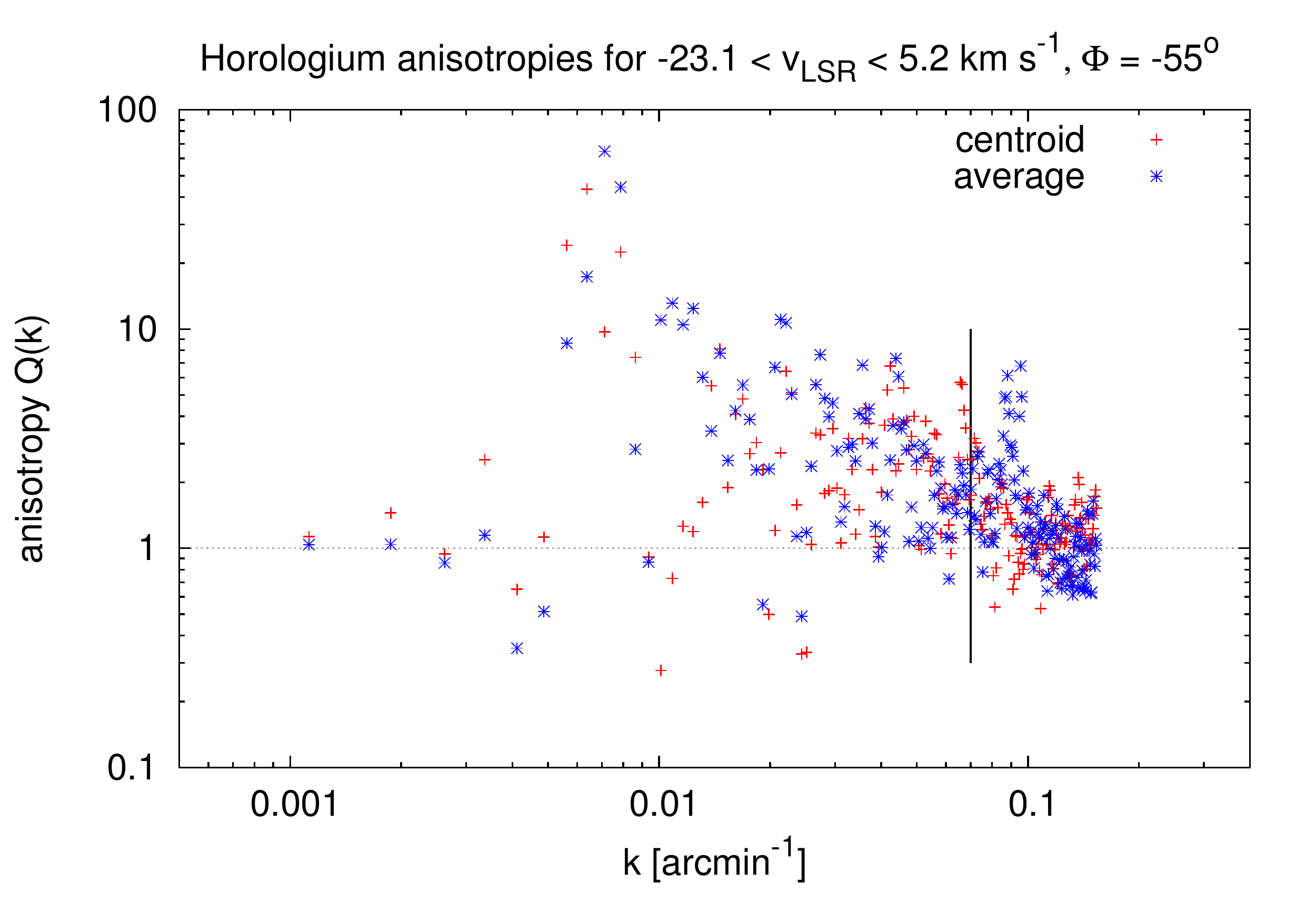}
   \caption{Anisotropies $Q(k)$ for the velocity centroid (red) and
     average emission (blue), both calculated for the velocity range $
     -23.1 < v_{\rm LSR} <5.25 $ \kms. }
   \label{Fig_ho_aver_cent}
\end{figure}

As an application of the VCA we calculate power spectra,
integrating for the Horologium field the \hi~emission over the velocity
range $-23.1 < v_{\rm LSR} < 5.2$ \kms. This is a thick slice. Figure
\ref{Fig_ho_aver_cent} displays the derived anisotropies that
can be attributed to the  3D  density distribution. For comparison we plot
also anisotropies derived from the velocity centroid
over the same velocity interval. In both cases, anisotropies at $k \sim
0.007$ arcmin$^{-1}$ are remarkable high with $80 \la Q \la 110$. For
higher spatial frequencies this plot confirms essentially the result
from Fig. \ref{Fig_Q_spec_HO_22} but anisotropies for centroid and thick
slice velocity average in Fig. \ref{Fig_ho_aver_cent} are there 
less pronounced.

\begin{figure}[thp] 
   \centering
   \includegraphics[width=6.5cm,angle=-90]{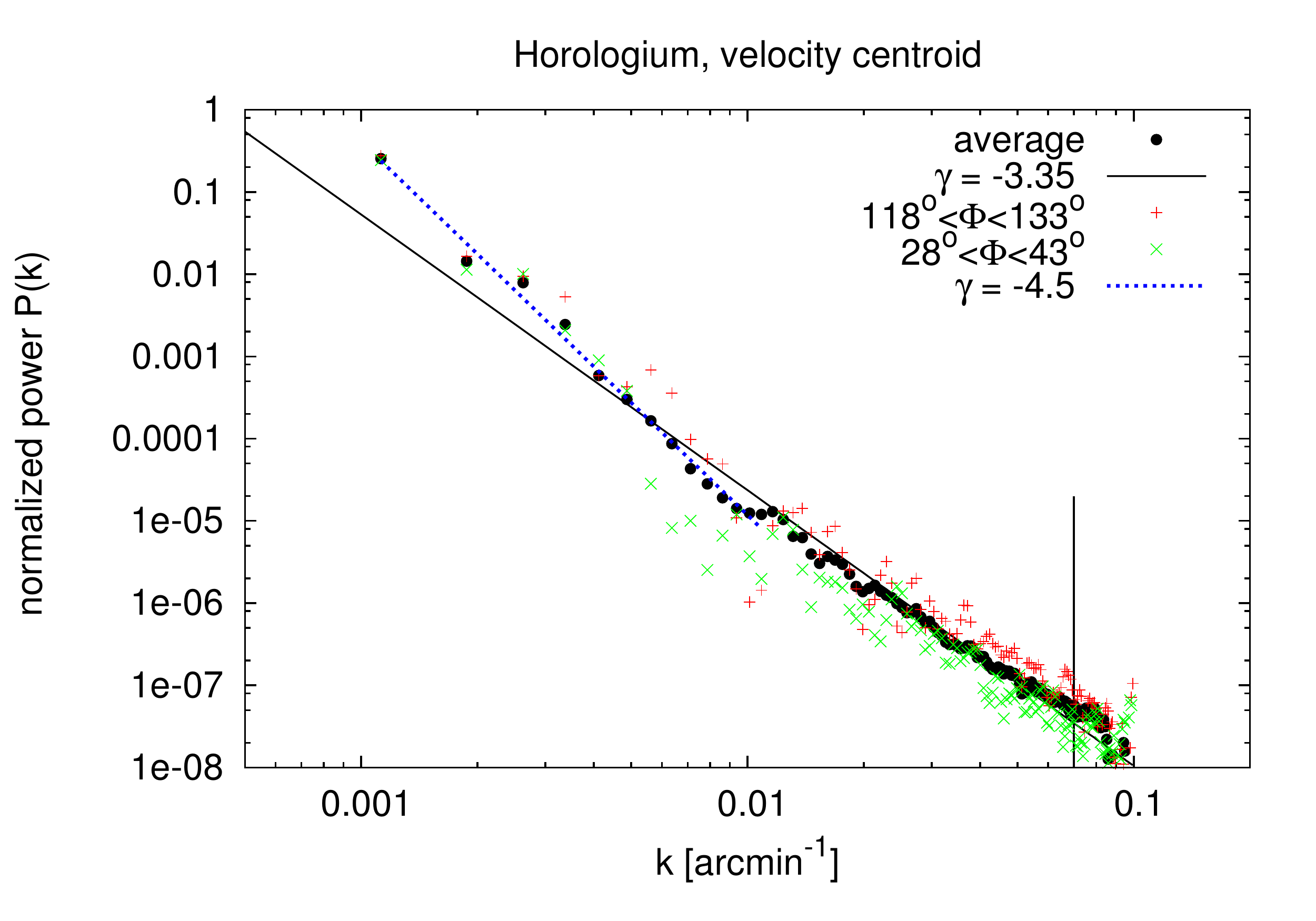}
   \includegraphics[width=6.5cm,angle=-90]{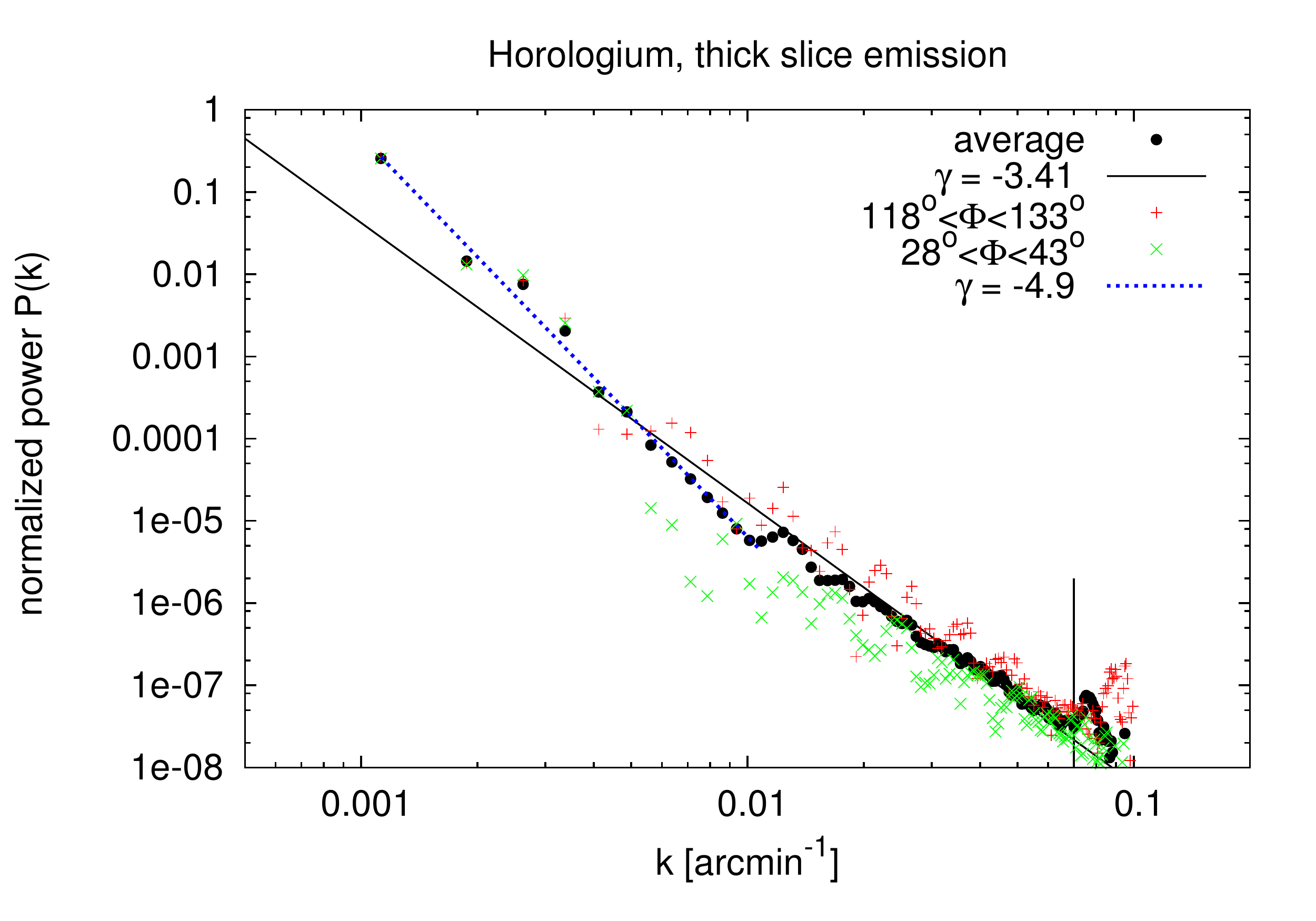}
   \caption{Power spectrum (black dots) calculated for the velocity
     centroid (top) and the thick slice emission (bottom) over the velocity
     range $-23.1 < v_{\rm LSR} < 5.2$ \kms. An average power law
     $\gamma = -3.35 \pm 0.05 $ for the centroid and respectively
     $\gamma = -3.41 \pm 0.05 $ for the thick velocity slice was fit for
     $ k < 0.07$ arcmin$^{-1}$ (vertical line). The power law fits for
     the restricted ranges $0.001 < k < 0.01$ arcmin$^{-1}$ in spatial
     frequency are indicated with blue dotted lines. In case of the
     centroid (top) $\gamma = -4.5 \pm 0.1 $ and $\gamma = -4.9 \pm 0.1
     $ for the average emission respectively.  In addition the power
     spectra for $118\degr < \Phi < 133\degr $ (red) and $28\degr < \Phi
     < 43\degr $ (green) are given. The average anisotropy factor for
     $0.007 < k < 0.07$ arcmin$^{-1}$ is $Q_{\rm aver} = 2.1$ for the
     centroid and $Q_{\rm aver} = 4.1 $ for the average emission. }
   \label{Fig_Horo_centroid}
\end{figure}

Figure \ref{Fig_Horo_centroid} shows power spectra for the same velocity
interval. For the average emission from the thick velocity slice
(bottom) we determine an average spectral index $\gamma = -3.41 \pm 0.05
$, derived for spatial frequencies $ k < 0.07$ arcmin$^{-1}$.
\citet[][Sect. 4.2]{Lazarian2000} caution that one should use very thick
slices for a reliable determination of the 3D spectral index in
density. We took this into account and repeated the analysis,
integrating over larger ranges in velocity. The derived spectral index
did not change, thus the velocity range $-23.1 < v_{\rm LSR} < 5.2$
\kmss can be considered as sufficiently broad under the very thick slice
condition (we note that the strong \hi~emission in
Fig. \ref{Fig_HO_overview} originates from regions with similar position
angle as shown in Fig.  \ref{Fig_HO_overview_angle}). Accordingly the
power index, derived from the thick slice emission over this velocity
range, may be considered as the power law exponent $\gamma_{\rm n} =
-3.4$ of the 3D density field. Such an index is considered to indicate
the case of a steep power spectrum \citep{Lazarian2000}.

Power spectra derived from the velocity centroid are shown in
Fig. \ref{Fig_Horo_centroid} at the top. The fit average power
distribution with $\gamma = -3.35 \pm 0.05 $ agrees within the
uncertainties with $\gamma = -3.41 \pm 0.05 $ from the thick slice. The
power spectra displayed in both panels of Fig. \ref{Fig_Horo_centroid}
are very similar. The implication from \citet{Lazarian2000} and
\citet{Lazarian2003} is that the 3D  turbulent density and
velocity fields in Horologium share similar properties.

\begin{figure}[htbp] 
   \centering
   \includegraphics[width=6.5cm,angle=-90]{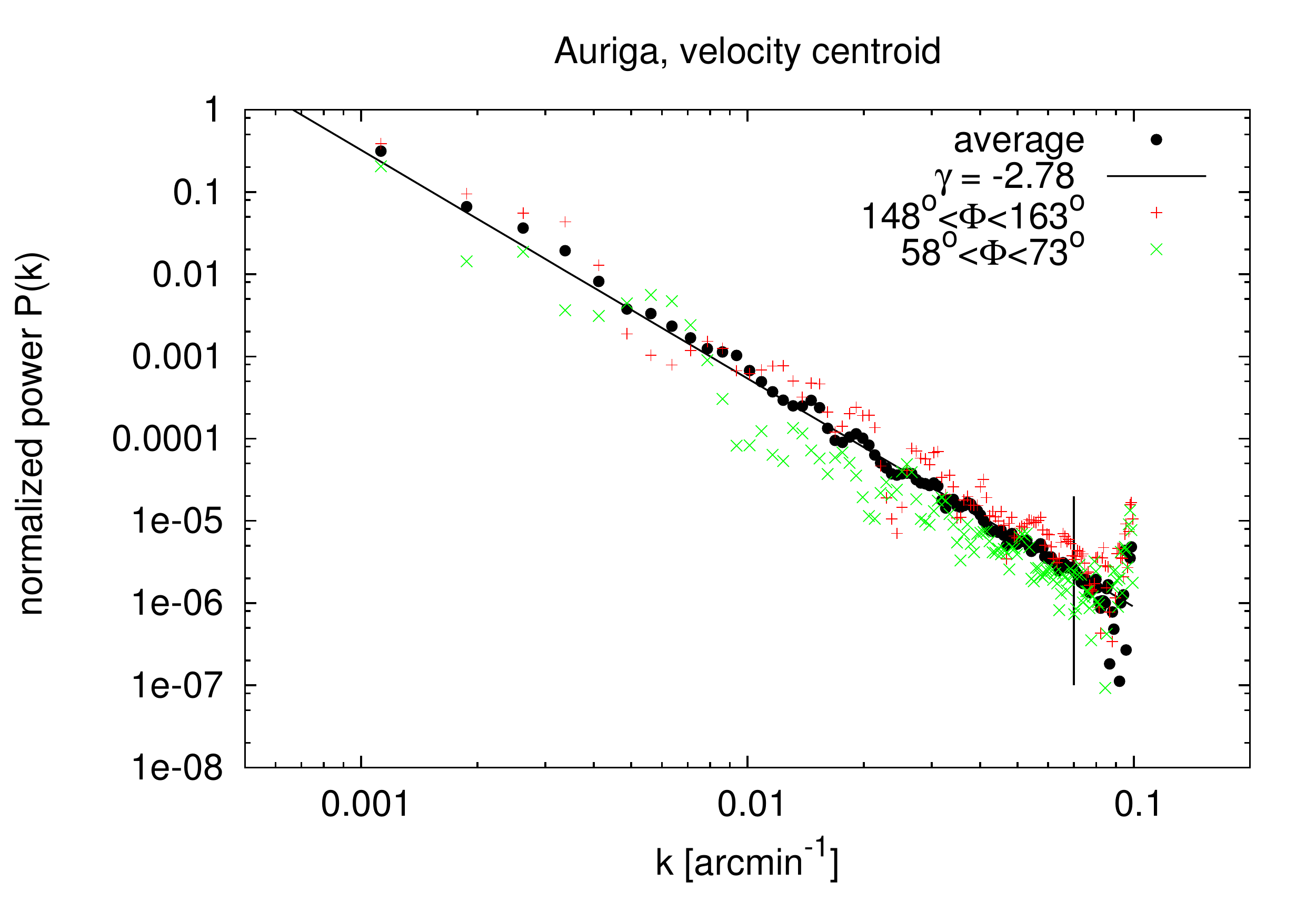}
   \includegraphics[width=6.5cm,angle=-90]{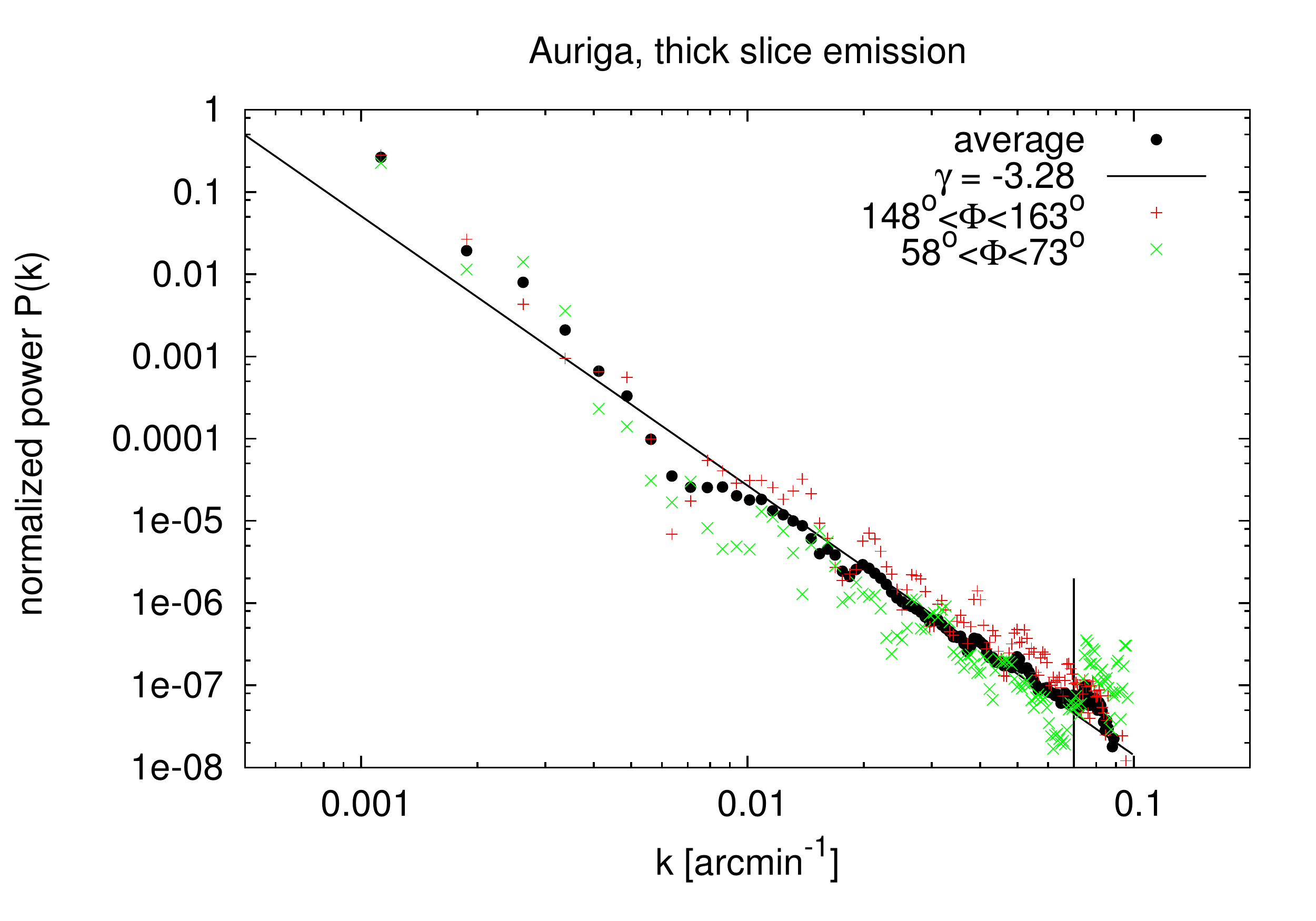}
   \caption{Power spectrum (black dots) calculated for the velocity
     centroid (top) and the thick slice emission (bottom) over the velocity
     range $ -10.2 < v_{\rm LSR} < 11.7 $ \kms. For the velocity
     centroid a power law with $\gamma = -2.78 \pm 0.03 $ was fit and
     respectively $\gamma = -3.28 \pm 0.04 $ for the average emission
     (black lines), in both cases for $ k < 0.07$ arcmin$^{-1}$
     (vertical line). In addition the power spectra for $148\degr < \Phi
     < 163\degr $ (red) and $58\degr < \Phi < 73\degr $ (green) are
     plotted. The average anisotropy factor for $0.007 < k < 0.07$
     arcmin$^{-1}$ is $Q_{\rm aver} = 3.2$ for the centroid and $Q_{\rm
       aver} = 2.6$ for the average emission.  }
   \label{Fig_Auriga_centroid}
\end{figure}

According to VCA in case of steep power spectra, thin velocity slices
can be used for a determination of the velocity power index. For
velocity fluctuations, described by a slope $m$, the expected 2D power
index for a thin velocity slice is $\gamma = -3 + m/2$. Here $m$ is
restricted to $ 0 < m < 2$ and $m = 2/3$ is distinct since it is the
Kolmogorov index \citep[][Sect. 2.4]{Lazarian2000}. In case of dominant
large-scale density fluctuations, the observed spectral slope is equal
to -8/3 if the observed velocity slice is thin and steepens to -10/3 for
thicker slices \citep[][Sect. 4.2]{Lazarian2000}. We note that in this case
in absence of a velocity shear caused by Galactic rotation turbulent
fluctuations will also produce a -8/3 spectrum in a thin slice. We
observe spectral indices close to this value for many thin velocity
channels when the \hi~gas is dominated by WNM, see Fig. \ref{Fig_Gauss}.

\begin{figure}[thp] 
   \centering
   \includegraphics[width=6.5cm,angle=-90]{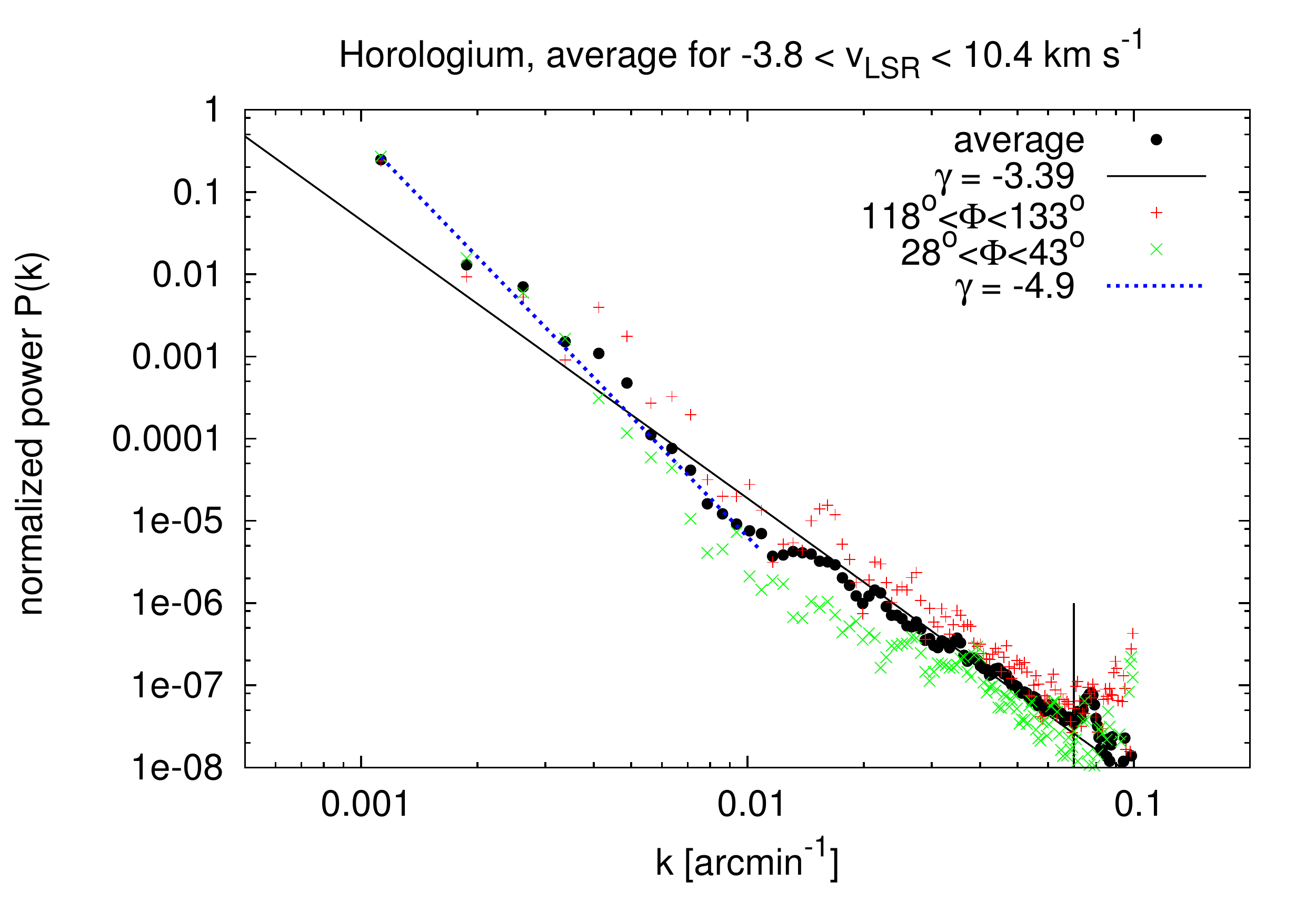}
   \caption{Power spectrum (black dots) calculated for the average
     emission over the velocity range $-3.8 < v_{\rm LSR} < 10.4$
     \kms. In addition the power spectra for $118\degr < \Phi < 133\degr
     $ (red) and $28\degr < \Phi < 43\degr $ (green) are given. An
     average power law $\gamma = -3.39 \pm 0.05 $ was fit for $ k <
     0.07$ arcmin$^{-1}$ (vertical line). For reference the power law
     fit for $ -10.2 < v_{\rm LSR} < 11.7 $ \kmss at a restricted range
     $0.001 < k < 0.01$ arcmin$^{-1}$ in spatial frequency from
     Fig. \ref{Fig_Horo_centroid} (bottom) is reproduced with the blue
     dotted line.  }
   \label{Fig_Horo_thick2}
\end{figure}

Unfortunately the VCA is only partly applicable to our targets.  An
inspection of Fig. \ref{Fig_Gauss} (bottom) shows that $m$ is
ill-defined for $-15.4 < v_{\rm LSR} < 6.5$ \kmss since $\gamma < -3$ in
this range. Interestingly, this is the velocity range where we observe
strong anisotropies and where we expect phase transitions; the \hi~gas
is cold and the WNM fraction is low, typically $T_{\rm G\, WNM}(v_{\rm
  LSR})/T_{\rm G\, WNM+CNM}(v_{\rm LSR}) \la 0.6$ for $T_{\rm D} < 1100
$ K. For a larger WNM ratio $T_{\rm G\, WNM}(v_{\rm LSR})/T_{\rm G\,
  WNM+CNM}(v_{\rm LSR}) \ga 0.7$, the 3D power index for velocity
fluctuations can be derived as $m = 0.6 \pm 0.1$, consistent with the
Kolmogorov index.

As pointed out in Sect. \ref{2d-3d}, the interpretation of the observed
2D spectral power may be affected by unknown distances and extensions of
the observed \hi~gas layer. For Horologium the \hi~gas at velocities
around $v_{\rm LSR} = -16.6 $ \kmss may be according to the rotation
curve at a distance of 1.5 kpc (Sect. \ref{Horo_aver}).  The gas at
$v_{\rm LSR} = 0 $ \kmss and (according to the rotation curve forbidden)
more positive velocities should be local. To test whether the thick
slice power spectrum, derived for $-23.1 < v_{\rm LSR} < 5.2$ \kmss and
displayed in Fig. \ref{Fig_Horo_centroid} could be biased from distance
problems we recalculated the thick slice power spectrum for a restricted
velocity range $-3.8 < v_{\rm LSR} < 10.4$ \kms. The result with the fit
$\gamma = -3.39 \pm 0.05$ is shown in Fig. \ref{Fig_Horo_thick2}. This
result is consistent with that derived for $-23.1 < v_{\rm LSR} < 5.2$
\kms. We conclude that for Horologium both velocity components share
similar properties, regardless of their distances.

Discrepancies between VCA predictions and observed spectral indices have
been noted previously by \citet{Dickey2001}. They observed significant
changes in spectral indices between regions that are considered to be
dominated by WNM or CNM. The question arises whether in our case the
Horologium field is special. But checking Fig. \ref{Fig_Gauss} (top) for
the Auriga field it is obvious that $m$ is also in this case ill-defined
for $-3.8 < v_{\rm LSR} < 5.3$ \kms. Again we find a low WNM fraction,
$T_{\rm G\, WNM}(v_{\rm LSR})/T_{\rm G\, WNM+CNM}(v_{\rm LSR}) \la 0.6$,
also strong anisotropies. For the remaining velocity channels with a
larger WNM ratio $T_{\rm G\, WNM}(v_{\rm LSR})/T_{\rm G\,
  WNM+CNM}(v_{\rm LSR}) \ga 0.7$, the power index for velocity
fluctuations is consistent with $m = 0.6 \pm 0.1$.

\subsection{Local deviations from the exponential power distribution}
\label{TI_deviations}

Turbulent power spectra consist typically of three ranges, the forcing
range, reflecting the energy input, the inertial range, reflecting the
turbulent decay, and the dissipation range. The isotropic (average)
power spectra derived by us belong to the inertial part and can usually
be fit well by a constant power law with indices that are compatible
with Kolmogorov turbulence. Our data cover a broad range in spatial
frequencies but we find no indications for a turn-over of the power
spectra at high spatial frequencies to the dissipation range. Such a
turn-over may also indicate that the observed \hi~distribution has a
limited depth along the line of sight
\citep[][Fig. 8]{Miville-Deschenes2003}.

However in case of the power spectra for velocity centroid and thick
slice emission in the Horologium field (Fig. \ref{Fig_Horo_centroid}) we
observe the opposite. We find a significant systematic steepening for
spatial frequencies $0.001 < k < 0.01$ arcmin$^{-1}$ only. This remains
valid if we restrict the velocity range used for the thick slice, see
Fig. \ref{Fig_Horo_thick2}.  The power index fit to the thick slice
emission in this range is $\gamma = -4.9 \pm 0.1$, for the velocity
centroid it is $\gamma = -4.5 \pm 0.1$. In both cases there is a step
like feature at $k \sim 0.01$ arcmin$^{-1}$ that is linking the steep
part of the power spectrum to the Kolmogorov type power law relation at
higher spatial frequencies. A similar effect, but less obvious, is
visible in the power spectrum for the average emission for Auriga, Fig.
\ref{Fig_Auriga_centroid} (bottom). The power spectra derived from the
velocity centroid (top) are rather straight with a shallower spectral
index.

Thin slice power spectra do not show a steep power law signature for
$0.001 < k < 0.01$ arcmin$^{-1}$, however we find in nearly all cases a
significant local increase of the power at $ k \sim 0.001$ arcmin$^{-1}$,
sometimes up to $ k \sim 0.002$ arcmin$^{-1}$. This effect is best
visible in Fig. \ref{Fig_spec_au_18}. 
 
A steepening of the power distribution, indicating dissipative
processes, occurs usually at high spatial frequencies at the end of the
inertial range. This does not happen in our case. In summary, we observe
the steepening at low spatial frequencies as a tilt that recovers with a
step at $ k \sim 0.01$ arcmin$^{-1}$. This change is strongest for
the density distribution (using thick velocity slices), less for the
velocity centroid, and least for thin velocity slice power spectra (Fig
\ref{Fig_spec_au_18}). A more detailed discussion of velocity centroids
is given in Appendix \ref{TI_centroid}.

\section{Phase transitions and changes in spectral power}
\label{TI_Power}

We interpret the steepening of the thin velocity slice power spectra in
a narrow velocity range, associated with a decrease of the WNM fraction
and the coexistence of cold anisotropic CNM filamentary structures as
caused by phase transitions. 

For a better understanding of the composition of the bi-stable \hi~gas
and dependencies of the power distribution at the line center
and adjacent velocities we compare similar to the analysis of absorption
lines ON and OFF data. We define the power spectrum $P_{\rm on}$ at the
velocity $v_{\rm on}$ of the steepest thin slice power spectrum. For the
OFF data we use close-by velocities $v_{\rm off}$ where the WNM fraction
is significantly larger. The selection of $v_{\rm off}$ is somewhat
arbitrary but does not affect the results significantly since we can in
any case only determine changes over a limited range of the WNM
fraction. We pick two velocities at the wings of $\gamma$ and $T_{\rm
  G\, WNM}(v_{\rm LSR})/T_{\rm G\, WNM+CNM}(v_{\rm LSR}) $, see
Fig. \ref{Fig_Gauss}.  The characteristic OFF power spectrum $ P_{\rm
  off}$ is derived then as the geometric mean of both OFF power spectra.

For the power spectra $P_{\rm off}(k)$ and $P_{\rm on}(k)$ we define
$P_{\rm on}(k) = P_{\rm off}(k) \cdot \mathfrak{T}(k) $ and accordingly
\begin{equation}
\mathfrak{T}(k) = P_{\rm on}(k) / P_{\rm off}(k),
\label{eq:TI}
\end{equation}
where $\mathfrak{T}(k)$ is the transfer function that describes changes
of the power distribution between ON and OFF. We understand
$\mathfrak{T}(k)$ as an approximation to the transfer of the 
power distribution caused by thermal instabilities. We note that we use in
Eq. \ref{eq:TI} power spectra $ P_{\rm on}$ and $ P_{\rm off}$ without
normalization.

\begin{figure}[tbp]  
   \centering
   \includegraphics[width=6.5cm,angle=-90]{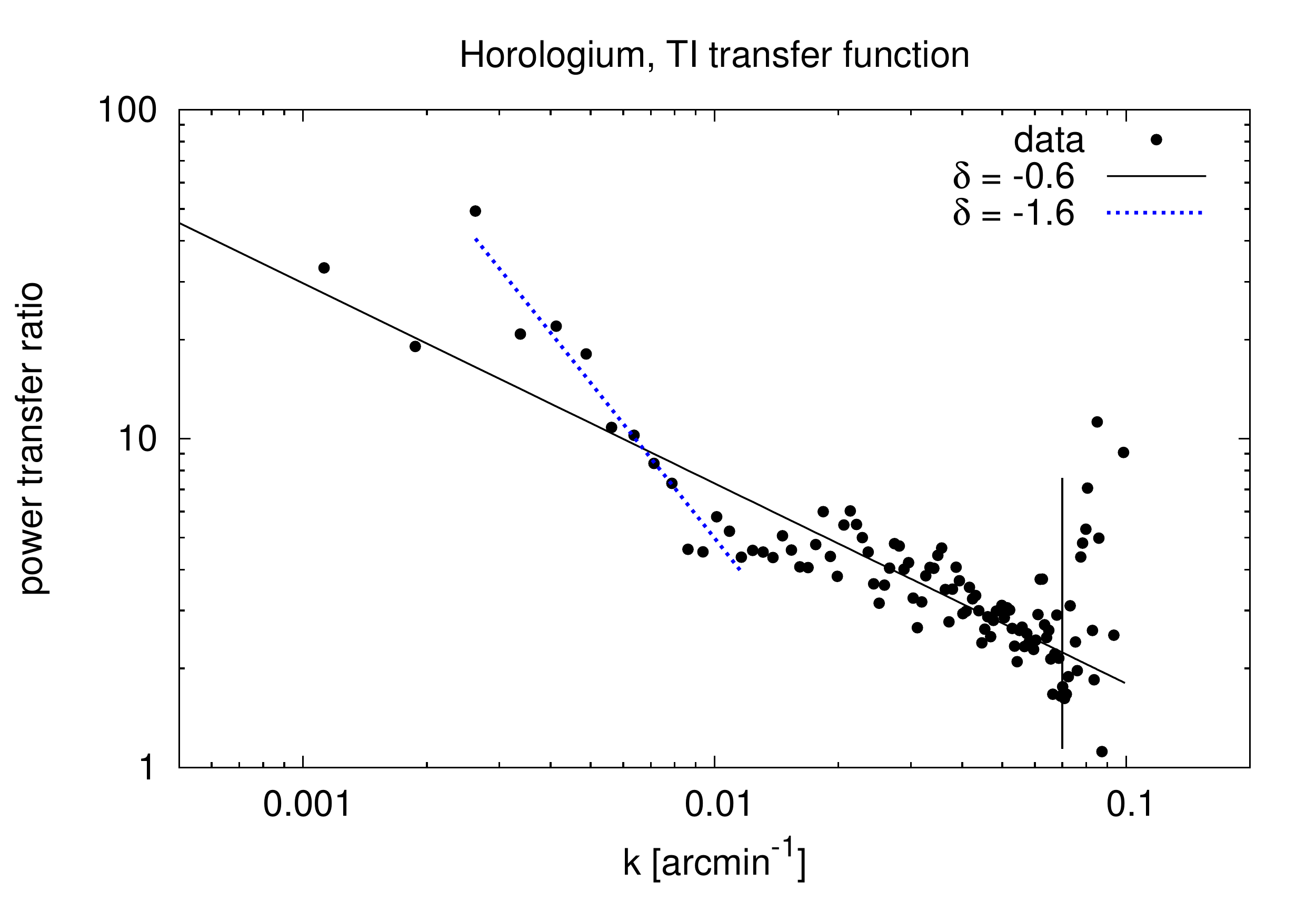}
   \caption{Changes in the spectral power distribution, most probably
     associated with phase transitions. Data points are derived
     according Eq. \ref{eq:TI}, the solid black line indicates the change 
     $\delta$ of the fit spectral indices. The blue dashed
     line represents a local fit for  $0.002 < k < 0.015$ arcmin$^{-1}$. }
   \label{Fig_Horologium_TI}
\end{figure}

\begin{figure}[thp]  
   \centering
   \includegraphics[width=6.5cm,angle=-90]{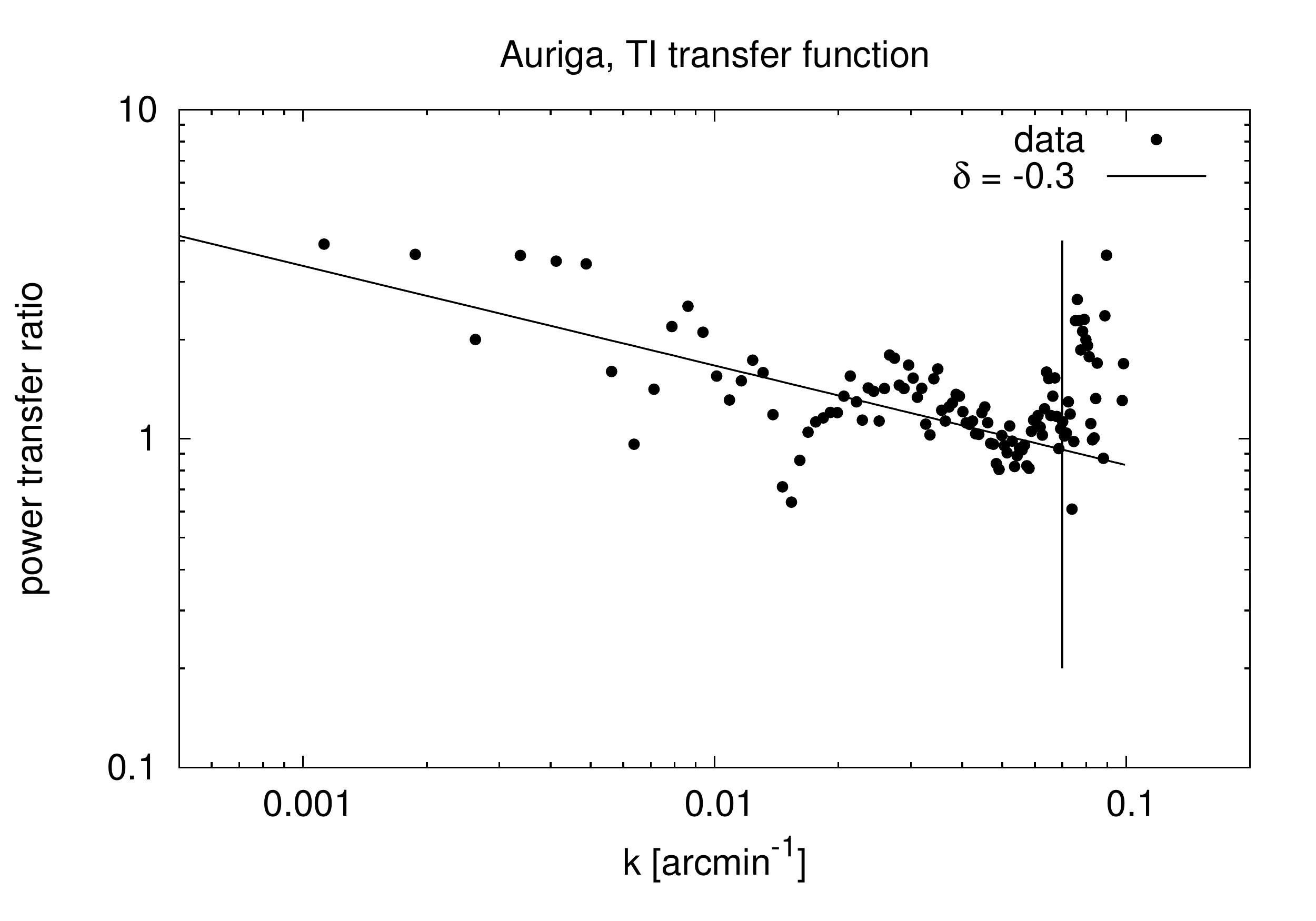}
   \caption{Changes in the spectral power distribution, most probably
     associated with phase transitions. Data points are derived
     according Eq. \ref{eq:TI}, the solid line indicates the change 
     $\delta$ of the fit spectral indices.  }
   \label{Fig_Auriga_TI}
\end{figure}

\begin{figure}[tbp]  
   \centering
   \includegraphics[width=6.5cm,angle=-90]{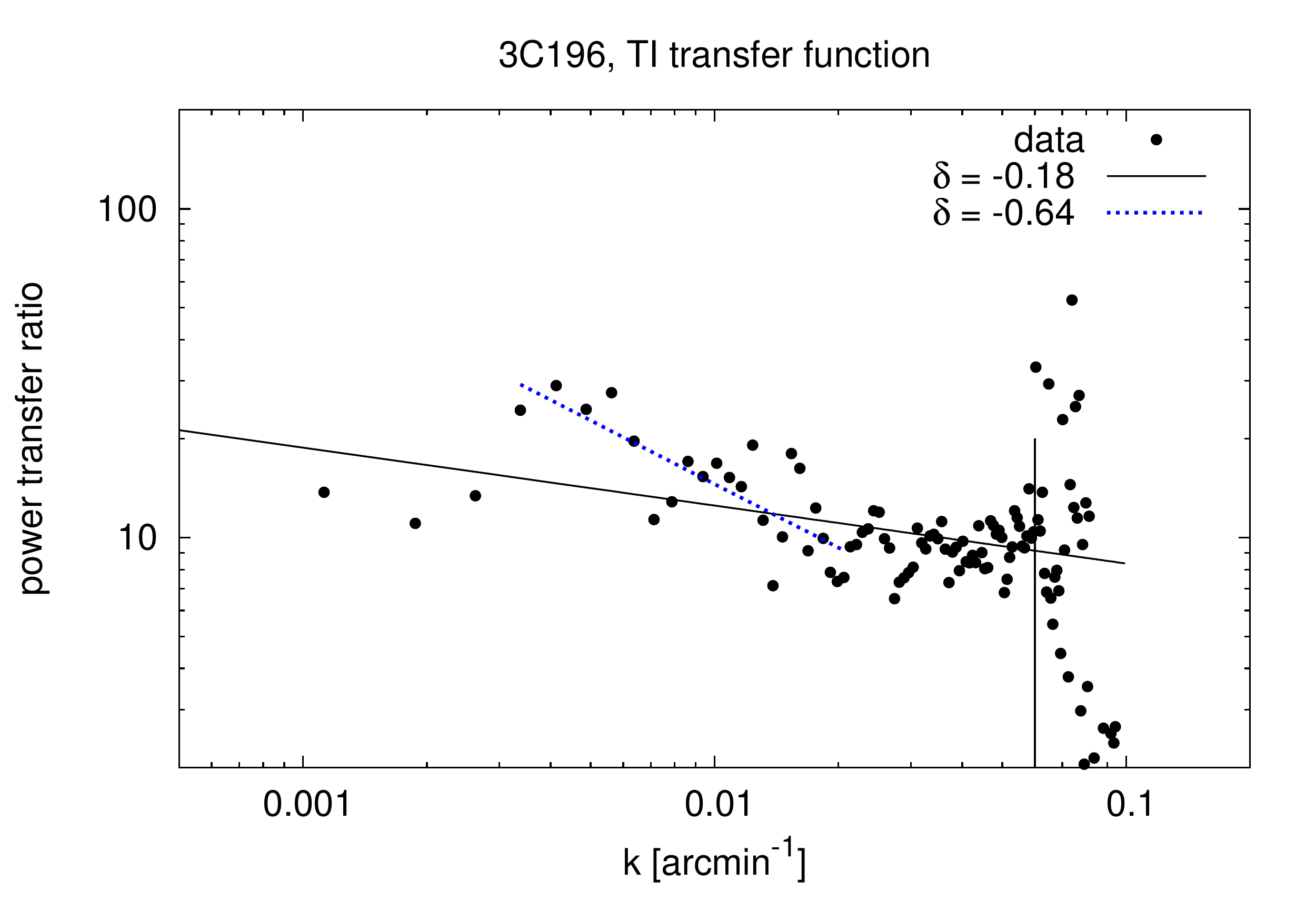}
   \caption{Changes in the spectral power distribution in direction to
     3C 196, most probably associated with phase transitions. Data
     points are derived according Eq. \ref{eq:TI}, the solid line
     indicates the change $\delta$ of the fit spectral indices. 
     The blue dashed line represents a local fit for $0.003 < k < 0.02$
     arcmin$^{-1}$.}
   \label{Fig_3C196_TI}
\end{figure}
 
To characterize the Horologium field we select power spectra at $v_{\rm
  on} = 2.7$ \kmss and $v_{\rm off} = -20.5$ and 13.0 \kms. Figure
\ref{Fig_Horologium_TI} displays $\mathfrak{T}(k)$. The change of the
average power law index is $\delta = -0.6$. For $0.002 < k < 0.015$
arcmin$^{-1}$ we find significant deviations from the average slope,
reflecting the local deviations from the power spectrum displayed in
Fig. \ref{Fig_Horo_centroid}. This part can be fit by an index $\delta =
-1.6$, the excess local steepening is comparable to the steepening
observed for the velocity centroid (Fig. \ref{Fig_Horo_centroid}). The
transition from the steep part of the $\mathfrak{T}(k)$ spectrum is at $
k \sim 0.01$ arcmin$^{-1}$, associated with strong power anisotropies,
see also Fig. \ref{Fig_ho_aver_cent}.

Remarkable is that we observe $\mathfrak{T}(k) > 1 $.  Extrapolating
$\mathfrak{T}$ we find $\mathfrak{T}(k) = 1 $ for $k \sim 0.3$
arcmin$^{-1}$. Interferometer observations would be necessary to observe
such spatial frequencies. Equal power for ON and OFF power spectra may
indicate a lower limit for the scale of eddies affected by phase
transitions.

The excess power can be explained with the observational finding that
absorption lines, indicating phase transitions, are frequently observed
close to the peaks of the \hi~emission. The power at low
spatial frequencies is there typically strongest. Consistently Fig.
\ref{Fig_HO_overview} shows that the \hi~emission is the strongest at
$v_{\rm on}$. We found also that the CNM signal, derived from the
Gaussian decomposition, has a peak at this velocity.

To derive $\mathfrak{T}(k)$ for Auriga we use $v_{\rm on} = 1.38$ \kmss
and $v_{\rm off} = -5.05$ and 6.54 \kms.  Figure \ref{Fig_Auriga_TI}
displays $\mathfrak{T}(k)$ as well as the average change in the power
spectral index, $\delta = -0.3$. For Auriga the transfer function
$\mathfrak{T}(k)$ is on average well represented by a power law,
although there is quite some scatter, in addition to a strong deviation
at $k \sim 0.015$ arcmin$^{-1}$.

We supplement here the discussion with data from \citetalias[][see their
Fig. 12]{Kalberla2016b} for the spectral index variation in the 3C 196
field. We use $v_{\rm on} = -2.47 $ \kms. Since in this case the
emission lines from the filamentary structures are affected by
line blending at positive velocities, we use only a single offset
channel at $v_{\rm off} = -11.5$ \kms to estimate the spectral power
distribution. Figure \ref{Fig_3C196_TI} shows the result. There is a
general trend for a steepening of the spectral index, but a closer look
indicates that $\mathfrak{T}(k)$ is approximately constant for $ k \ga
0.02$ arcmin$^{-1}$. This is the spatial frequency range where we
observed in \citetalias{Kalberla2016b} little anisotropies, see there
Fig. 25. $\mathfrak{T}(k)$ in Fig.  \ref{Fig_3C196_TI} shows a
significant gradient for $0.003 \la k \la 0.02$, this is the range where
strong anisotropies were detected, see Fig. 24 of
\citetalias{Kalberla2016b}.

We conclude that all three fields discussed here show significant
gradients in $\mathfrak{T}(k)$ over a broad range in spatial
frequencies. We interpret this as an indication that turbulent phase
transitions affect a range of scales.

\section{Discussion}
\label{Discussion}

\subsection{Sheets or filaments, along the  mean local magnetic
  field}

The \hi~maps in the Horologium field show numerous crowded filamentary
structures for $-23.1 < v_{\rm LSR} < 5.2$ \kms, almost all of them are
running parallel to the Galactic plane, see
Figs. \ref{Fig_Overlay_HO_2.7}, \ref{Fig_Overlay_HO-16.6} and
\ref{Fig_Horo_gal}. This is also the orientation of the
radio-polarimetric depolarization canals and the mean magnetic field
according to \citet[][Fig. 5]{PlanckXIX}. Auriga is a different case
since we have two or three \hi~features at different radial velocities
with different position angles that match to the radio-polarimetric
depolarization canals.  The dominating \hi~structure at $v_{\rm LSR} =
-2.5 $ \kmss that matches the polarized filaments well is parallel to
the Galactic plane. This is also the observed mean direction of the
magnetic field \citep[][Fig. 5]{PlanckXIX}.  In the case of 3C~196
\citepalias{Kalberla2016b}, \hi~anisotropies share clearly the
orientation of the magnetic field \citep{Zaroubi2015}.  In summary, the
best defined \hi~filamentary structures are all within a scatter of
$\sim 6\deg$ parallel to the Galactic plane and anisotropies in the
spectral power distribution are oriented along the observed mean
magnetic field direction.

While radio continuum observations in polarization map the Stokes
parameters I, Q, and U and rotation measures integrated along the line
of sight, \hi~observations are sensitive to the velocity
distribution. Each individual velocity channel reflects the average
properties at a particular velocity interval. Distinct features at
different velocities are expected to be spatially separated if the
differences in radial velocity are large enough. Hence \hi~observations
potentially allow to disentangle different \hi~layers along the line of
sight. This property can be exploited for polarized filaments correlated
with the \hi. For example, the \hi~filaments at $v_{\rm LSR} = -19.2 $
\kmss that align with the polarized filaments have an orientation 
  not parallel to the Galactic plane. This indicates that along this
line of sight some magnetic field components are on average directed
along the Galactic plane, while others are not. The correspondence
between \hi~and polarized filaments, pioneered by \citet{Clark2014},
thus yields (under the assumption of perfect alignment between
\hi~filamentary structures and magnetic fields) a method to
differentiate the different polarized filaments in distance, based on
\hi~velocity measurements.

However, projection effects are not easy to disentangle.  Filamentary
structures in \hi, associated with magnetic fields, are sheets observed
edge-on, according to \citet{Heiles2005} and \citet{Kalberla2016}, which
can be uniquely identified only if they are predominantly stretched out
along the line of sight, see Appendix \ref{Anatomy}. However,
filamentary \hi~structures may also be organized in fibers
\citep{Clark2014}. Correspondingly phase transitions may occur in these
fibers \citep{Inoue2016} or sheets \citep{Vazquez-Semadeni2006},
possibly even on angular scales of eddies embedded in a thermally
bistable turbulent medium as discussed by \citet{Vazquez-Semadeni2012}.

All of the most dominant \hi~anisotropies observed by us are limited to
narrow velocity intervals, covering typically only two channels or $
\Delta v_{\rm LSR} \sim 3$ \kms. Associated Doppler temperatures are
$T_{\rm D} \la 200$ K, spin temperatures may be as low as $T_{\rm spin}
\sim 50$ K. For a broader range in velocity we find in the $T_{\rm B}$
and USM data filamentary structures that are approximately parallel to
the prominent filaments displayed in Figs. \ref{Fig_Overlay_HO_2.7},
\ref{Fig_Overlay_HO-16.6}, \ref{Fig_Overlay_HO_2.5} and
\ref{Fig_Overlay_HO-19.2} (also Fig. 2 of
\citetalias{Kalberla2016b}). Changing the velocity of the \hi~channel
maps causes for these filamentary structures apparent position shifts
perpendicular to the filaments hence also perpendicular to the
orientation of the mean magnetic field projected on the plane of the
sky, see Appendix \ref{Anatomy}.

Such systematic velocity channel gradients are expected in the framework
of the \citet{Heiles2005} sheet model and are according to
\citet{Kalberla2016} frequently observed. \hi~sheets are coherent
structures in position-velocity space. From observations it is not
possible to disentangle the turbulent 3D density and velocity field
structure.  We emphasize that these sheets must be very cold and thin to
be observed in projection as filamentary structures.

\subsection{The correlation between \hi~and polarized filaments}

In both Horologium and Auriga fields, the directional correspondence
between polarized filaments and \hi~filaments is striking. The
correspondence in direction can be explained by the alignment of both
\hi~and polarized filaments with the mean local Galactic magnetic field,
mostly directed along the Galactic plane. However, the tight
correspondence between the widths of the filaments is more puzzling. In
particular, Figs. \ref{Fig_Overlay_HO_2.7}, \ref{Fig_Overlay_HO-16.6},
\ref{Fig_Overlay_HO_2.5} and \ref{Fig_Overlay_HO-19.2} seem to show an
anti-correlation between \hi~and polarized intensity along a
filament, usually bordered by a radio-polarimetric depolarization canal.

According to \citet{Heiles2012}, the electron density in the CNM is
insufficient to produce a measurable Faraday rotation along the line of
sight for the extent of the filamentary CNM structure. This implies that
the filamentary \hi~structures need to be almost co-located with ionized
gas which gives rise to the observed Faraday rotation from the
magneto-ionic medium.

If the correlation can be explained by CNM filaments wrapped in an
ionized envelope, the envelope needs to have substantial width and
electron density to explain the observed RMs in these polarized
filaments which are typically a few radians~m$^{-2}$
\citep{Haverkorn2003a, Haverkorn2003b}. (However, we note that care should
be taken in interpreting mixed synchrotron-emitting and Faraday-rotating
gas in terms of traditional RMs, see \cite{Brentjens2005}.) These low
RMs could be explained by a line-of-sight through the ionized gas of
$\sim1$~pc, combined with an electron density of $\sim 1$~cm$^{-3}$ and
a mean magnetic field of a few microgauss, which are very reasonable
values.

On the other hand, the anti-correlation between \hi~and ionized gas
within a filament may indicate that these filaments are the locations of
phase transitions from warm ionized gas to neutral medium.
\citet{Heiles2012} argue for an intermediate gaseous phase, the warm
partially ionized medium (WPIM), producing and perhaps dominating the
ISM's Faraday Rotation. It appears promising to explore the relations
between filamentary CNM structures and magneto-ionic structures in more
detail but this is beyond the scope of the current paper.

\subsection{Comparison to earlier observations of the 3C 147 field}
\label{3C_147}

Our analysis is based on EBHIS data with a moderate resolution of
10\farcm8. It is worth to revisit observations of 3C 147, making use of
combined data from the Effelsberg and WSRT telescopes at a resolution of
1 to 3 arcmin \citep{Kalberla1985}. The WSRT observations on 3C 147
sampled the UV plane completely without tapering and needed no
cleaning of the data. After self-calibration a dynamical range of 38 dB
was obtained.  For a field of view of $36\arcmin$ \citet{Kalberla1983}
have determined a spectral index of $\gamma = -2.5 \pm 0.3 $ for $-2 <
v_{\rm LSR} < 3 $ \kms, the velocity of the main emission line. Other
results from this data set, which are unpublished, are $\gamma = -2.75
\pm 0.3 $ for $-14 < v_{\rm LSR} < -11 $ \kms, $\gamma = -3 \pm 0.3 $
for $-9 < v_{\rm LSR} < -7 $ \kms. In all of these cases the \hi~gas is
cold and has a considerable optical depth. All thin slice power spectra
are steeper than the thick velocity slice power spectrum for $-14 <
v_{\rm LSR} < 4 $ \kmss with $\gamma = -2.33 \pm 0.3$.

The \hi~with the steepest spectral index at a velocity of $v_{\rm LSR}
= -8.1 $ \kmss is organized in a weak but well defined filamentary
structure that is aligned parallel to the Galactic plane \citep[][their
Figs. 2d and 4d]{Kalberla1985}. EBHIS data show that this \hi~component
is clearly associated with a minimum in the WNM fraction. The
determination of the WNM fraction from $T_{\rm G\, WNM}(v_{\rm
  LSR})/T_{\rm G\, WNM+CNM}(v_{\rm LSR}) $ depends critically on the
Doppler temperature of $T_{\rm D} < 1100 $ K that we have chosen for a
selection of CNM components, but this assumption leads in case of 3C~147
to consistent results. The spin temperature of this \hi~component with
an optical depth $ \tau = 0.87 \pm 0.01$ was determined to $T_{\rm spin}
= 32 \pm 18$K from Gaussian decomposition of WSRT data, and to $T_{\rm
  clump} = 34 \pm 17$ K taking self-absorption in a clumpy medium into
account. This is in good agreement with the EBHIS data, which indicate
for the CNM at this velocity a brightness temperature of $ T_{\rm G} =
30$ K. 

VLBI observations show for this feature with the highest optical depth
the most significant structures in column densities, with strong
temporal and spatial fluctuations at spatial frequencies corresponding
to scales as low as 10 AU \citep{Diamond1989,Faison2001,Lazio2009}.  The
high resolution \hi~data at $v_{\rm LSR} = -8.1 $ \kmss in direction to
3C 147 are consistent with the power spectra for our EBHIS targets
on scales of 10\arcmin. A more detailed discussion on 3C 147 is beyond
the scope of this publication but ``mysterious'' optical depth
structures at AU scales \citep{Deshpande2000} may find a simple
explanation if MHD anisotropies in presence of a magnetic field are
taken into account.

The steepening of the 3C 147 power spectrum, $\gamma = -3 \pm 0.3 $ for
$-9 < v_{\rm LSR} < -7 $ \kms, relative to the power spectra at other
velocities implies a decrease of turbulent fluctuations in the source
structure at high spatial frequencies, hence a low VLBI signal is
expected. Yet, observations by \citet{Diamond1989,Faison2001,Lazio2009}
show significant opacity and column density fluctuations that are
apparently in conflict with such an assumption. \citet{Deshpande2000}
explains small scale structure at AU scales with a single constant power
law over all observed scales. Highly over-dense \hi~small-scale
structures are proposed not to be real but a misinterpretation of a
turbulent source distribution. For this argumentation anisotropies have
not been taken into account.

\hi~opacity fluctuations on AU scales are observed perpendicular
to the direction of the filamentary structure observed with the
WSRT. According to \citet{Goldreich1995} the power of the turbulence
spectrum is strongest in direction perpendicular to the observed
  mean magnetic field direction, one or two orders of magnitude are
possible, much in favor of high VLBI visibilities.

\subsection{Possible physical explanations}
\subsubsection{The McKee and Ostriker (1977) model}

The \citet{McKee1977} model of the interstellar medium predicts an
association between cold \hi~gas and ionized medium: ``whenever
  there is a cold cloud along a given line of sight, at a given
  velocity, there should be warm ionized material as well at the same
  velocity''\footnote[4]{Inverting this argument, there may be
  particular cases where warm ionized material could indicate the
  presence of associated CNM. This is the conjecture that led to the
  current investigations.}. Around CNM cores they sketch an onion skin
structure for the transitions between CNM, WNM and surrounding ionized
gas \citep[][Fig. 1]{McKee1977}. In case of a regular magnetic field
anisotropies are expected and onion-skins should get deformed. For the
model of cold \hi~sheets, proposed by \citet{Heiles2005} and advocated
by \citet{Kalberla2016}, we should obtain instead of onion-skins
flattened and elongated structures, observed nearly edge-on. 
``Edge-on sheets should be edge-on shocks in which the field is
  parallel to the sheet'' \citep{HeilesCrutcher2005}. Then, the
\citet[][Fig. 1]{McKee1977} model remains valid if we replace the
simplified spherical clouds with scale dependent anisotropic eddies, see
Figs. 2, 28 and 29 of \citetalias{Kalberla2016b} .

Such a geometry is complicated and line-of-sight effects are
important. One should not expect always to observe a clear association
between \hi~and Faraday depth structures. However, a close alignment of
anisotropies is mandatory and such anisotropies are observed by us.
\hi~channel maps at other velocities show some more cases with
alignments but we can demonstrate here only the most prominent cases. A
detailed modeling of EBHIS against LOFAR or WSRT data is not available
but we believe that our results are consistent with the
\citet{McKee1977} model with a sheet geometry as proposed by
\citet{Heiles2005}, see Appendix \ref{Anatomy}.

\subsubsection{Ideal MHD turbulence}

Postulates for ideal 3D MHD turbulence are quite different. Density and
velocity fields are assumed to be Gaussian and independent. 
``\hi~data cubes exhibit a lot of small-scale emissivity
  structure. The question is what part of them is real, i.e., is
  associated with density enhancements in Galactic coordinates, and what
  part is produced by velocity fluctuations''
  \citep[][Sect. 6.3.5]{Lazarian2000}.  Accordingly density
fluctuations with Gaussian distribution and power spectra result in
filamentary structures that do not need to be ``real''.  To our
understanding neither the observed \hi~structures, nor the polarized
structures, are random or independent in density and velocity, see
Appendix \ref{Anatomy}. \hi~sheets are coherent structures in
position-velocity space with distinct velocity gradients orthogonal to
the observed orientation of filament and  mean magnetic field.  Steep thin
slice power spectra reflect that the usual model assumptions of
independent Gaussian distributions in density and velocity are in our
case not applicable. In our case the VCA is not useful for a
determination of the spectral power distribution of the turbulent
velocity field.

\subsubsection{Anisotropies according to Goldreich-Sridhar}

A turbulence model allowing directional anisotropies is the
\citet{Goldreich1995} model, based on the assumption of a critically
balanced shear Alfv{\'e}nic cascade.  It is questionable whether in case
of shocks and induced thermal instabilities such a balance can
develop. The predicted systematic increase of anisotropies with spatial
frequency \citep{Goldreich1995} is only observed at a few velocity
channels in the 3C 196 field.  Anisotropies appear to have considerable
fluctuations, with local enhancements at different spatial frequencies.
The balance depends on the Alfv{\'e}n velocity, $V_{\rm A} \propto
\langle |\mathbf{B}| \rangle/\sqrt{\langle \rho \rangle}$, usually
defined as the average over the field of view. In case of phase
transitions the density $\rho$ can change by orders of magnitude
\citep{Wolfire2003}. The magnetic field strength may also be affected
but according to \citet{Crutcher2010} the magnetic field in the diffuse
ISM does not scale with density for $\rho < 300$ cm$^{-3}$. Thus
turbulent anisotropy in the \hi~may be more complex than indicated by
these models.  However, more advanced MHD turbulence simulations are
able to reproduce, at least qualitatively, the observed filamentary
\hi~and polarized structures, see for example \citet{Mallet2017}. 

Here we do not analyze velocity anisotropies. But such anisotropies are
predicted by \citet{Goldreich1995} and have been analyzed for example by
\citet{Heyer2008}. They identify a velocity anisotropy that is aligned
within 10\degr of the mean local magnetic field direction derived from
optical polarization measurements.

 \subsubsection{Filamentary structures from colliding flows}

Recent simulations of molecular cloud formation by \citet{Gomez2014}
have demonstrated that filaments can develop self-consistently by
colliding flows. The WNM, affected by thermal instabilities, initially
forms planar structures (or sheets) which become later Jeans-unstable
and break into filamentary structures. Thus the collapse occurs as a
cascade where structures at different scales have different
morphologies: from sheets to filaments and from filaments to clumps. As
discussed by \citet[][their Fig. 5]{Hartmann2001} clouds tend to form at
bends or ``kinks'' in the magnetic field and are also associated with
rapid changes in the field direction. Similar, along the filamentary
structures there are rapid changes in the velocity field. Thus changes
in density, velocity, and magnetic field direction are found to be
correlated. The compression of the parallel magnetic field in such a
configuration can cause a delay, but cannot prevent post-shock gas from
compressing to high densities with phase transitions. This scenario fits
well to our observations, provided that the filamentary CNM structures
are associated with ionized gas, giving rise to the observed Faraday
polarization filaments.

\subsubsection{Filaments from turbulent velocity strain }
\label{strain}

In incompressible turbulence, some structures tend to be filamentary and
it was shown by \citet[][Figs. 6 and 7]{Moisy2004} that those structures
with the highest rate-of-strain and dissipation are in the form of
sheets or ribbons.

\citet{Hennebelle2013} performed a series of numerical simulations to
study the formation of clumps in a turbulent ISM. The result was that
clumps in MHD simulations tend to be more filamentary than clumps in
hydrodynamical simulations which are preferentially sheet-like.
Such linear filamentary structures result from the stretch induced by
turbulence and the filaments are in general preferentially aligned with
the strain. As the filaments are getting stretched, the magnetic
field is amplified and becomes largely parallel to the filament.

\citet{Inoue2016} study the formation of CNM in a shock-compressed
layer. A magnetized thermally unstable gas layer develops behind the
shock wave. Thermal instabilities cause fragments of filamentary CNM
structures which align with the magnetic field if the shear flow is
strong enough. For a weak shear strain along the magnetic field they
observe an increase of the number of linear structures that lie
perpendicular to the magnetic field. 

\subsubsection{Thermal instabilities and shocks}

An other possibility for the formation of filamentary magneto-ionic
structures is through thermal instabilities, as was previously discussed
by \citet{Jelic2015}. According to \citet{Choi2012} the evaporation of
cold gas from the surfaces of dense structures is strongly suppressed
in regions where the field is parallel to the interface. This
attenuates turbulence generated by thermal instabilities and supports
the alignment between filamentary CNM and associated ionized
structures. 

Finally, we consider the possibility of low-Mach number shocks as cause
for the filaments.  For all of our velocity channel maps with
anisotropies we observe a series of CNM filaments, roughly parallel and
structured similar to washboard waves. \citet{Fletcher2007} consider the
case that radio-polarimetric depolarization canals are produced by a series of shocks with
low Mach numbers, $M \sim 1.2$. As the primary source of shocks they
assume supernovae. When these shocks encounter gas clouds they will be
reflected, generating a population of secondary shocks. For such a
foreground Faraday screen they estimate discontinuities to have an
average distance of 15\arcmin, corresponding to a spatial frequency of
$k \sim 0.07$ arcmin$^{-1}$. We find for Horologium peak anisotropies at
the same spatial frequency, for Auriga at $k \sim 0.09$ arcmin$^{-1}$,
and for 3C 196 at $k \sim 0.06$ arcmin$^{-1}$. This is a reasonable
agreement and a series of shock waves may according to \citet{Saury2014}
explain the observed filamentary \hi~structures. Such shocks must lead in
the presence of a magnetic field to the observed strong anisotropies
\citep{Goldreich1995}.
  
A steepening of the slope of the power spectrum implies that the power
distribution is predominantly affected by dissipative events at small
scales although we do not find any significant break in the power
distribution. The filamentary CNM structures at low Doppler temperatures
imply recent phase transitions. \citet{Saury2014} have shown that for
typical values of the density, pressure and velocity dispersion of the
WNM turbulent motions of the \hi~cannot provoke the phase transition
from WNM to CNM. This is found to be valid regardless of amplitude and
distribution for turbulent motions in solenoidal and compressive modes
and implies that the WNM must be pressurized for phase transitions. Such
a model is consistent with the observed shell stucture (see Appendix
\ref{Anatomy}) but \citet[][Figs. 18 and 19]{Saury2014} find shallow
spectral indices after phase transition from WNM to CNM.

\subsection{MBM 16 - shear flow and thermal instabilities}

Our comparison field MBM 16, see Appendix \ref{MBMpower}, differs from
the targets with a filamentary radio-polarimetric structure. In
radio-continuum this source does not show any structure of particular
interest. MBM 16 contains molecular gas but there are no indications for
star formation \citep{Pingel2013}. Furthermore, this field has an
almost perfect isotropic \hi~column density distribution. Some
filamentary structure of the molecular gas does exist but the
orientation of the mean magnetic field direction, determined from
optical polarization data \citep{Gomez1997}, is aligned in perpendicular
direction. The steepest spectral index for MBM 16 is found at a velocity
which is close to the peak of the average CO line emission
\citep[][Fig. 2]{Pingel2013}.

From H$_2$CO observations \citet{Magnani1993} suggest that the molecular
gas is caused by a large-scale shear instability. We observe that the
velocity for the coldest \hi~gas, containing the least amount of WNM, is
shifted systematically with respect to the velocity of the gas with the
steepest spectral index. \citet{LaRosa1999} argue that coherent
molecular gas structures may originate from an externally driven
turbulent shear flow, causing both MHD and fluid turbulence on small
scales. Our observations give some evidence for dynamical interactions
and support this proposal.

\section{Summary and Conclusions}
\label{Summary}

Using methods introduced in \citetalias{Kalberla2016b}, we study here
two WSRT fields with well defined straight polarization filaments and
associated radio-polarimetric depolarization canals, named Horologium and Auriga
\citep{Haverkorn2003a,Haverkorn2003b,Haverkorn2003c}.

EBHIS \hi~observations show \hi~filaments aligned with the polarized
filaments in both fields. Figures \ref{Fig_Overlay_HO_2.7},
\ref{Fig_Overlay_HO_2.5}, \ref{Fig_Overlay_HO-19.2}, and in particular
Fig. \ref{Fig_Overlay_HO-16.6} show \hi~features that are well aligned
with the polarized filaments. A similar good alignement was found in
\citetalias{Kalberla2016b}, see there Fig. 2. The \hi~power spectra in
both fields discussed here show significant anisotropy, with the
steepest power spectra aligned along the polarized filaments. Most
prominent filaments are aligned with the Galactic plane, which is
presumed to be the direction of the mean (local) Galactic magnetic
field. In Auriga, one prominent \hi~filament aligns with a polarized
filament but not with the Galactic plane, taken to indicate a mean 
local Galactic magnetic field direction which deviates from parallel to
the Galactic plane.

The anisotropic filamentary \hi~structures are cold.  For the Auriga
field the spin temperature can be estimated from 3C 147, at $v_{\rm LSR} = -8.1 $
and 0.2 \kms, to be $T_{\rm spin} = 34 \pm 17$ or $ 37 \pm 24 $ K
respectively. In case of 3C 196 \citepalias{Kalberla2016b} $T_{\rm spin}
= 80.9$ K. No spin temperature can be derived for the Horologium field
but also here the derived geometrical mean Doppler temperature, an upper
limit to the spin temperature, is low, $T_{\rm spin} < T_{\rm D} \sim
167$ K.  Anisotropies in \hi~power-spectra  tend in all cases to be
strongest for velocity channels with low Doppler temperatures.

We find that local minima in $T_{\rm D}$ are associated with minima in
the WNM ratio $T_{\rm G\, WNM}(v_{\rm LSR})/T_{\rm G\, WNM+CNM}(v_{\rm
  LSR})$, determined from a Gaussian analysis. We conclude that
\hi~anisotropies must be associated with phase transitions.  We also
find evidence for a distinct steepening of the average spectral power
distribution.

To our best knowledge, steep thick slice \hi~power spectra with indices
as low as $\gamma = -4.9 $ and associated steep thin velocity slice
power spectra with indices $\gamma \la -3.2 $ have not been
reported before. To study this steepening in more detail, we compared
the steepest power spectra with more shallow power spectra at velocities
close-by where the WNM fraction is significantly larger.  We define the
power transfer function $\mathfrak{T}(k)$ (Eq. \ref{eq:TI}) to
characterize spatial changes in the power distribution between
the coldest \hi~gas with the least WNM content and its environment.

$\mathfrak{T}(k)$ shows a steepening over a broad range in spatial
frequencies and we interpret this as an indication that also a broad
range of scales in the image plane must be affected by phase
transitions.  Model calculations by \citet{Hennebelle2013} indicate that
phase transitions occur predominantly at scales around 1 pc. The strong
steepening on scales $ k \la 0.01 $ or 0.02 arcmin$^{-1}$ imply that
phase transitions are most dominant on larger scales. Phase transitions
obviously do not occur locally in isolated spots but rather
simultaneously in a correlated way on larger scales, perhaps in larger
eddies. \citet{Heiles2005} propose transitions in sheets, Appendix
\ref{Anatomy} supports such structures. This is consistent with
\citet{Hennebelle2013}, who found that the primary cause of the
  existence of filaments (or elongated clumps) is the stretching of the
  fluid particles induced by turbulent motions. He argues that
the magnetic field helps to keep filamentary structures coherent and is
playing a determining role in the formation of filamentary structures.

Steep power spectra for thin velocity slices were detected also for two
other targets that are not associated with radio-polarimetric
depolarization canals.  Firback North 1 \citep[FN1,][]{Dole2001} was
studied in \citetalias{Kalberla2016b}. Steep thin velocity slice power
spectra in FN1 are associated with anisotropies and cold CNM at a
Doppler temperature $T_{\rm D} \sim 250$ K. For MBM 16, discussed in
detail in Appendix \ref{MBMpower}, we find no significant anisotropies
but the steep spectral index is associated with molecular gas at a
similar velocity. Remarkable is that the cold CNM with $T_{\rm D} \sim
120$ K is offset from that in velocity by a few \kms, indicating
dynamical interactions.

The radio-polarimetric targets discussed here, Horologium and Auriga,
but also 3C 147 are located within the Fan Region. Here the mean field
is remarkably well aligned with the Galactic plane, as indicated by both
the dust polarization maps from \citet[][Fig. 5]{PlanckXIX} and and the
synchrotron polarization known from much earlier work since
\citet{Brouw1976}. We find a qualitative agreement of the mean magnetic
field direction with anisotropies in the \hi~distribution, caused by
filamentary \hi~structures that are associated with magneto-ionic
structures (canals).  We can not provide a proof that these structures
are physically related to each other. However, coincidences exist for
several targets: Horologium at $v_{\rm LSR} = 2.7 $ and --16.6 \kms,
Auriga at $v_{\rm LSR} = -2.6 $ \kms, also 3C 196 at $v_{\rm LSR} = -3.8
$ \kms, and 3C 147 at $v_{\rm LSR} = -8.1$ \kmss (only \hi~data
available). Together, all these cases suggest that an ordered magnetic
field, giving rise to polarized filamentary structures and associated
anisotropic \hi~emission  may play a significant role for phase
transitions of compressed cold \hi~gas in a sheet-like geometry. We find
\hi~gas with steep thin slice power spectra that is associated with cold
gas. For our targets this coincidence is detected on scales
between 10 AU (3C 147) up to tens of pc.

\begin{acknowledgements}
  We acknowledge the referee for careful reading, concise and
  constructive criticism. P. K. thanks A. Lazarian for discussions at
  the ISM~2017 conference in Cologne. B. Winkel helped us with the
  generation of EBHIS maps in 1950.0 equatorial coordinates. We thank
  C. Heiles for stimulating comments. J. K. thanks the Deutsche
  Forschungsgemeinschaft (DFG) for support under grant numbers
  KE757/7--1, KE757/7--2, KE757/7--3, and KE757/11--1.
  U. H. acknowledges the support by the Estonian Research Council grant
  IUT26-2, and by the European Regional Development Fund
  (TK133). M. H. acknowledges the support of the research program
  639.042.915, which is (partly) financed by the Netherlands
  Organisation for Scientific Research (NWO). This research has made use
  of NASA's Astrophysics Data System.  EBHIS is based on observations
  with the 100-m telescope of the MPIfR (Max-Planck-Institut f\"ur
  Radioastronomie) at Effelsberg.
   \end{acknowledgements}

\appendix

  \section{Processing of power spectra: apodization versus median}
\label{MBMpower}

Our data processing methods, introduced in \citetalias{Kalberla2016b},
were developed with the aim to allow an unbiased derivation of
anisotropies in the \hi~spectral power distribution. Usually the
analyzed region has a rectangular shape. The response of the window
function is recognizable as a strong central cross in the power
distribution (see Fig. 4 left in \citetalias{Kalberla2016b} or Fig. 6 in
\citet{Pingel2013}). The observing process implies further that the data
are convolved by the beam function, in addition instrumental noise is
added to the genuine signal. To avoid observational biases we apodize
the data by a circular taper, deconvolve for the beam function and
subtract the statistical noise bias $N(k)$, see Eq. \ref{eq:Pav},
\citetalias[][Sect. 4 and there the presentation in Figs. 3 to
6]{Kalberla2016b}. When fitting spectral indices we use a cut-off at
high spatial frequencies to ensure that data below a S/R of three are
excluded.

According to the convolution theorem apodization leads to a smoothing of
the data in the Fourier transformed domain. This process is equivalent
to the smoothing caused by the primary beam of a synthesis telescope
\citep{Green1993} but in our case the equivalent ``apodization beam'' is
defined by the observed region. The formal FWHM diameter of the
apodization function is in our case 10\degr~ for the Auriga and
Horologium fields and 9\degr~ for the MBM 16 field considered below. In
the UV plane this smoothing function has 20 dB sidelobes however we do
not expect a significant degradation of the derived power spectra since
we matched the sampling in the UV plane to the FWHM diameter of the
smoothing function. Binning of the power spectra in spatial frequencies
was chosen to limit the FWHM size of the smoothing function to 70\%
(80\% in case of MBM 16) of the selected bin. The smoothing in the UV
plane, caused by apodization, is independent of $k$. Accordingly we have
chosen a constant sampling and all $k$ samples of the power spectra may
be considered as nearly independent. In case of position dependent power
spectra there is an exception. For sectors with a width of 15\degr~
samples at $ k < 0.003$ arcmin$^{-1}$ are not independent. This is only
important for the derivation of anisotropies and we do not interpret
anisotropies for such values.

When deriving power spectra we find significant changes in the derived
spectral indices, see Figs. 12, 13, and 18 in
\citetalias{Kalberla2016b}, also Figs. 1, 8 and the discussion in
Sect. \ref{Spectral_index}. Such a steepening was not observed before
and could therefore be related to a systematic effect in our
methodology. For clarification we use here a case study and compare
results from our data processing with recently published results
obtained from more ``standard'' data processing methods that are so far
considered to be robust against systematic effects, for discussion see
for example \citet{Martin2015,Blagrave2017}. These methods rely on median
filtering and the results are assumed to be equivalent to those using
apodization for power distributions with azimuthal symmetry in the UV
plane \citep[e.g.,][]{Miville-Deschenes2002}.

To evaluate differences in data processing we consider the starless
molecular cloud MBM 16 (suggested by the referee) and compare our
results with those of \citet{Pingel2013}. These authors used the Arecibo
telescope to derive the turbulent properties of \hi~in this region for
the range 42\fdg5 < RA < 56\fdg25 and 4\deg < DEC < 16\deg (J2000). We
select EBHIS observations for the same region.

\begin{figure}[htbp]  
   \centering
   \includegraphics[width=6.5cm,angle=-90]{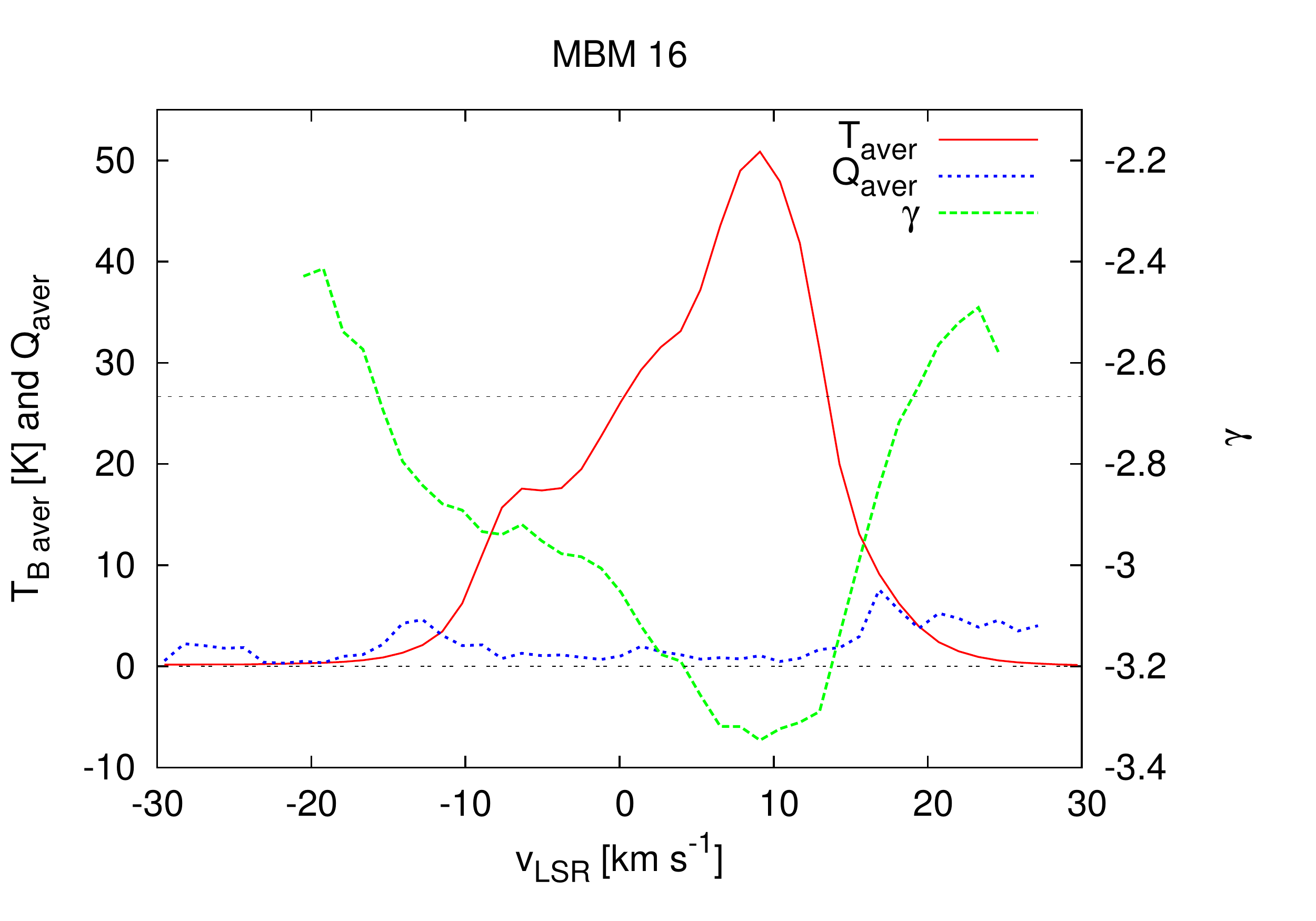}
   \caption{Average brightness temperature $T_{\rm B\, aver}$ (red),
     anisotropy factor $Q_{\rm aver}$ (blue) for $0.007 < k < 0.06$
     arcmin$^{-1}$, and spectral index $\gamma$ (green) for the MBM 16
     field as derived from EBHIS observations. The upper horizontal
     black dotted line indicates the Kolmogorov spectral index of $\gamma =
     -8/3$, the lower dash-dotted line $Q_{\rm aver} = 0$. }
   \label{Fig_MBM_overview}
\end{figure}

\begin{figure}[thp]  
   \centering
   \includegraphics[width=6.5cm,angle=-90]{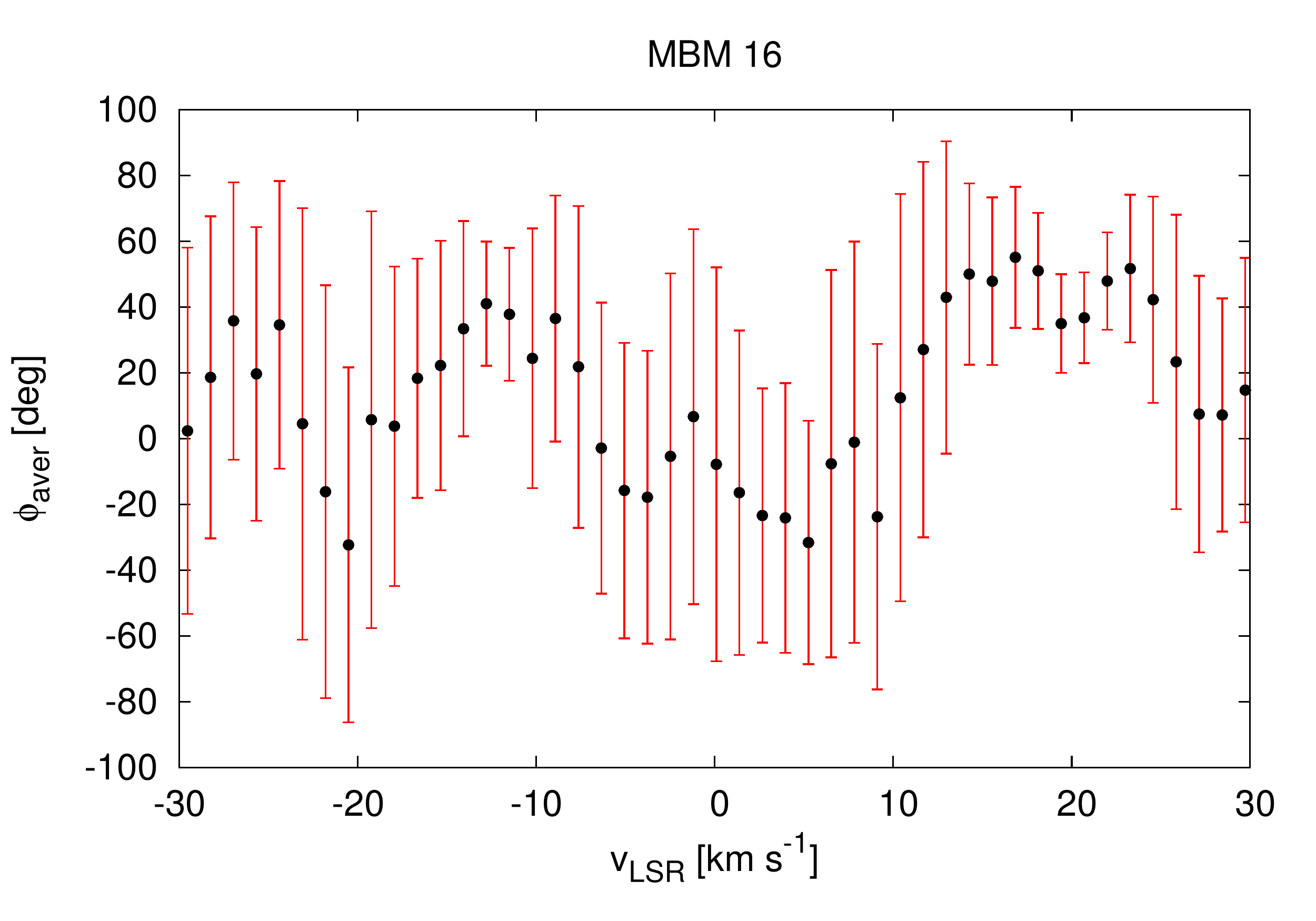}
   \caption{Average position angles $\Phi_{\rm aver}$ calculated for $
     0.007 < k < 0.06 $ arcmin$^{-1}$ and associated one $\sigma$ 
       rms scatter for the MBM 16 field. }
   \label{Fig_MBM_overview_angle}
\end{figure}

The data processing is identical to that for the targets described in
Sects. \ref{Horologium} and \ref{Auriga}.  Figure \ref{Fig_MBM_overview}
displays the average brightness temperature $T_{\rm B\, aver}$,
the anisotropy factor $Q_{\rm aver}$ for $0.007 < k < 0.06$
arcmin$^{-1}$, and the derived spectral index $\gamma$ for the
MBM 16 field. Figure \ref{Fig_MBM_overview_angle} shows the
corresponding average position angles for $Q_{\rm aver}$. The scatter of
$Q_{\rm aver}$ is large and a visual inspection of the data cube confirms
that, except for velocities $ v \ga 14 $ \kms, anisotropies are
insignificant. Complications from source anisotropies are thus not
expected and we conclude that the MBM 16 field is an ideal case for a
comparison of different data processing methods.

\begin{figure}[thp]  
   \centering
   \includegraphics[width=6.5cm,angle=-90]{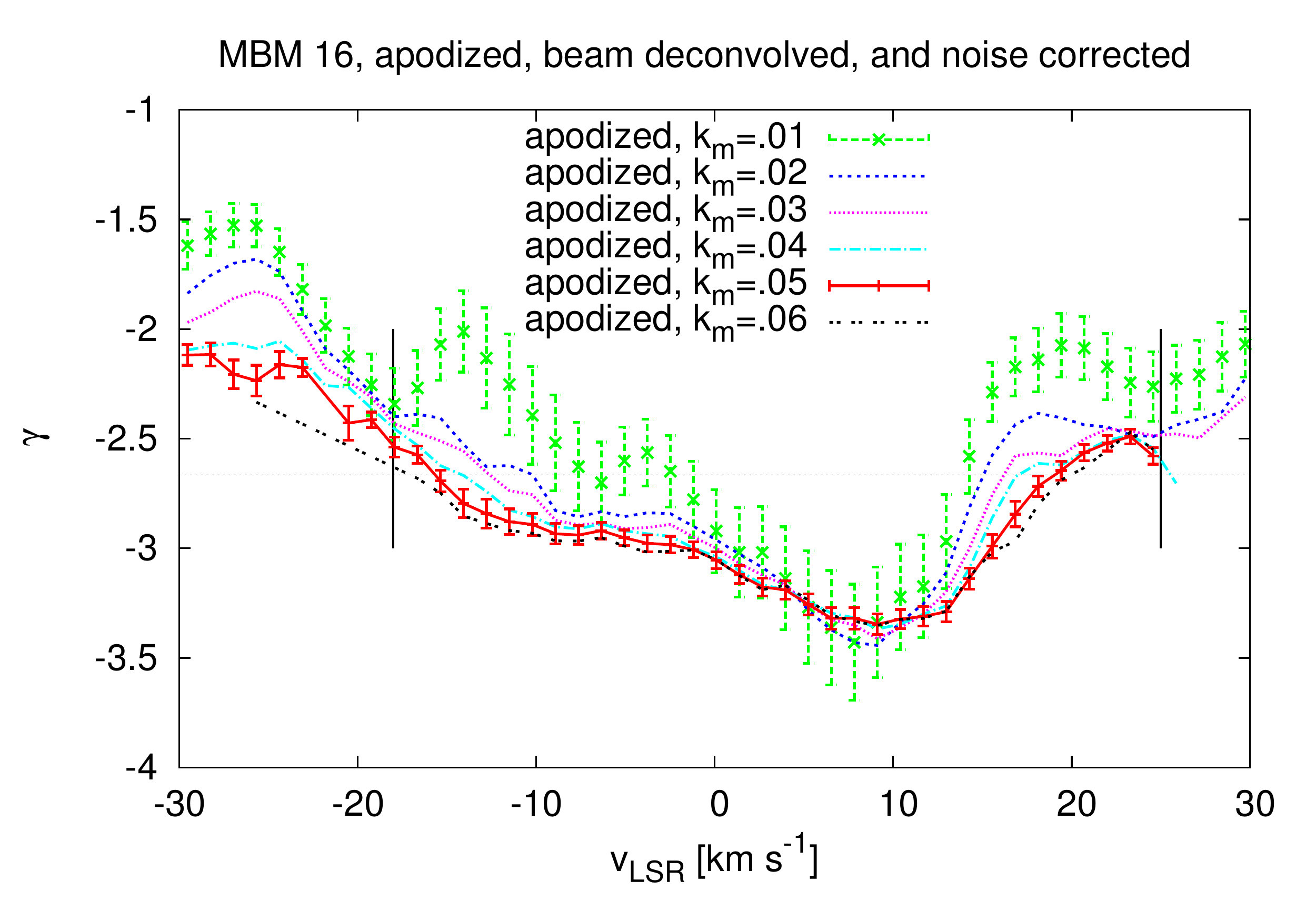}
   \caption{Comparison of results according to the method proposed in
     \citetalias{Kalberla2016b} (using apodization, beam correction, and
     noise subtraction) obtained from fits with different upper limits
     $0.01 < k_{\rm m} < 0.06$ arcmin$^{-1}$ in spatial frequency. Error bars are
     given in two cases to provide an impression of formal uncertainties
     of the least square fits. The horizontal black dotted line
     indicates the Kolmogorov spectral index $\gamma = -8/3$, the solid
     vertical lines bracket the velocity range with average brightness
     temperatures $T_{\rm B\, aver} > 0.5$ K.  }
   \label{Fig_MBM_compare22}
\end{figure}

Using the procedures introduced in \citetalias{Kalberla2016b}
(apodization, beam deconvolution and noise bias correction) we derive
mean power spectra for the MBM 16 field and fit the spectral indices
$\gamma$. To avoid biases from instrumental noise we usually restrict
the least square fits to the power spectra by applying a constant three
sigma limit $k_{\rm m}$ at high spatial frequencies. To demonstrate the
effect of such an upper limit we consider here several cases in the
range $0.01 < k_{\rm m} < 0.06$ arcmin$^{-1}$. The results are shown in
Fig. \ref{Fig_MBM_compare22}. With the horizontal line we
indicate the expected Kolmogorov spectral index, $\gamma = -8/3$, the
vertical lines bracket the velocity range with average brightness
temperatures $T_{\rm B\, aver} > 0.5$ K. We consider derived spectral
indices for lower $T_{\rm B\, aver}$ values as very uncertain and
possibly systematically biased despite the fact that the formal
uncertainties from the least squares fit do not change significantly for
lower $T_{\rm B\, aver}$.

\begin{figure}[htp]  
   \centering
   \includegraphics[width=6.5cm,angle=-90]{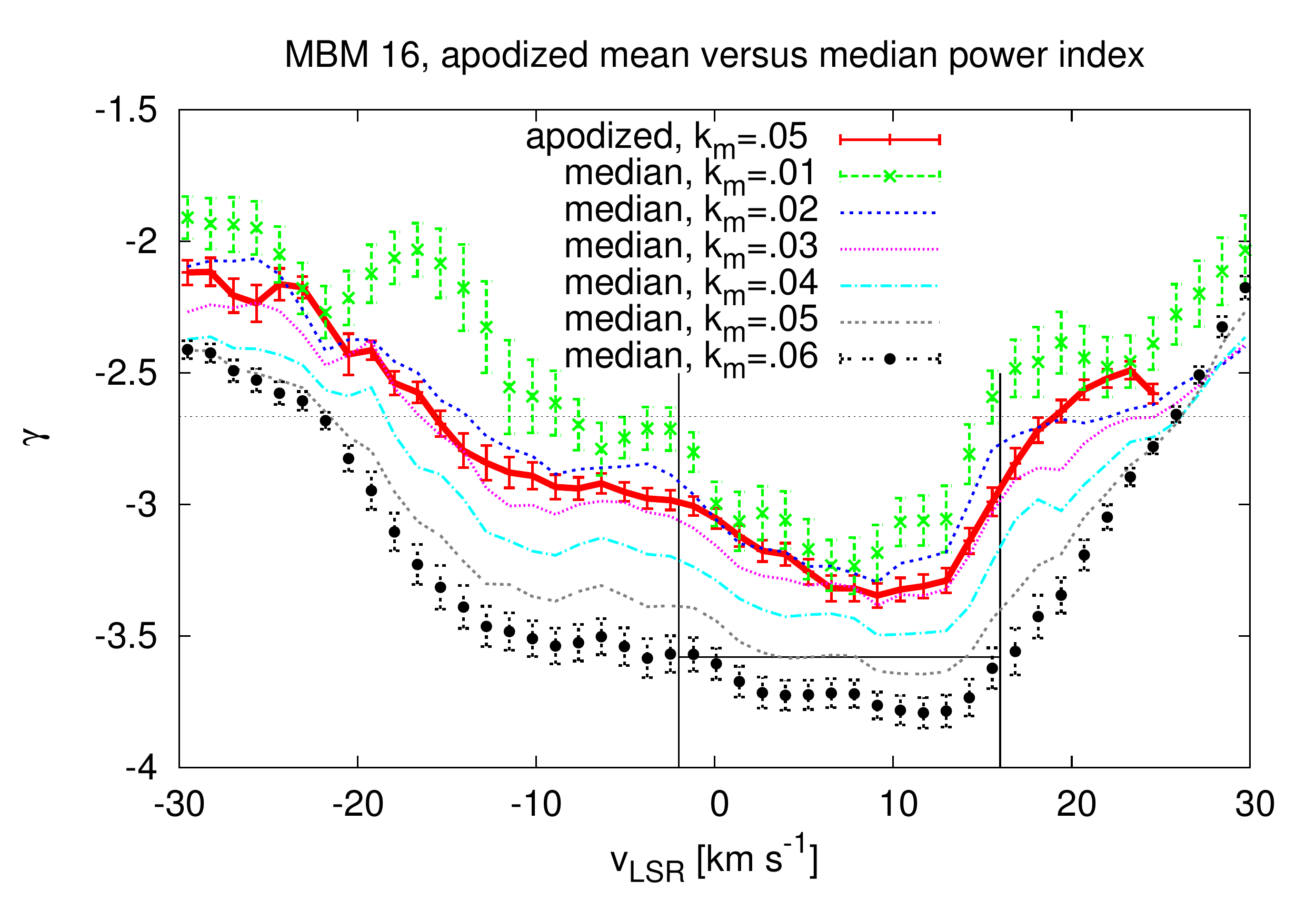}
   \caption{The thick red line represents results obtained after
     apodization, beam correction, and noise subtraction. The other
     graphs show results obtained without apodization by fitting the
     median power distribution.  $k_{\rm m}$ is in each case the upper
     limit for the spatial frequencies used for the fit. Error bars are
     given in three cases to provide an impression of formal
     uncertainties of the least square fits. The horizontal black
     dotted line represents the expected Kolmogorov index $\gamma =
     -8/3$. The vertical solid black lines bracket the velocity range
     analyzed by \citet{Pingel2013}, the horizontal solid black line
     gives the weighted average slope $\gamma = -3.58 \pm 0.09$ obtained
     by them for this velocity range. }
   \label{Fig_MBM_compare}
\end{figure}

We repeat the derivation of spectral indices but use this time a
processing similar to \citet{Pingel2013}. No apodization, beam
correction or noise bias correction is applied to the data and we
determine for each annulus in the UV plane the median power instead of
the mean power before we fit the spectrum. The different cases in
Fig. \ref{Fig_MBM_compare} are as before results obtained for upper
limits $0.01 < k_{\rm m} < 0.06$ arcmin$^{-1}$ in spatial
frequencies. For comparison we plot the spectral
index distribution from Fig. \ref{Fig_MBM_compare22} obtained by our
method for $k_{\rm m} = 0.05$ arcmin$^{-1}$.

Comparing Fig. \ref{Fig_MBM_compare} with Fig. \ref{Fig_MBM_compare22}
it is obvious that for our apodization method the upper limit $ k_{\rm
  m}$ is not a critical parameter\footnote[5]{It is more critical to
  determine an accurate three sigma limit in case of anisotropy
  studies. In this case only 1/12 of the samples are available.}. For
$0.04 < k_{\rm m} < 0.06$ arcmin$^{-1}$ the spectral index profiles in
Fig. \ref{Fig_MBM_compare22} are similar with systematic deviations
close to the formal uncertainties of the least square fits for $ k_{\rm
  m} = 0.05$ arcmin$^{-1}$. In particular, Fig. \ref{Fig_MBM_compare22}
shows that the steep spectral index near $v_{\rm LSR} \sim 9.1$ \kms is
well defined and within the errors identical for all $ k_{\rm m}$ values
used.

For the processing without apodization, beam deconvolution, and noise
bias correction (Fig. \ref{Fig_MBM_compare}) the spectral index profiles
depend strongly on the applied spatial frequency cutoff $ k_{\rm m}$.
The weighted average slope for the six panels in Fig. 9 of
\citet{Pingel2013} is $\gamma = 3.58 \pm 0.09$, compatible with our
result for the median fit with $ k_{\rm m} = 0.05$ arcmin$^{-1}$ but is
incompatible with the apodized fit from Fig. \ref{Fig_MBM_compare22},
using the same constraint $k_{\rm m} = 0.05$ arcmin$^{-1}$.

\begin{figure}[thp]  
   \centering
   \includegraphics[width=6.5cm,angle=-90]{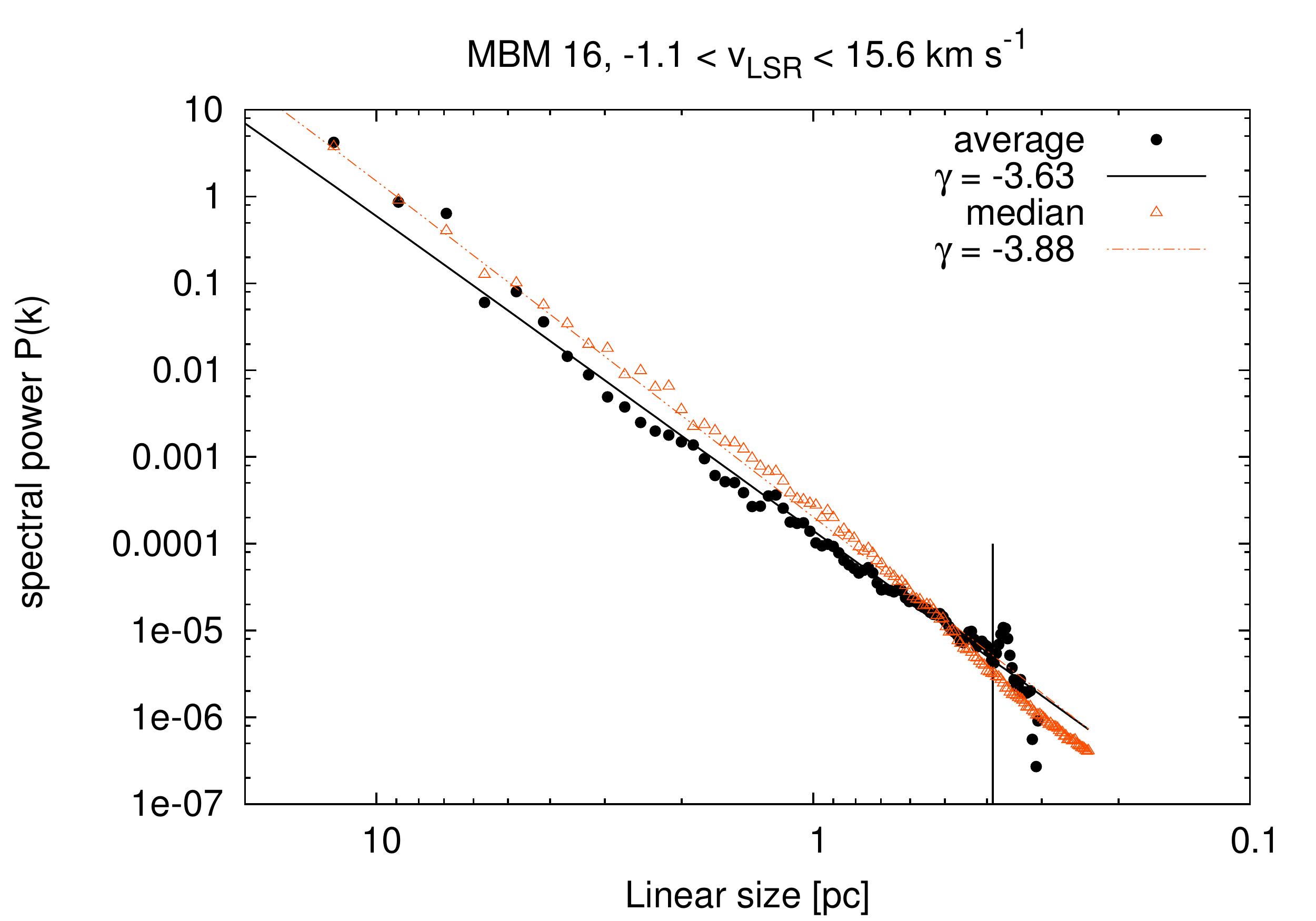}
   \caption{Unnormalized thick velocity slice power spectra for MBM 16,
     integrating over $-1.1 < v_{\rm LSR} < 15.6 $ \kmss and spectral
     indices from least squares fits. High spatial frequency limits
     $k_{\rm m} < 0.06$ arcmin$^{-1}$, corresponding to a linear size of
     0.39 pc were applied and are indicated by the vertical black
     line. Black dots represent data derived after apodization, beam
     correction, and noise subtraction. In this case we fit $\gamma =
     -3.63 \pm 0.04$. The red triangles show results obtained without
     apodization and beam correction by fitting the median power
     distribution. Here we fit $\gamma = -3.88 \pm 0.04$. }
   \label{Fig_MBM_broad}
\end{figure}

We continue our comparison of different data processing methods by
deriving spectral indices for thick velocity slices.  To calculate the
spectral indices we select the velocity range $-1.1 < v_{\rm LSR} < 15.6
$ \kms as defined by \citet{Pingel2013}.  We apply in both cases high
spatial frequency limits corresponding to a linear size of 0.39 pc (at a
source distance of 80 pc), to exclude instrumental noise for $k_{\rm m}
> 0.06$ arcmin$^{-1}$. The results are shown in
Fig. \ref{Fig_MBM_broad}.  For the data derived after our apodization,
beam correction, and noise bias correction, we fit $\gamma = -3.63 \pm
0.04$. The triangles represent results obtained without apodization by
fitting the median power distribution. We fit $\gamma = -3.88 \pm 0.04$
(to be compared with $\gamma = -3.63 \pm 0.04$ from apodization) and
below we will argue that the median is biased.

\begin{figure}[thp]  
   \centering
   \includegraphics[width=6.5cm,angle=-90]{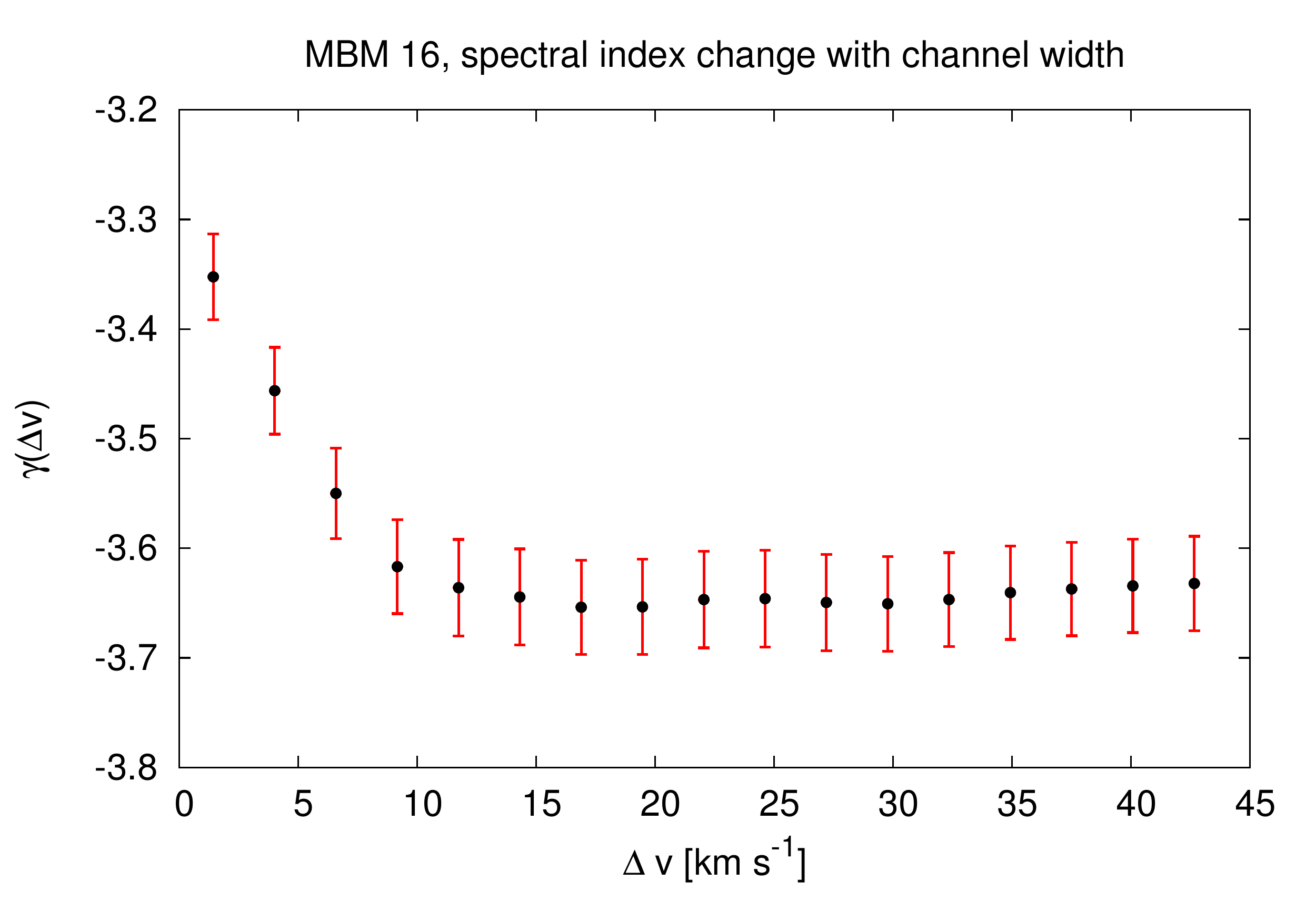}
   \caption{Spectral index changes with velocity slice thickness for the
     MBM 16 \hi~gas centered at $v_{\rm LSR} = 9.1$ \kms. A high spatial
     frequency limit $k_{\rm m} = 0.06$ arcmin$^{-1}$, corresponding to
     a linear size of 0.39 pc, was applied when fitting the spectral
     index. The bars indicate uncertainties of the power law indices. }
   \label{Fig_MBM_VCA}
\end{figure}

\citet{Pingel2013} derived $\gamma = -3.7 \pm .2$, within the errors
consistent with our result of $\gamma = -3.63 \pm 0.04$. We repeat our
analysis (with apodization) for the full velocity range $-30 < v_{\rm
  LSR} < 30 $ \kms. Within the errors we get the same result, $\gamma =
-3.63 $, with only slightly lower uncertainties. For the normalized
velocity centroid over $-30 < v_{\rm LSR} < 30 $ \kmss we determine
$\gamma = -3.65 \pm 0.03$. This result indicates that the turbulent
density and velocity fields share within the uncertainties similar
properties, compatible with a 3D Kolmogorov -11/3 slope expected for
incompressible turbulence. For comparison with
\citet[][Fig. 10]{Pingel2013} we calculate the variation of the 2D power
spectrum slope, $\gamma (\Delta v_{\rm LSR})$, with velocity slice
thickness $\Delta v_{\rm LSR}$. Figure \ref{Fig_MBM_VCA} shows the
change in spectral index for the \hi~gas centered at $v_{\rm LSR} = 9.1$
\kms. The slice may be considered as thick for $\Delta v_{\rm LSR} \ga
15$ \kmss but the \hi~emission extends over a total range of $\Delta
v_{\rm LSR} \sim 30$ \kms, see Fig. \ref{Fig_MBM_overview}.

Controversial results, apodization versus median, need to get explained
but we do not intend to discuss subtle differences between the Arecibo
and Effelsberg telescopes, nor differences in the data processing
pipeline. A major difference is however the sampling in spatial
frequencies. \citet{Pingel2013} use 13 samples with equal spacings in
${\rm log}(k)$. We use approximately ten times more samples with linear
spacing in $k$. The main reason for our choice is that it allows a
simple but reliable determination of the noise bias $N(k)$, demonstrated
with Fig. 6 of \citetalias{Kalberla2016b}. The noise bias depends on the
S/R, the fine grain linear spacing allows an accurate determination and
subsequent correction for this noise bias, see Sect. 4 of
\citetalias{Kalberla2016b}. This correction, however, is a statistical
correction for the average power level and does not result in
``noise-free'' power distributions as erroneously assumed by
\citet{Green1993}. Uncertainties remain and are amplified by the beam
correction (see Figs. 5 and 6 in \citetalias{Kalberla2016b}). It is
therefore necessary to restrict the interpretation of the power spectra
to those spatial frequencies where the signal is stronger than the
instrumental noise. For this purpose we use a three sigma upper limit
$k_{\rm m}$.

It is important to realize that for any reasonable instrumental setup
the power at the highest spatial frequencies is dominated by
instrumental noise. The transition from the genuine astrophysical signal
to instrumental noise depends on the S/R and the telescope beam
function.  Without beam correction instrumental effects remain
hidden. Accordingly the uncorrected signal in Fig. \ref{Fig_MBM_broad}
does not give the slightest hint where the transition from astrophysical
signal to instrumental noise takes place. Figure 6 of
\citetalias{Kalberla2016b} shows that this transition is sharp and well
defined in spatial frequency.

Our approach leads to power spectra that are on average straight as
expected for the inertial range of turbulence spectra. Power spectra
without correction for beam and noise bias have a concave shape, caused
by the response to the telescope beam as demonstrated by \citet[][their
Figs. 3, 4, and 6]{Dickey2001}. Such a concave bending is visible in
Fig. \ref{Fig_MBM_broad}. It was already demonstrated with Fig. 22 of
\citetalias{Kalberla2016b} that the missing beam correction, together
with the missing noise bias removal, leads to such a bending. In turn
the spectral index is biased. We notice that all of the power spectra in
Fig. 7 and 8 of \citet{Pingel2013} are concave, similar to Figs. 3,4,
and 6 of \citet{Dickey2001}. Fitting such spectra with a straight power
law is difficult, in particular without a detailed evaluation of the
significance of the data. Biases and uncertainties in derived spectral
indices are unavoidable, the results depend on selection effects, such
as details of the sampling method.

When deriving and comparing thin with thick slice power spectra, changes
in the noise bias must be taken into account.  \citet{Pingel2013} used
for this comparison channel widths between $\Delta v_{\rm LSR} = 0.18$ and $\Delta
v_{\rm LSR} = 18$ \kms. This implies that the noise power bias for the
corresponding power spectra changes by a factor of 10. Disregarding this
effect must lead to biases in the spectral index but it is difficult to
estimate such an effect quantitatively (EBHIS does not provide data at a
bandwidth of $\Delta v_{\rm LSR} < 1.44$ \kms).

\begin{figure}[thbp]  
   \centering
   \includegraphics[width=6.5cm,angle=-90]{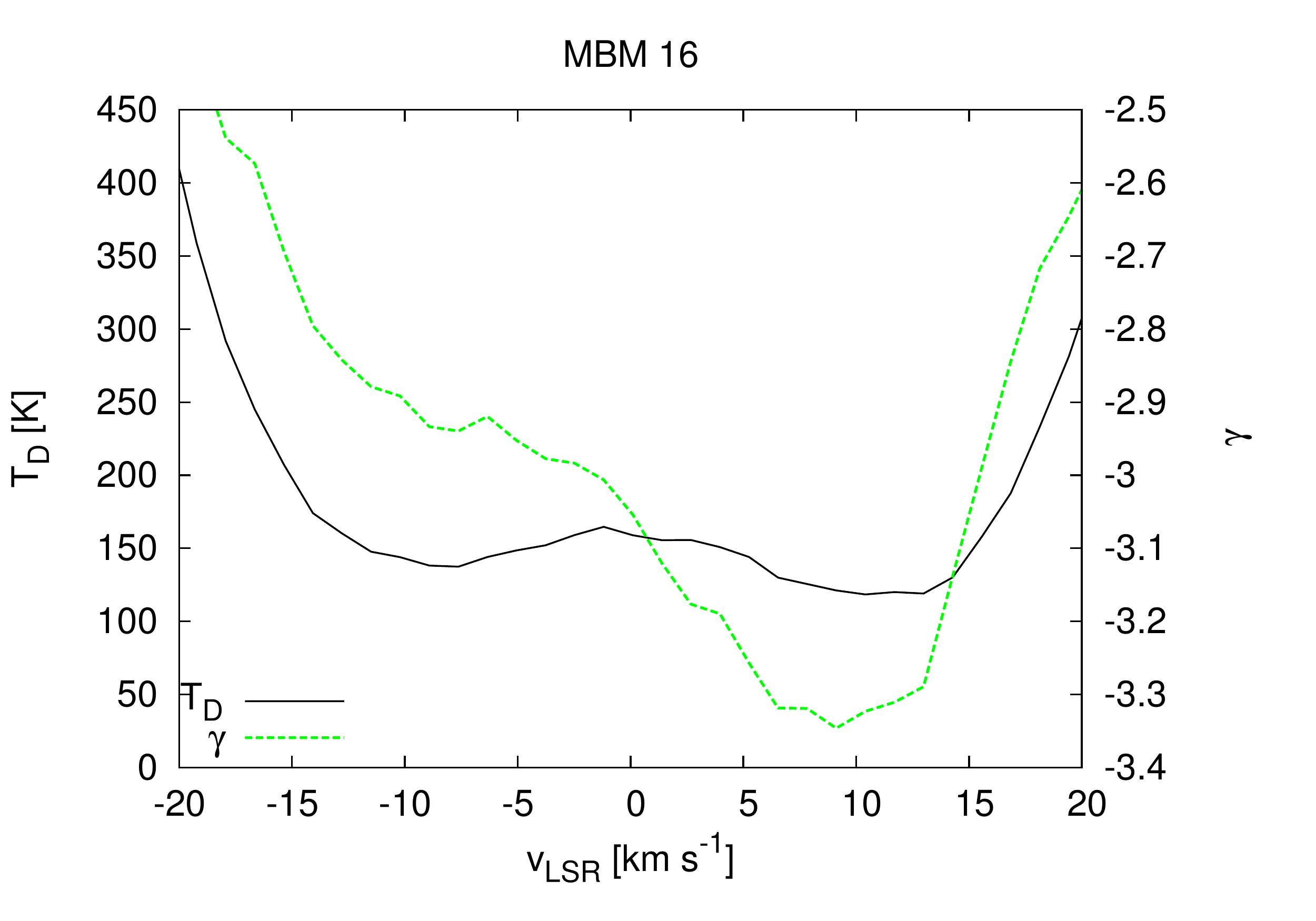}
   \caption{Comparison between the geometric mean Doppler temperature
     $T_{\rm D}$ (black) and the spectral index $\gamma$ (green dashed)
     for MBM 16. }
   \label{Fig_MBM_overview2}
\end{figure}

\begin{figure}[tbp]  
   \centering
   \includegraphics[width=6.5cm,angle=-90]{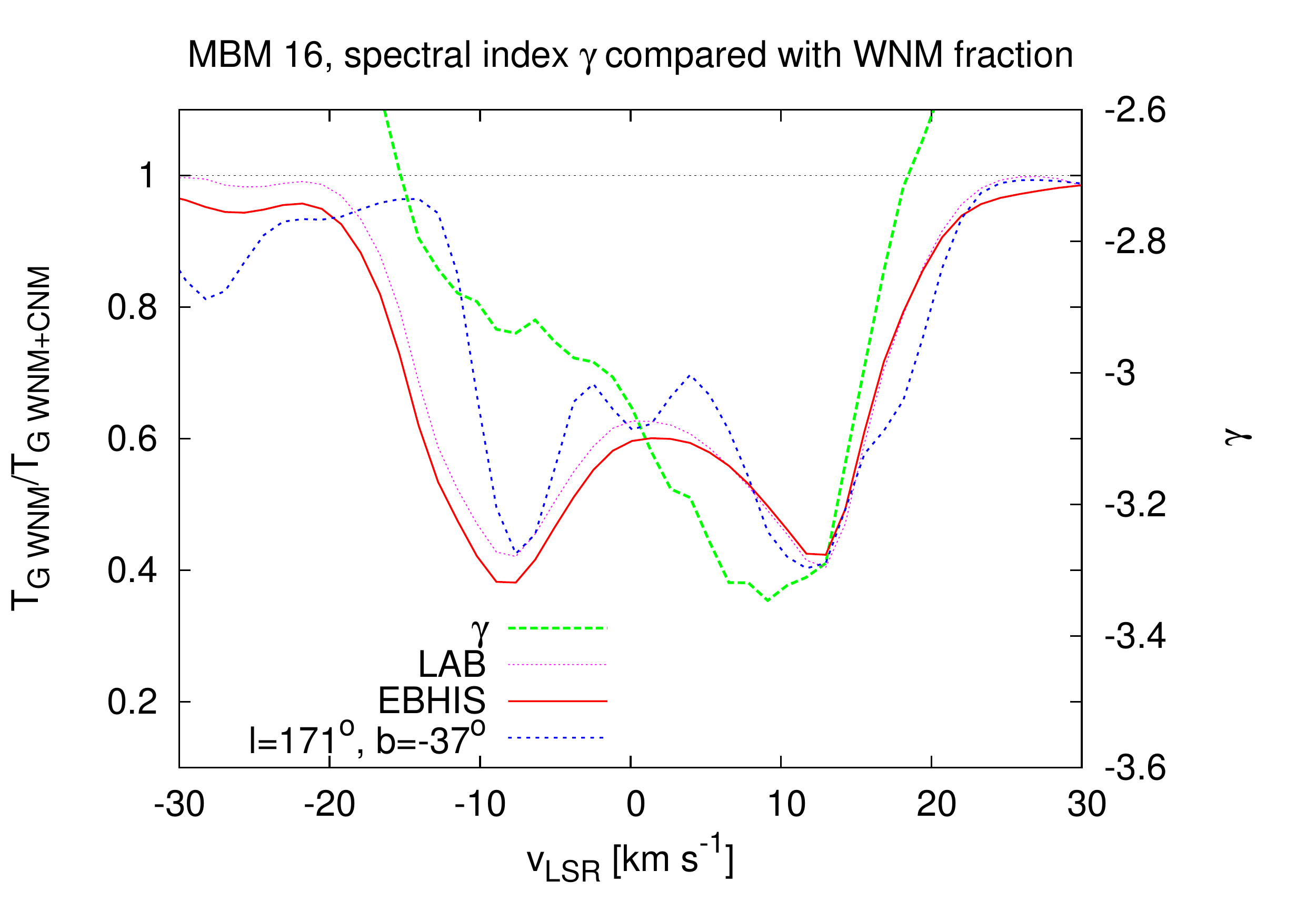}
   \caption{Spectral index $\gamma(v_{\rm LSR})$ (green) compared with
     the WNM fraction $T_{\rm WNM}(v_{\rm LSR})/T_{\rm WNM+CNM}(v_{\rm
       LSR}) $ derived from LAB (pink) and EBHIS Gaussians (red) for the
     MBM 16 field. An upper limit for the Doppler temperature of the CNM
     of $T_{\rm D} < 1100 $ K, corresponding to a turbulent  CNM 
     Mach number of $M_{\rm T} =3.7$ was applied. We display also with a
     blue dotted line the WNM fraction $T_{\rm WNM}(v_{\rm LSR})/T_{\rm
       WNM+CNM}(v_{\rm LSR})$ derived from EBHIS, using $T_{\rm D} <
     1100 $ K, for a central position $l = 171\deg, b = -37\deg$ within a
     radius of 1\fdg5. This corresponds to the region with CO emission
     observed by \citet[][Fig. 1]{LaRosa1999}.  }
   \label{Fig_MBM_Gauss}
\end{figure}

\begin{figure*}[htp]  
   \centering
   \includegraphics[width=7cm,angle=-90]{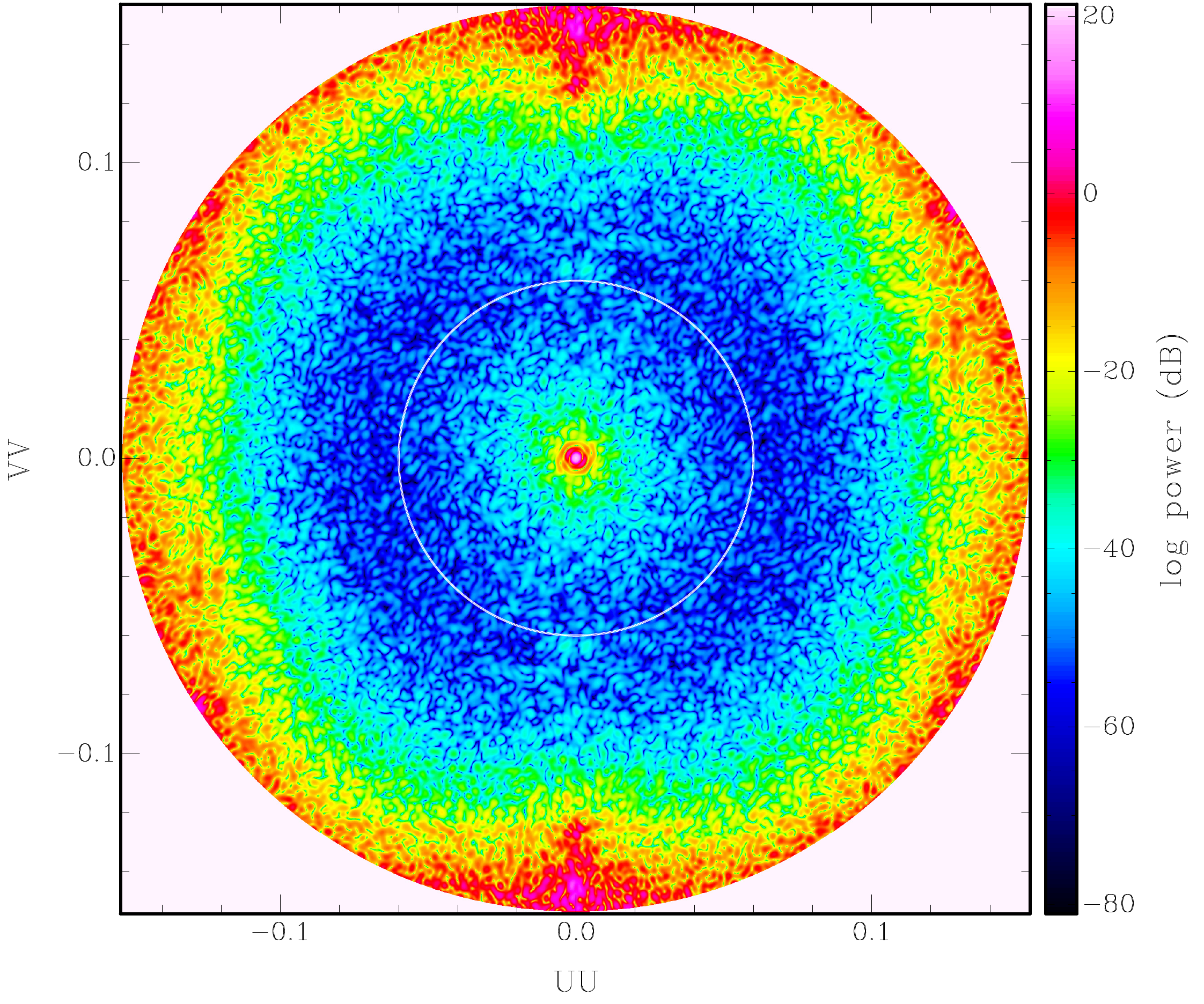}
   \includegraphics[width=7cm,angle=-90]{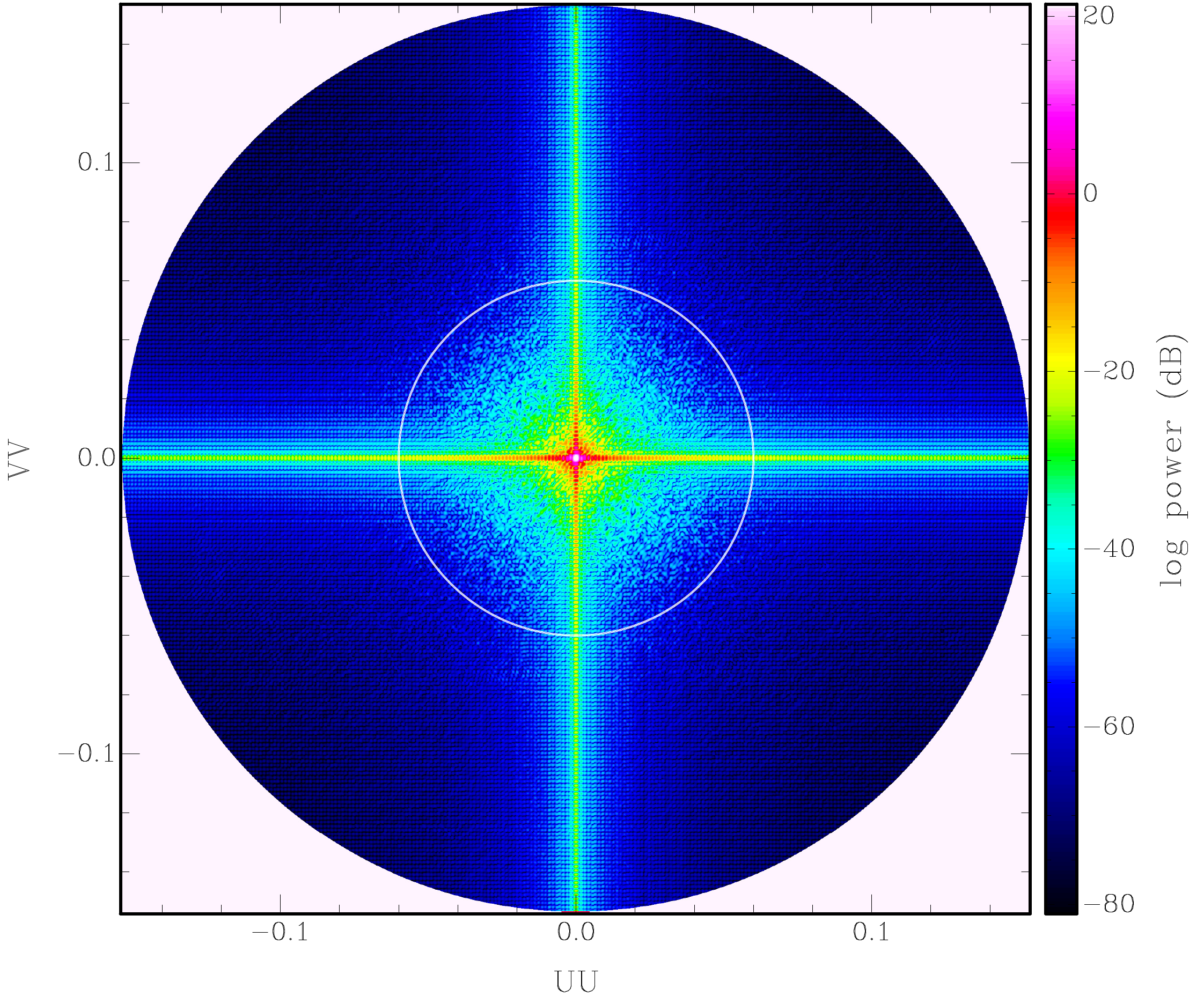}
   \caption{Thick velocity slice power distributions for
     MBM 16, integrating over $-1.1 < v_{\rm LSR} < 15.6 $ \kms. Left:
     apodized distribution, corrected for the beam function. Right: the
     same data without apodization and beam correction. The scales are
     logarithmic in dB. The circle indicates the 3 $\sigma$ spatial
     frequency limit $k_{\rm m} = 0.06$ arcmin$^{-1}$, corresponding to
     a linear size of 0.39 pc. Data for higher spatial frequencies are
     dominated by the instrumental noise. }
   \label{Fig_UV_MBM}
\end{figure*}

The data processing has also consequences on the interpretation of power
spectral distribution. \citet{Pingel2013} considered primarily
velocities $-1 < v_{\rm LSR} < 16 $ \kmss but mention in their Sect. 3
that the results remain valid for the full velocity range $-20 <
v_{\rm LSR} < 20 $ \kms.  Within the uncertainties they find little
fluctuations for the spectral index ($\gamma = 3.58 \pm 0.09$). We
interpret the full velocity range with $T_{\rm B\, aver} > 0.5$ K and
get significant spectral index fluctuations in the range $ -2.7 > \gamma
> -3.35$.

We demonstrate in Sect.  \ref{Spectral_index} that the steepest part of
the spectral index distribution for the Horologium and Auriga fields is
associated with cold gas at low Doppler temperatures with a low fraction
of the WNM (Figs. \ref{Fig_AU_overview2} and \ref{Fig_Gauss}).  In
Fig. \ref{Fig_MBM_overview2} we compare for the MBM 16 field the
spectral index distribution with the geometric mean Doppler temperature
$T_{\rm D}$. In Fig.  \ref{Fig_MBM_Gauss} we show how $\gamma$ is
related to the WNM fraction $T_{\rm WNM}(v_{\rm LSR})/T_{\rm
  WNM+CNM}(v_{\rm LSR})$. Consistent with Sect. \ref{Spectral_index} we
use for the CNM an upper limit for the Doppler temperature of $T_{\rm D}
< 1100 $ K and plot the results for LAB and EBHIS.

Similar to Horologium and Auriga, the steepest part of the spectral
index distribution in the MBM 16 field is found to be associated with
low values in $T_{\rm D}$ and $T_{\rm WNM}(v_{\rm LSR})/T_{\rm
  WNM+CNM}(v_{\rm LSR})$ but in both cases there is for MBM 16 a shift
for the minima by about 3 to 4 \kms. Cold gas is also found at $v_{\rm
  LSR} \sim -10 $ \kmss but interestingly the spectral index is here
less steep and only little molecular gas is associated with this part of
the \hi~distribution.

The MBM 16 field, used by \citet{Pingel2013}, is offset from the field
center observed by \citep{LaRosa1999}. We repeat the EBHIS calculation
of $T_{\rm WNM}(v_{\rm LSR})/T_{\rm WNM+CNM}(v_{\rm LSR})$ for a center
position of $l = 171\deg, b = -37\deg$ within a limited radius of
1\fdg5. This covers most of the region with CO as observed by
\citet[][Fig. 1]{LaRosa1999}. The result is shown in Fig.
\ref{Fig_MBM_Gauss}. The positive wing of the
double horned distribution for $T_{\rm WNM}(v_{\rm LSR})/T_{\rm
  WNM+CNM}(v_{\rm LSR})$ has shifted somewhat closer to the velocity of
the steepest spectral index. The negative part peaks now at $v_{\rm LSR}
= -9 $ \kmss, consistent with a weak secondary peak of the CO
distribution \citep[][Fig. 2]{Pingel2013} at this velocity. $T_{\rm
  WNM}(v_{\rm LSR})/T_{\rm WNM+CNM}(v_{\rm LSR})$ contains now a central
component (originating from \hi~gas at low Galactic latitudes). From the
almost symmetrical shape of this distribution relative to the central
component we suspect that some dynamical interactions are responsible
for the $T_{\rm WNM}(v_{\rm LSR})/T_{\rm WNM+CNM}(v_{\rm LSR})$ and
$T_{\rm D}$ distribution.

CO observations in the MBM 16 field suggest that the molecular gas is
driven by an external shear flow \citep{LaRosa1999}. The velocity shifts
observed by us may imply systematical internal motions in the \hi~gas,
with shock induced phase transitions, leading to a separation of cold
and warm gas. Whether there is a detailed relation to the velocity
structure of the CO gas is an interesting question but far beyond the
scope of this Appendix.

To summarize this part of the Appendix, we prefer a Bayesian approach
over the so-called robust processing using the median. We exclude
instrumental biases by apodization and beam correction and eliminate
outliers caused by the system noise, using only data above a S/R of
three.\footnote[6]{The data analysis of the 3C 147 observations followed
  the same approach. Fig. 1 of \citet{Kalberla1983} shows at high
  spatial frequencies the instrumental noise of the WSRT. } The
probability that data with significant instrumental biases belong to the
true power distribution is close to zero, hence these data should be
corrected and excluded from the analysis, see \citet{Jaynes2007} for
detailed discussions of Bayesian and robust approaches in probability
theory.

Figure \ref{Fig_UV_MBM} compares the data used for our Bayesian approach
(left) with those from the robust data analysis (right). Obviously, both
distributions are very different. The image displayed on the right hand
side contains in comparison to the apodized distribution excess power
from the source distribution outside the apodized
region. Discontinuities in the \hi~column density distribution at the
rectangular field boundaries cause the prominent cross structure in the
Fourier transformed domain. The second problem with the power
distribution on the right hand side is that the attenuation caused by
the beam function of the telescope remained uncorrected. Quite
surprisingly, the power spectra, given in Fig. \ref{Fig_MBM_broad}, show
a remarkable good agreement between both methods. However without
apodization the power at intermediate spatial frequencies (scales
corresponding to 0.8 -- 8 pc) is too high and the power at the smallest
scales (below 0.5 pc) is too low.


\section{Horologium anisotropies at high velocities}
\label{appendix}

\begin{figure}[htp]  
   \centering
   \includegraphics[width=6.5cm,angle=-90]{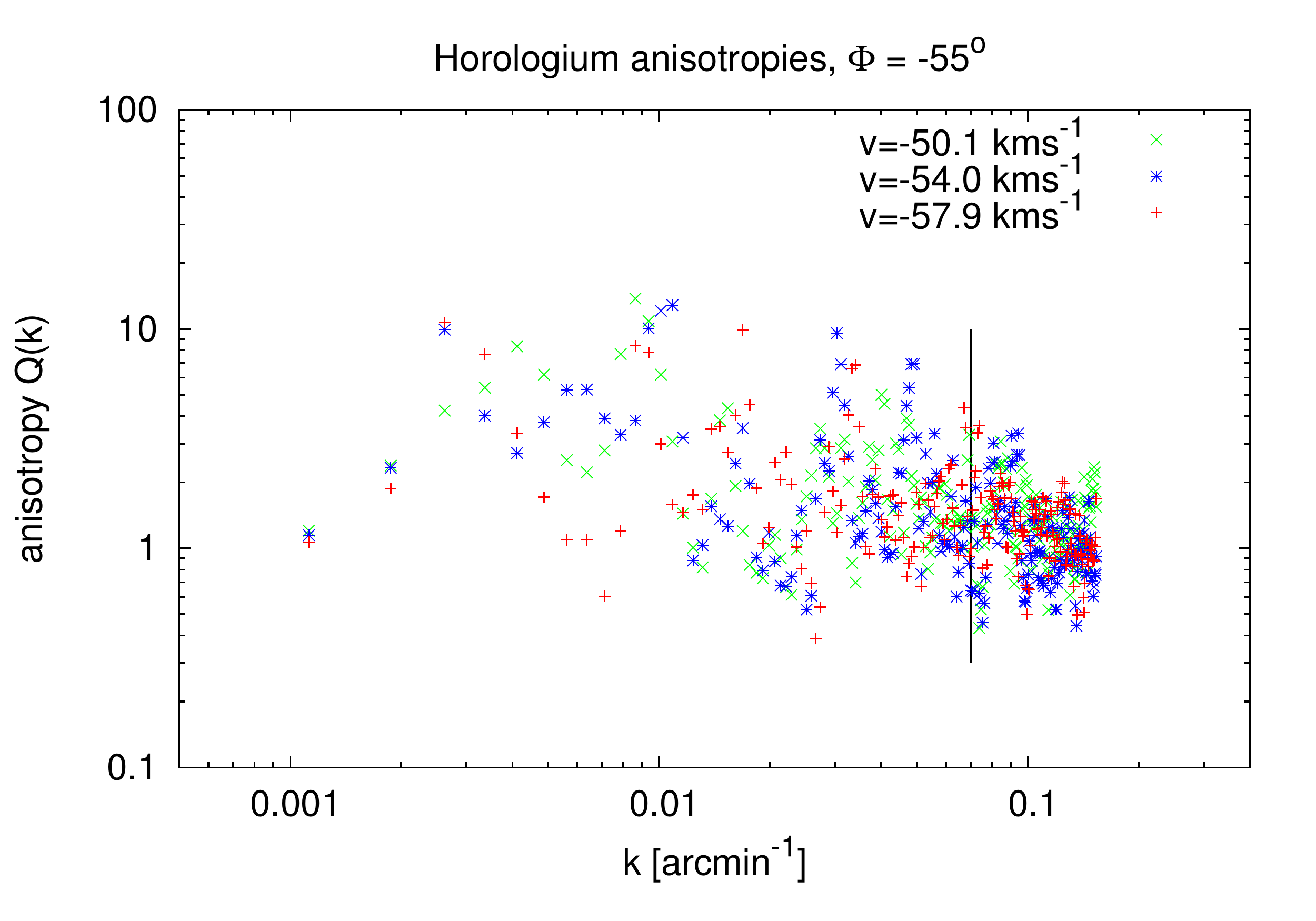}
   \caption{Anisotropies $Q(k)$ in the Perseus arm at velocities 
     $v_{\rm LSR} = -50.1 $ \kmss (green),  $v_{\rm LSR} = -54.0 $ \kmss
     (blue), and $v_{\rm LSR} = -57.9 $ \kmss (red). 
}
   \label{Fig_HV_49_55}
\end{figure}

\begin{figure}[htp]  
   \centering
   \includegraphics[width=6.5cm,angle=-90]{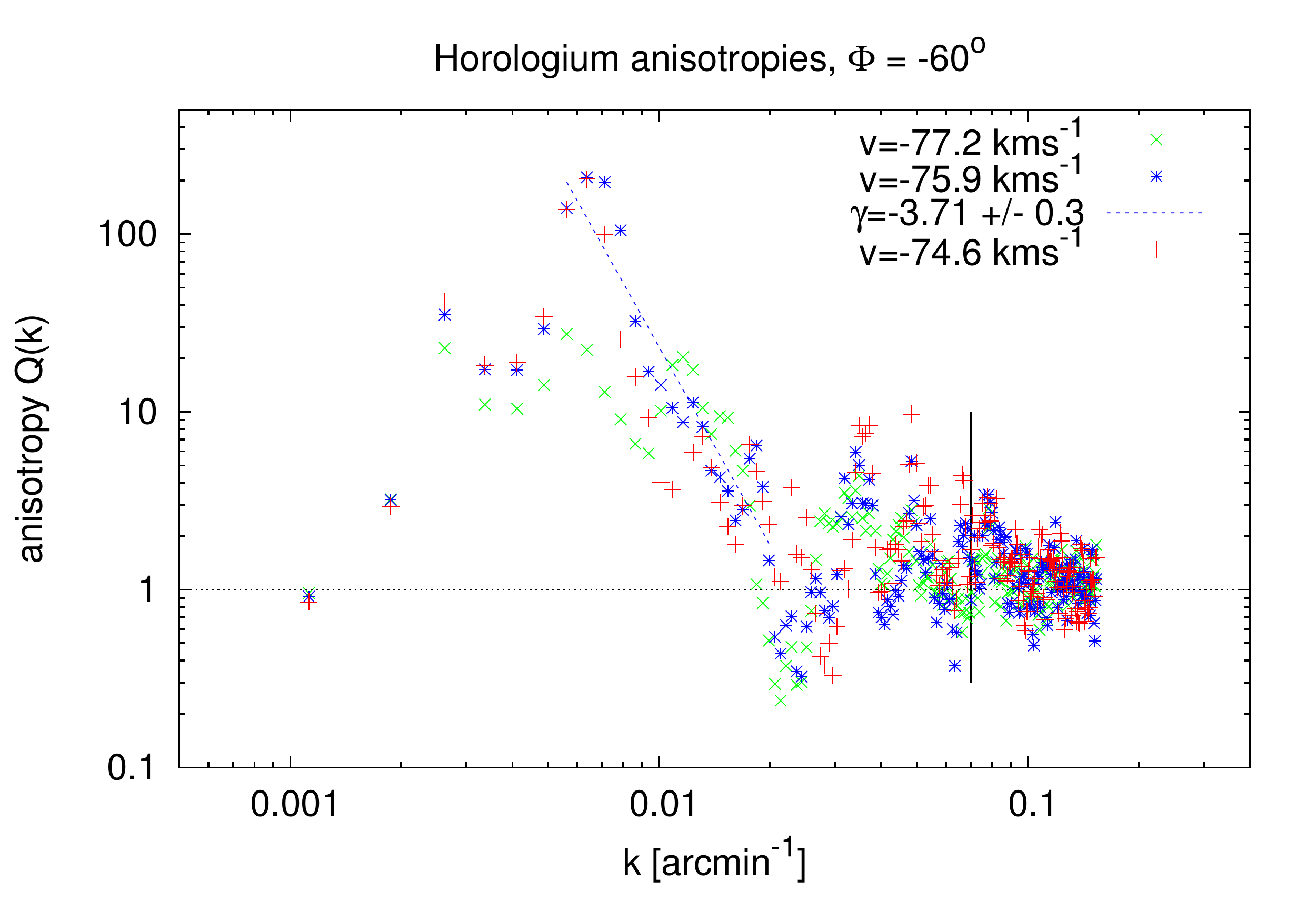}
   \caption{Anisotropies $Q(k)$ in the outer arm at velocities $v_{\rm
       LSR} = -77.2 $ \kmss (green), $v_{\rm LSR} = -75.9 $ \kmss
     (blue), and $v_{\rm LSR} = -74.6 $ \kmss (red). The blue line
     reproduces a fit to 20 data points at $v_{\rm LSR} = -75.9 $ \kms,
     tentatively indicating the decay of the strong local anisotropy
     with $Q = 209$. }
   \label{Fig_HO_HV_34}
\end{figure}

Searching for anisotropies in the \hi~component at a velocity of $v_{\rm
  LSR} = -54 $ \kmss we found also a position angle of $\Phi \sim
-55\degr$, corresponding to an alignment parallel to the Galactic
plane. Figure \ref{Fig_HV_49_55} displays for three velocity channels
moderate anisotropies up to $Q \sim 10$ at $k \sim 0.01$ arcmin$^{-1}$,
corresponding to a linear scale of 175 pc at a distance of 6 kpc.

Figure \ref{Fig_HO_overview} displays a local minimum in the spectral
index distribution $\gamma$ at $v_{\rm LSR} = -76 $ \kms. This feature
is associated with a well defined position angles $\Phi = -65\degr \pm
8\degr$ in Fig. \ref{Fig_HO_overview_angle} although there is no obvious
emission component visible in the average brightness temperature profile
from Fig. \ref{Fig_HO_overview}. An inspection of the EBHIS \hi~maps
shows a bright (up to 66 K peak brightness temperature) but segregated
filamentary structure at low latitudes with a length of $\sim
2\fdg5$. This feature is part of a more extended structure outside the
field of view, parallel to the Galactic plane. The kinematic distance
for this \hi~gas is about 10 kpc, the Galactocentric distance 17
kpc. This feature fits well to arm 1 of the four-armed logarithmic
spiral proposed by \citet[][Fig. 4a]{Levine2006}.

Figure \ref{Fig_HO_HV_34} displays anisotropies for three channels, at
$v_{\rm LSR} = -77.2, -75.9 $, and $v_{\rm LSR} = -74.6 $ \kms. For two
of the channels we find strong anisotropies of $Q \sim 210$ at $k =
0.0064$ arcmin$^{-1}$. These anistropies with a scale length of about
450 pc appear to decay with a well defined steep slope of $\gamma =
-3.71$. For comparison, the spectral index of the average power spectrum
is here $\gamma = -2.89 \pm 0.07$.

\section{Velocity centroids}
\label{TI_centroid}

\begin{table*}
\caption{Comparison of fit spectral indices}              
\label{table:1}      
\centering                                      
\begin{tabular}{c c c c c c c }          
\hline\hline                        
Source & $v_{LSR}$ range & Thick slice & VC & UVC & VVC & Steepest thin slice \\    
\hline                                   
Auriga & -10.2 to 11.7 km s$^{-1}$ & $-3.28 \pm 0.04$ & $  -2.78 \pm 0.03$ & $-2.70 \pm 0.03$ & $ -3.28 \pm 0.04$  &$ -3.22 \pm 0.04 $ \\     
Horologium & -23.1 to 5.2 km s$^{-1}$ & $-3.41 \pm 0.05$ &$ -3.35 \pm 0.04$ &$ -3.18 \pm 0.05$ &$ -3.24 \pm 0.04  $ &$-3.25 \pm 0.03$ \\     
3C 196 \tablefootmark{a}& -8.9 to 6.5 km s$^{-1}$ &$ -3.07 \pm 0.04$ &$ -2.75 \pm 0.03$ &$ -2.70 \pm 0.03$ & $-3.12 \pm 0.04$ & $-2.85 \pm 0.03$  \\     
\hline                                             
MBM 16& -29.5 to 29.7 km s$^{-1}$ & $-3.63 \pm 0.04$ & $-3.65 \pm 0.03$ & $-3.71 \pm 0.03$  &$ -3.78 \pm 0.03$ &$  -3.35 \pm 0.04$  \\     
FN1 \tablefootmark{a}& -86.2 to 47.8 km s$^{-1}$ & $-3.00 \pm 0.03$ & $-2.80 \pm 0.03$ & $-2.75 \pm 0.02$  &$ -2.53 \pm 0.03$ &$  -3.07 \pm 0.03$  \\     
\hline                                             
\end{tabular}
\tablefoot{For definition of the different centroids see
  Eqs. \ref{eq:VC} to \ref{eq:VVC}.\\
  \tablefoottext{a}{Using data from \citetalias{Kalberla2016b}, the FN1
    field was used as reference with the velocity range as defined by
    \citet{Miville-Deschenes2007}}. 
}
\label{table1}
\end{table*}

\subsection{Normalized velocity centroid (VC)}
\label{VC}

Using fractional Brownian motion simulations,
\citet{Miville-Deschenes2003} used normalized velocity centroids
(VC) according to Eq. \ref{eq:VC}.  They demonstrated that the power
spectrum of the velocity centroid map has in case of an isotropic and
optically thin distribution the same spectral index as that of the
velocity field. \citet{Esquivel2005}, using MHD simulations, confirm
these results but caution that the results may get questionable for high
turbulent Mach numbers $M_{\rm T} \ga 2.5$. For 3C 147 we derived
$M_{\rm T} \sim 2.7$, else we assume $M_{\rm T} \sim 3.7$. Thus, an
application of the velocity centroid method may be somewhat questionable
with our assumptions but it is certainly not applicable in case of
$M_{\rm T} =7.7$ \citep{Chepurnov2010}. In comparison with VCA or the
velocity coordinate spectrum (VCS) method \citep{Lazarian2006},
\citet{Esquivel2005} report advantages for velocity centroids in case
that velocity statistics is not a straight power law. They mention that
velocity centroids can better pick up the dissipation and injection
energy scales.

Observed \hi~line profiles at low and intermediate latitudes are
usually complex since they contain several independent components.  It
is then necessary to restrict the observations to the velocity range
dominated by the \hi~cloud or \hi~layer under investigation
(Eq. \ref{eq:VC}). For velocity centroids an appropriate window is
easily applicable. For the VCS method on the other hand the velocity
window causes unwanted side effects which need to be taken into
account. Such an elaborate analysis is beyond the scope of this paper.

VCA is not applicable for steep thin velocity slice power spectra
observed by us, we therefore calculate the Horologium velocity centroid
over the velocity range $-23.1 < v_{\rm LSR} < 5.2$ \kms (see
Fig. \ref{Fig_HO_overview}), the same range as applied for the
determination of the spectral index for the density field.  Figure
\ref{Fig_Horo_centroid} (top) shows the result, the derived average
spectral power index for the Horologium velocity field is within the
errors identical to the index $\gamma = -3.4 \pm 0.05 $ for the density
field.

For comparison, in case of Auriga (Fig. \ref{Fig_Auriga_centroid}
bottom) we derive a thick slice spectral index of $\gamma = -3.28 \pm
0.04$ for the density field and from the velocity centroid $\gamma =
-2.78 \pm 0.03 $ (top) for $ k < 0.07$ arcmin$^{-1}$. In both cases we
used a velocity range of $-10.2 < v_{\rm LSR} < 11.7$ \kms (see Fig
\ref{Fig_AU_overview}). The width of the velocity window of 22 \kmss
fits formally to VCA thick slice condition $\Delta v_{\rm LSR} > 17$ \kms.  In
this case we have however at least two \hi~layers along the line of
sight, blending may cause unwanted contributions and the thick slice
window $-10.2 < v_{\rm LSR} < 11.7$ \kmss does probably not match the
criteria of a very thick slice. The derived spectral indices may suffer
from such contributions.

\subsection{Unnormalized velocity centroid (UVC)}
\label{UVC}

For normalized velocity centroids the denominator in Eq. \ref{eq:VC}
introduces an algebraic complication for the analytical treatment of
turbulence spectra or power distributions, we refer to the discussion by
\citet{Esquivel2005}. To allow a better comparison of our results with
model calculations we calculate UVCs in parallel to VCs, using also the
prescription given in \citet[Eq. 33][]{Esquivel2005} to verify that the
velocity centroids trace the turbulent velocity statistics,
\begin{equation}
  UVC(x,y) = \int_{v1}^{v2} T_{\rm B}(x,y,v_{\rm LSR}) v_{\rm LSR} \Delta v_{\rm LSR}.
\label{eq:UVC}
\end{equation}
A visual comparison of centroid maps shows that both methods are nearly
equivalent, except that UVC images look slightly more
diffuse. Correspondingly UVC spectral indices are slightly shallower
than VC indices, see Table \ref{table1}.

\subsection{$v^2$ centroid (VVC)}
\label{VVC}

This centroid is analogous to Eq. \ref{eq:VC}, but replacing $v$ with
$v^2$ and was introduced by \citet{Burkhart2014} for anisotropy studies.
The VVC measures the second moment, making the isotropy degree more
sensitive to velocity and less sensitive to density. Alternatively VVC
may be considered as the first moment of the line width distribution,
hence telling us about turbulent fluctuations of Doppler temperatures, 

\begin{equation}
  VVC(x,y) = \frac {\int_{v1}^{v2} T_{\rm B}(x,y,v_{\rm LSR}) v{^2}_{\rm LSR} \Delta v_{\rm LSR}}{\int_{v1}^{v2} T_{\rm B}(x,y,v_{\rm LSR}) \Delta v_{\rm LSR}}.
\label{eq:VVC}
\end{equation}

\subsection{Comparing VC, UVC, and VVC spectral slopes}
\label{tab1}

Table \ref{table1} lists spectral indices for thick slices, centroids
and also indices for the steepest thin slice power spectra. We include
3C 196 and FN1 from \citetalias{Kalberla2016b}. MBM 16 and FN1 are
reference fields and therefore not considered in the following
discussion.

It is easy to verify from Table \ref{table1} that VC and UVC are similar
but have slight systematic differences. VVC spectral slopes tend to be
steep, in case of Auriga and Horologium with a spectral index that is
comparable to the steepest thin slice spectral index. Anisotropies in
the VVC maps were found in all cases to be relatively weak, only
slightly less than those of the VC or UVC maps and restricted to limited
ranges in spatial frequency. We confirm the model calculations by
\citet{Burkhart2014} who found that VVC anisotropies are similar to the
VC case.

The VVC maps represent the spatial distribution of squared line widths,
hence Doppler temperatures, and the inspection of these maps shows that
they are dominated by broad WNM lines and correspondingly by high
Doppler temperatures. The relative steep VVC power spectra imply that
line widths and Doppler temperatures decrease strongly with increasing
spatial frequency, compatible with the low geometric mean Doppler
temperatures determined previously. However steep VVC power spectra are
not associated with anisotropies. As pointed out by
\citet{Burkhart2014}, VVC was introduced to enhance the sensitivity of
centroids with respect to the velocity field. Enhanced anisotropies are
not observed by us and correspondingly there is no evidence that
anisotropies are driven by the velocity field.

Along with this interpretation is our finding that anisotropies for the
coldest gas components increase with decreasing velocity width of the
slices under investigation. VC, UVC, and VVC maps show intermediate
anisotropies. This is exactly opposite to the results reported by
\citet{Kandel2016}. Last, but not least, our observations
contradict VCA results as proposed by
\citet{Lazarian2000,Lazarian2004}. All these theoretical investigations
are based on the assumption that turbulent velocity and density fields
can be represented by independent Gaussian fields. In Appendix
\ref{Anatomy} we demonstrate that the filamentary \hi~stuctures are
caused by projection effects. The observed thin slice column densities
depend on velocity and direction of the mean magnetic field. We
find a well defined velocity gradient pependicular to the magnetic
field.

\section{Anatomy of a filament }
\label{Anatomy}

In common language the term filament is used to describe ``a single
thread or a thin flexible threadlike object'' (Merriam-Webster). In this
paper we use ``filamentary structure'' to describe such an object.  But
what is the 3D distribution of the \hi~gas that gives rise to the
observed filamentary structure, projected onto the plane of the sky?
Referring to \citep{HeilesCrutcher2005,Heiles2005,Kalberla2016} we argue
for sheets, seen almost edge-on. A different geometry is proposed by
\citet{Clark2014}, they interpret filamentary structures as fibers.

We intend to disclose here as an example the structure of one of the CNM
filaments that appears to be associated with radio-polarimetric
depolarization canals, the feature observed in the Horologium field at
RA = 3$^{\rm h}$, DEC = 66\deg (B1950.0) with a velocity $v_{\rm LSR} =
-16.6 $ \kmss, see Fig. \ref{Fig_Overlay_HO-16.6}.

\begin{figure*}[htp] 
   \centering
   \includegraphics[width=6.5cm,angle=-90]{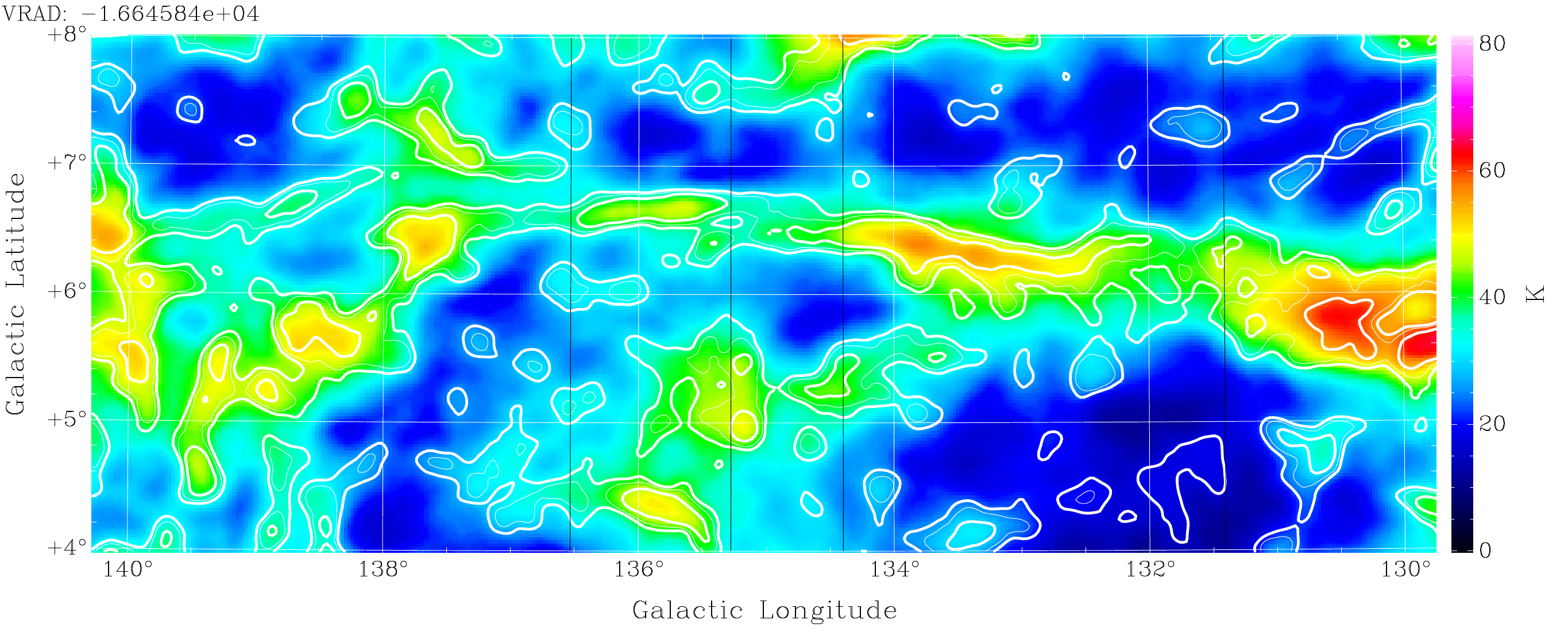}
   \includegraphics[width=6.5cm,angle=-90]{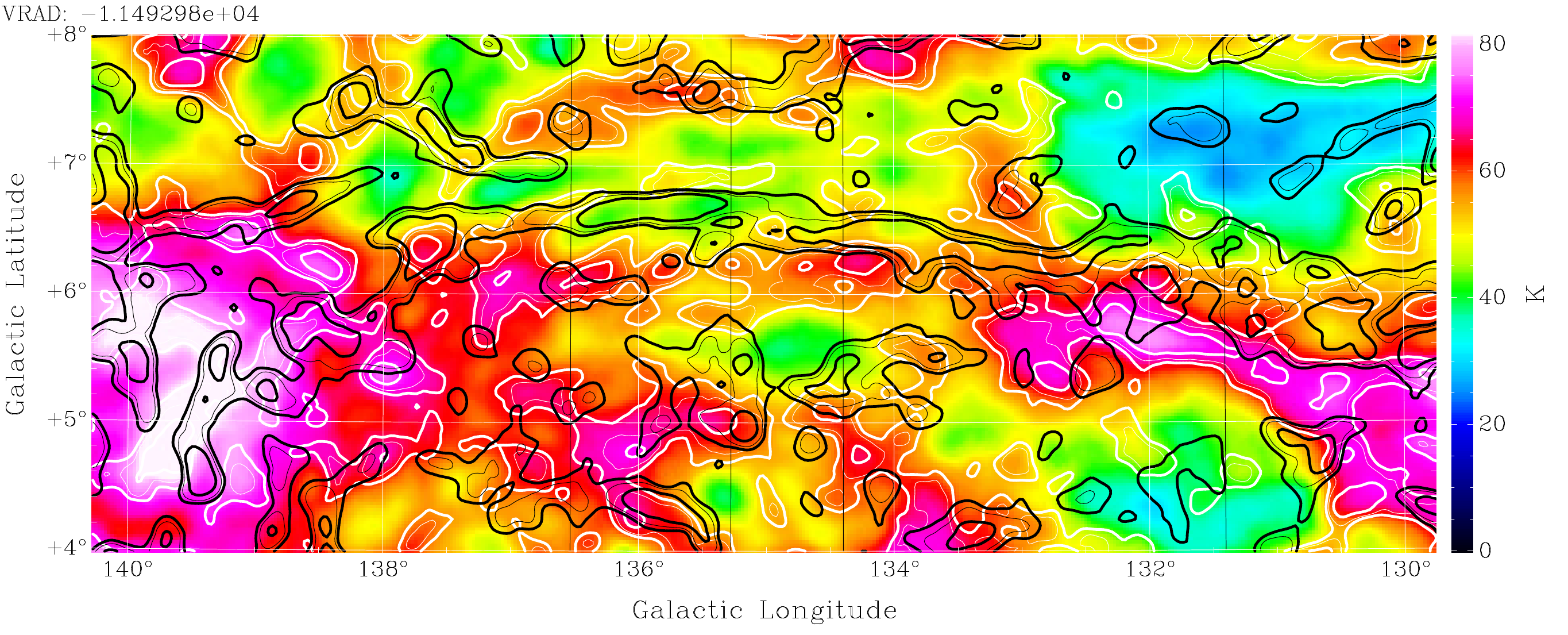}
   \caption{Top: color coded brightness temperature distribution for the
     filamentary structure in the Horologium field at $v_{\rm LSR} =
     -16.6 $ \kms. The white isophotes display the USM temperatures at
     the same velocity.  Bottom: the brightness temperature distribution
     at $v_{\rm LSR} = -11.5 $ \kmss is shown in color, the white
     isophotes are USM temperatures at the same velocity while black
     isophotes are at $v_{\rm LSR} = -16.6 $ \kms for comparison with
     the top panel.  Isophote levels are 1, 2.5, and 5 K Galactic
     coordinates are used, the black lines indicate longitudes of the
     velocity-position plots in Fig. \ref{Fig_Horo_cuts}. }
   \label{Fig_Horo_gal}
\end{figure*}

\begin{figure*}[htbp] 
   \centering
   \includegraphics[width=3.7cm,angle=-90]{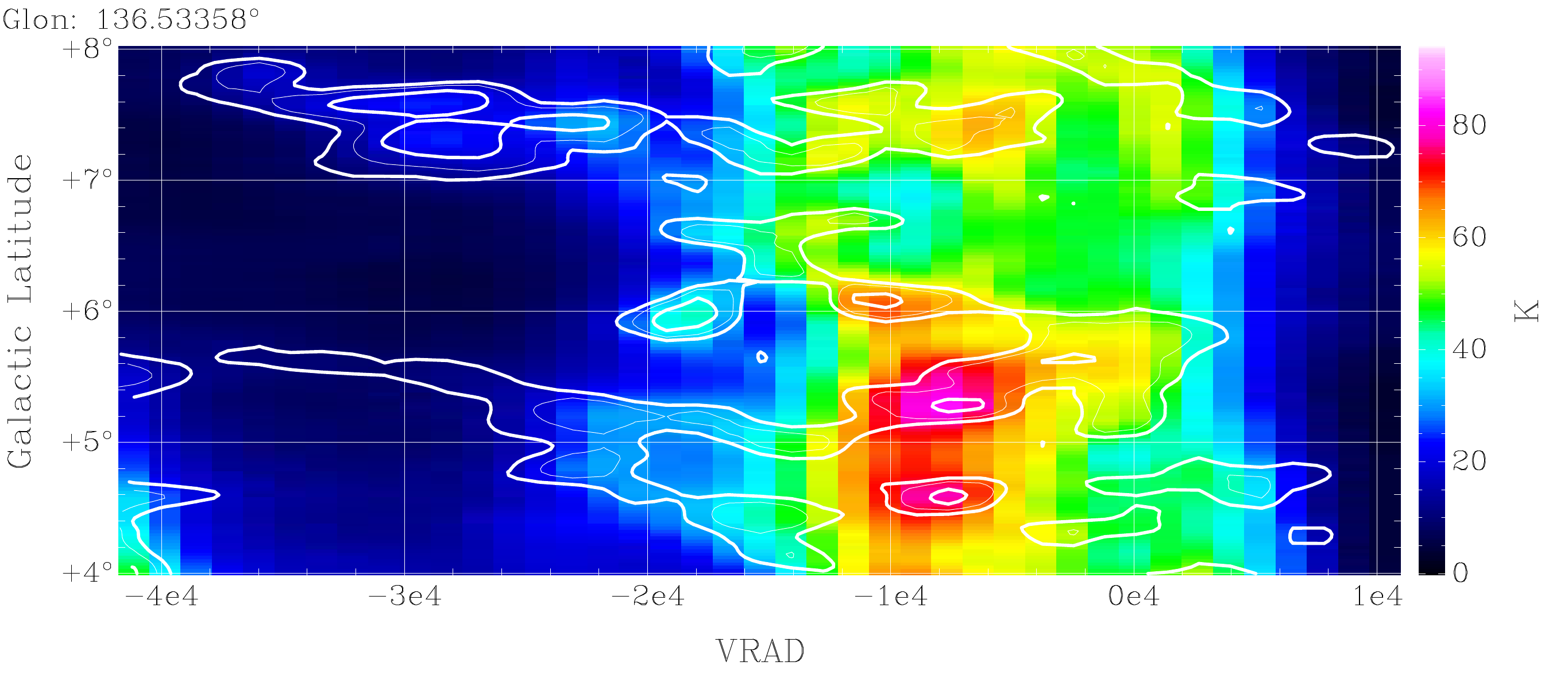}
   \includegraphics[width=3.7cm,angle=-90]{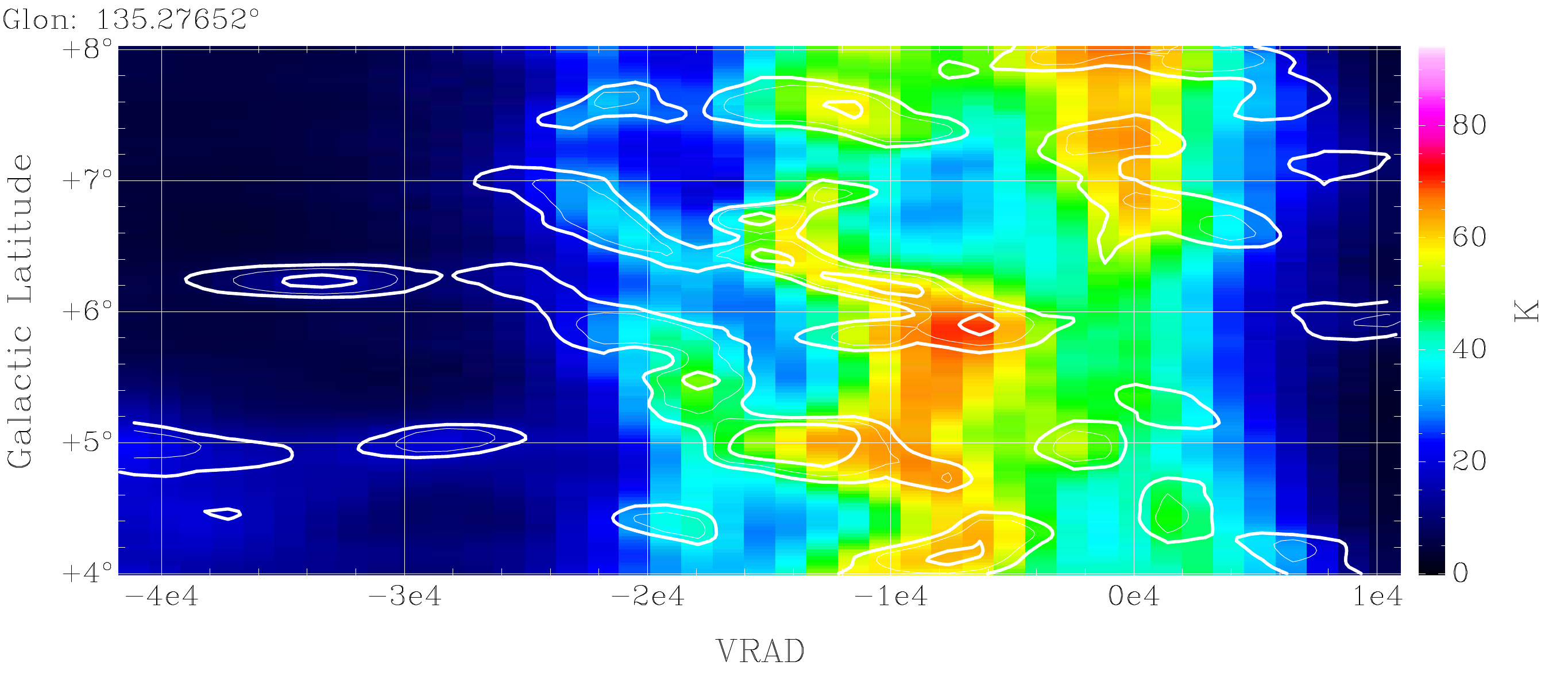}
   \includegraphics[width=3.7cm,angle=-90]{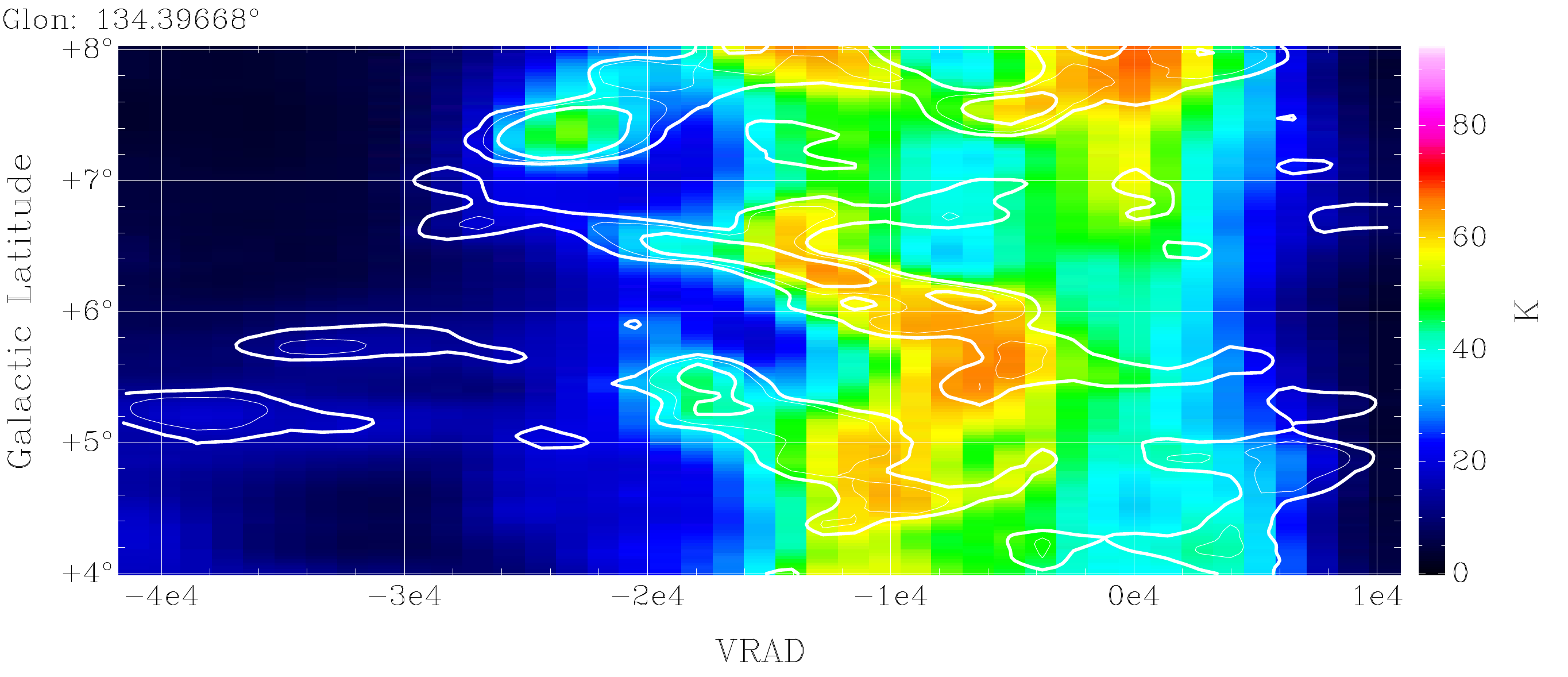}
   \includegraphics[width=3.7cm,angle=-90]{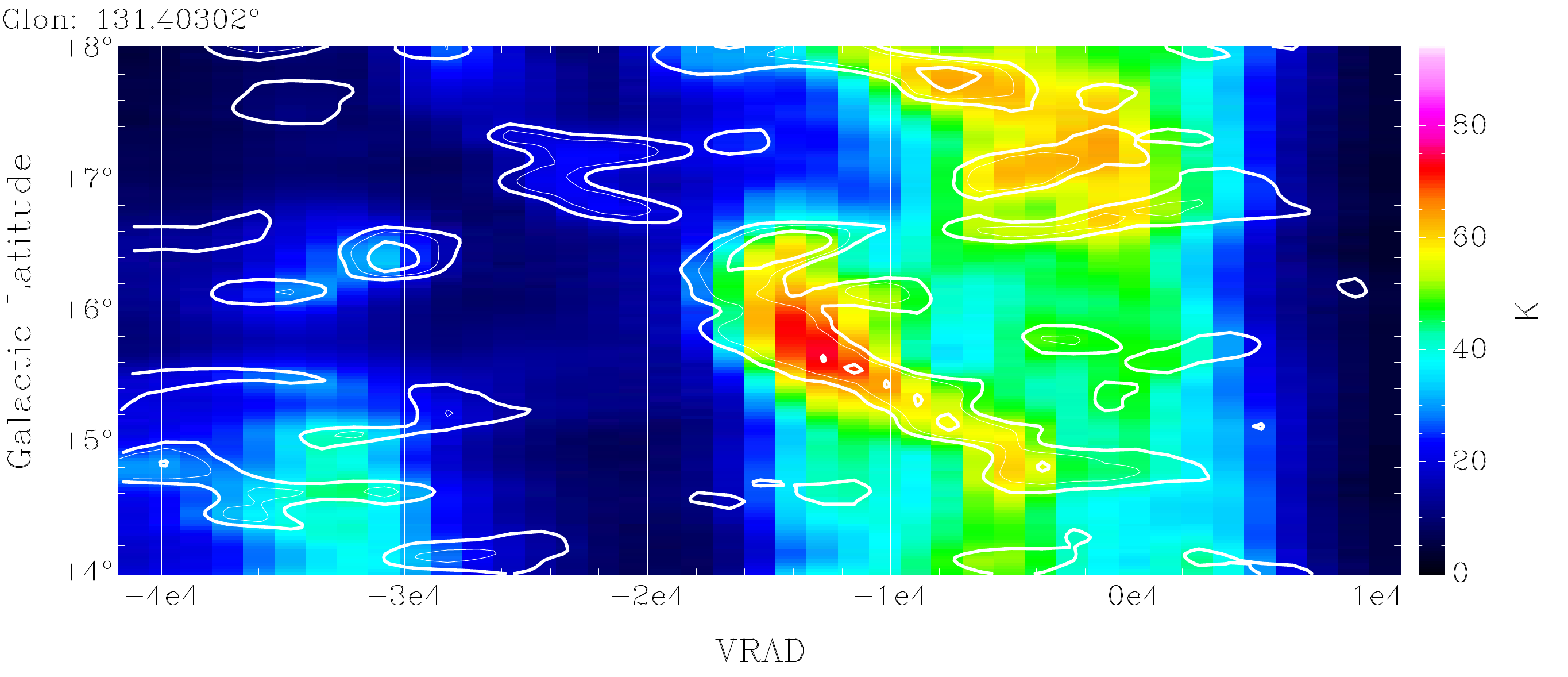}
   \caption{Velocity-position diagrams, brightness temperatures (color
     coded) for cuts at constant longitudes $l = $ 136\fdg534,
     135\fdg277, 134\fdg397, and 131\fdg403 (top left to bottom right),
     indicated in Fig. \ref{Fig_Horo_gal} by black lines. Radial
     velocities are in m\,s$^{-1}$, isophotes are USM temperatures of 1,
     2.5, and 5 K. }
   \label{Fig_Horo_cuts}
\end{figure*}

\begin{figure*}[htbp] 
   \centering
   \includegraphics[width=5.cm,angle=-90]{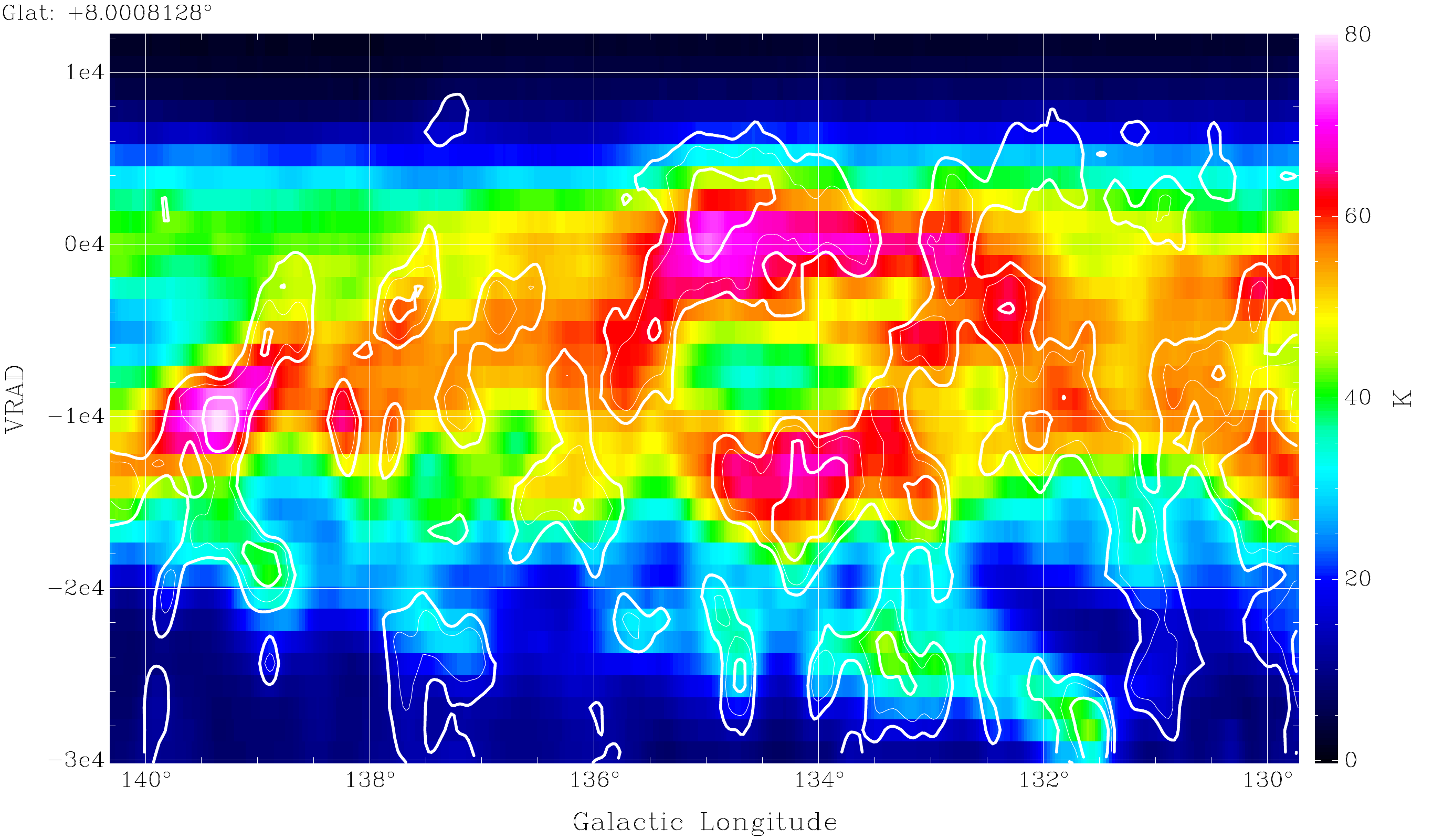}
   \includegraphics[width=5.cm,angle=-90]{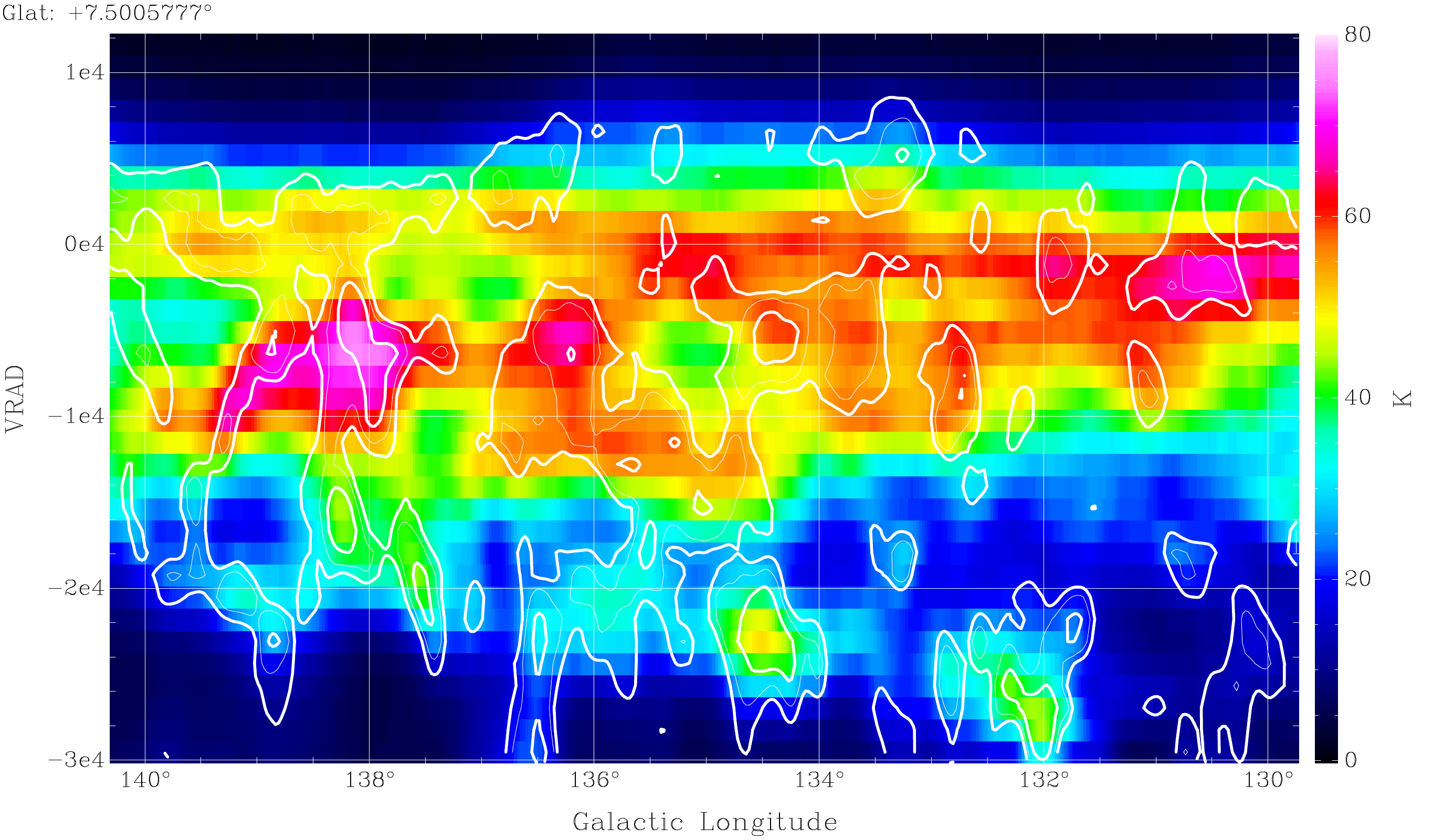}
   \includegraphics[width=5.cm,angle=-90]{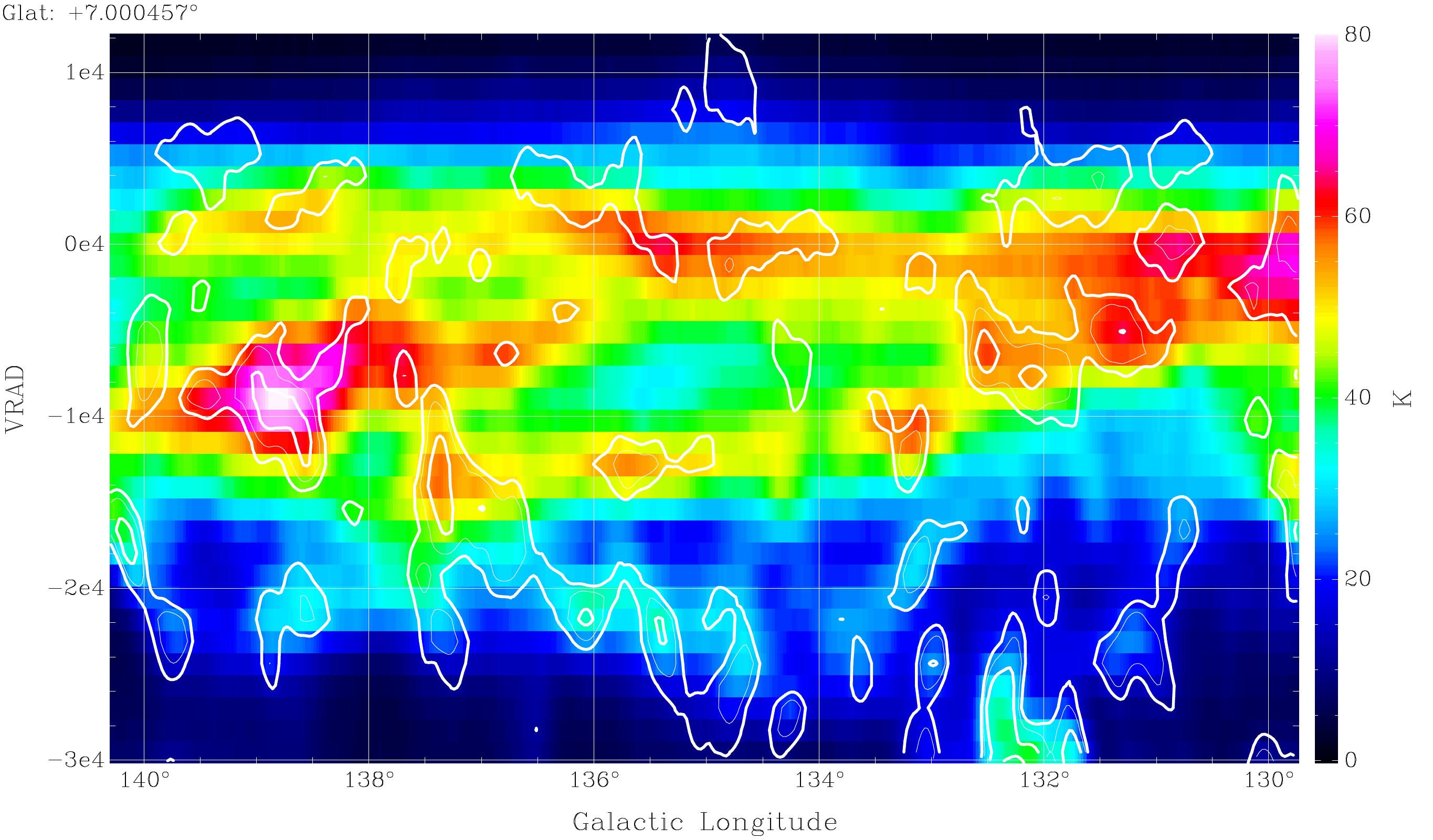}
   \includegraphics[width=5.cm,angle=-90]{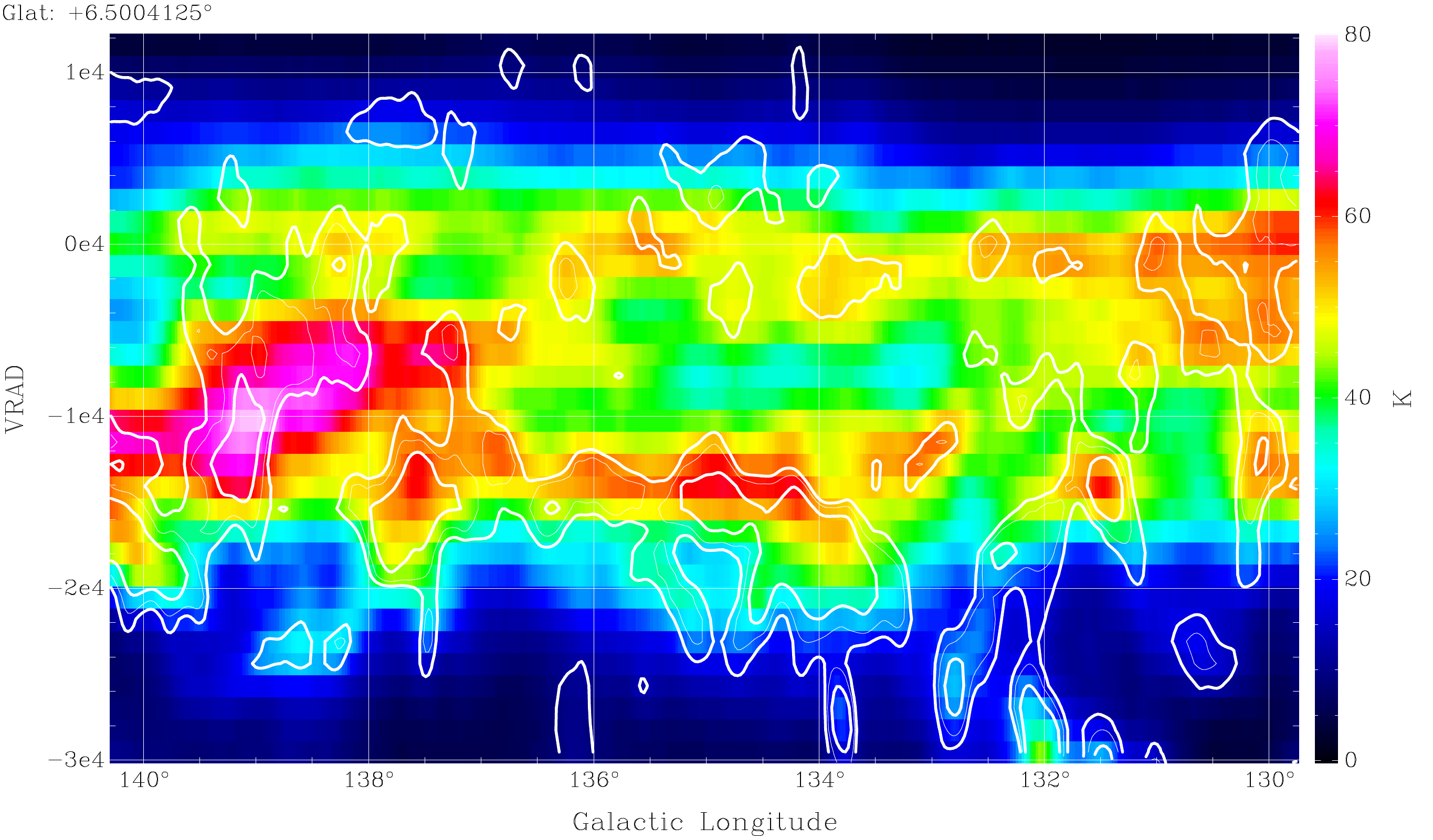}
   \includegraphics[width=5.cm,angle=-90]{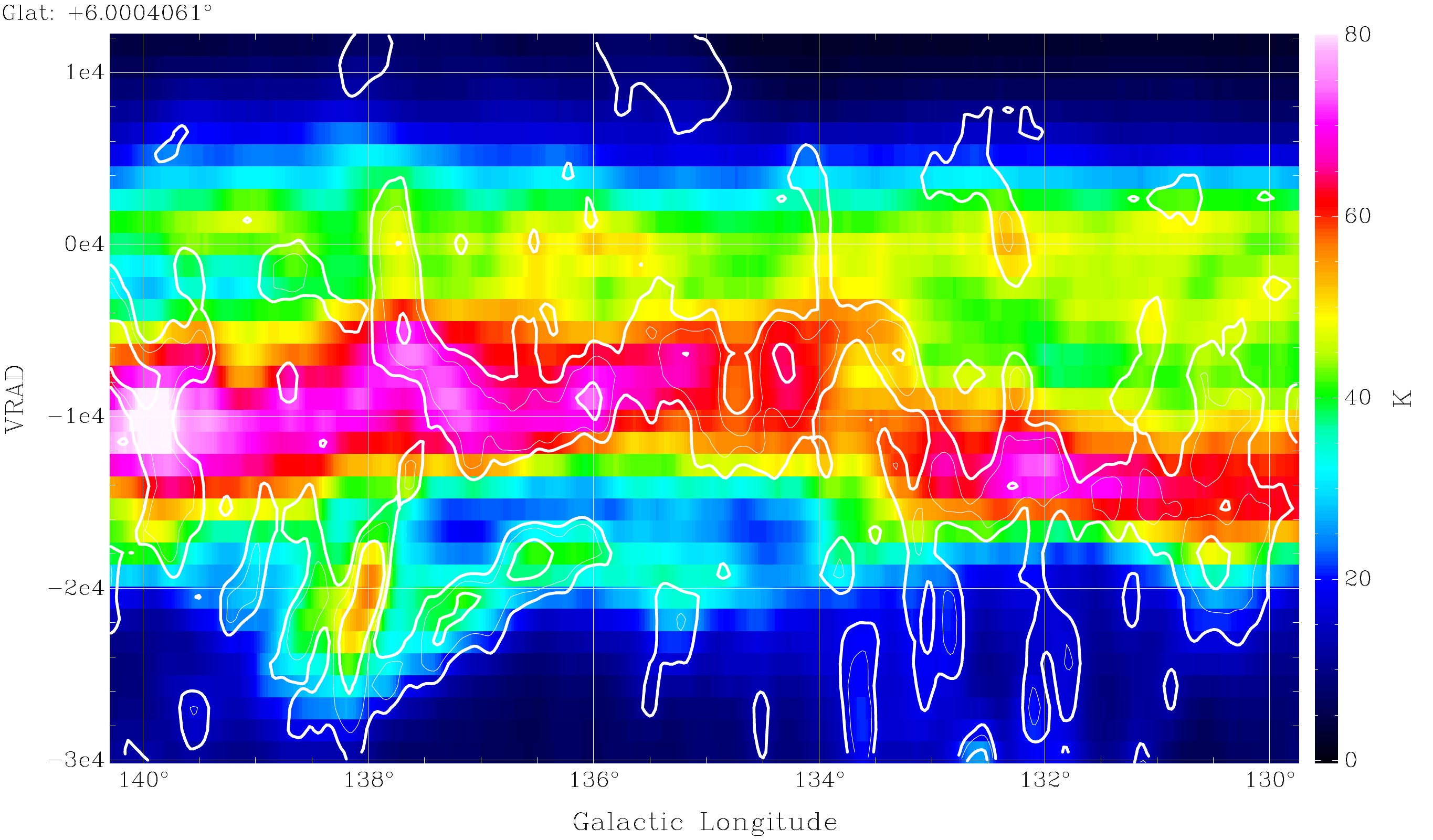}
   \includegraphics[width=5.cm,angle=-90]{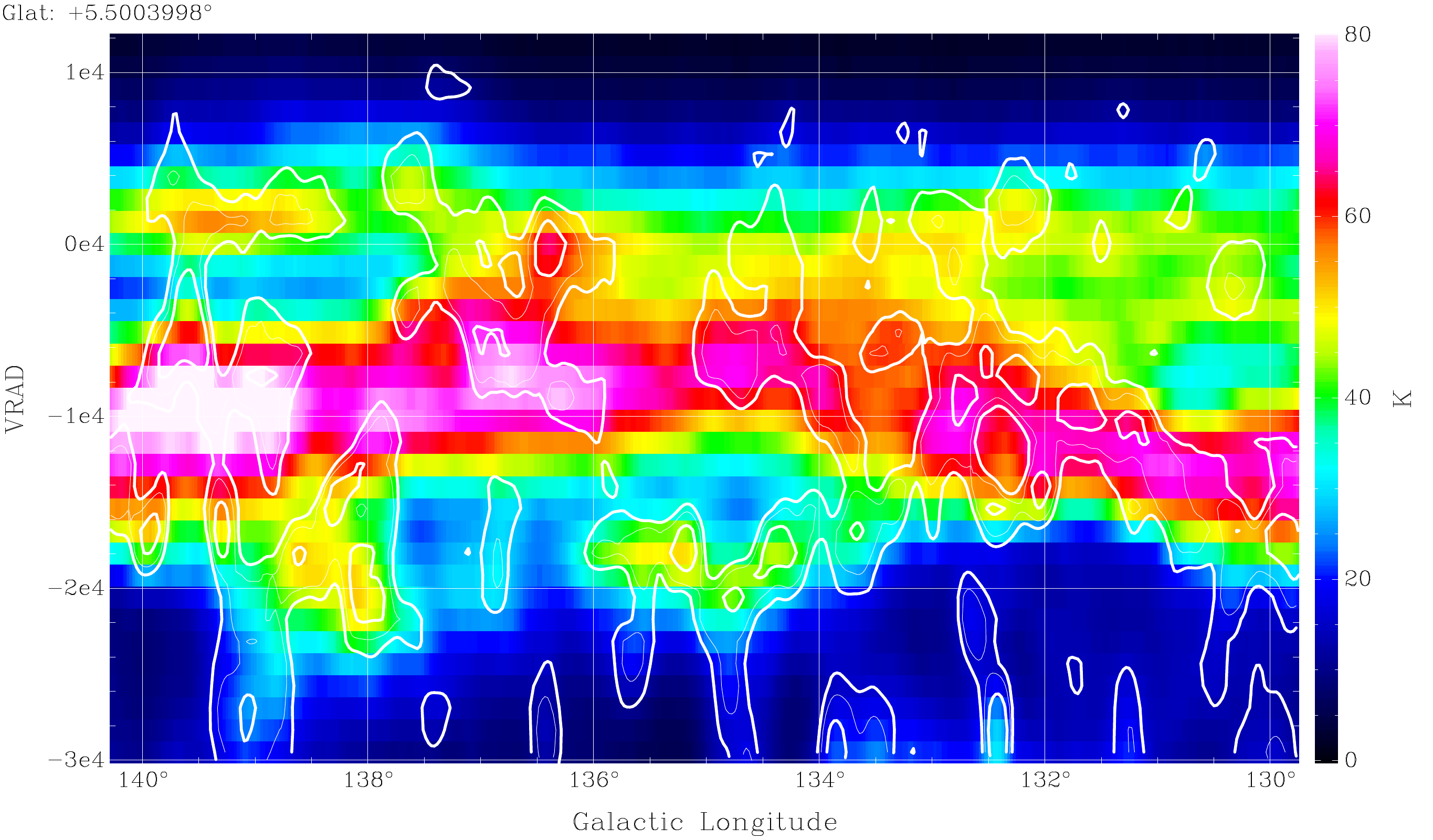}
   \includegraphics[width=5.cm,angle=-90]{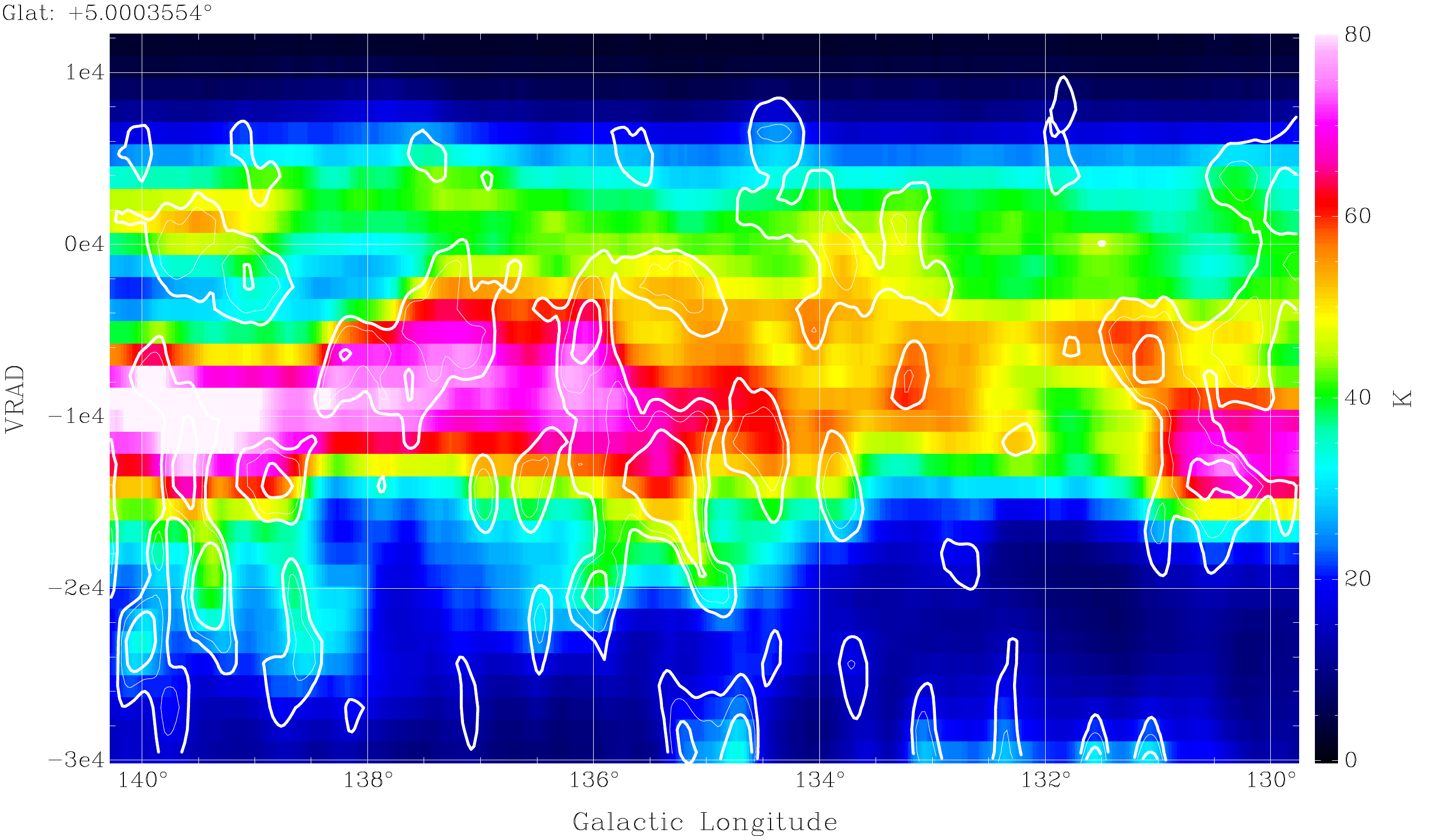}
   \includegraphics[width=5.cm,angle=-90]{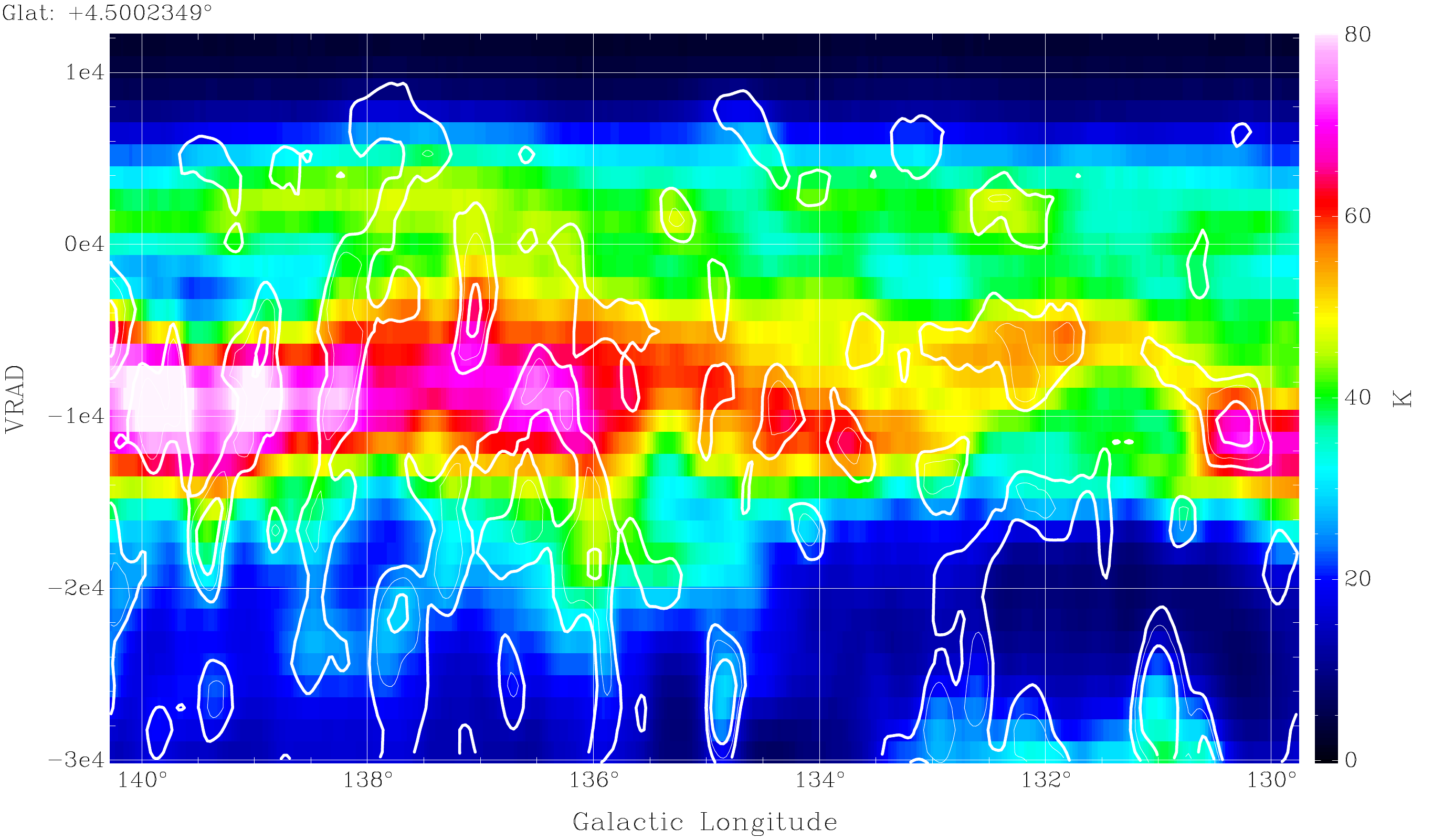}
   \caption{Position-velocity diagrams, color coded brightness
     temperatures for cuts at constant latitudes $b = $ 8\fdg0 to 4\fdg5
     in steps of -0\fdg5 (top left to bottom right), overlaid with
     isophotes displaying USM temperatures of 1, 2.5, and 5 K.  Radial
     velocities are in m\,s$^{-1}$.}
   \label{Fig_Horo_cuts_lat}
\end{figure*}

This filamentary feature is oriented parallel to the Galactic plane. To
determine details of the large scale structure we calculate EBHIS maps
in Galactic coordinates. Figure \ref{Fig_Horo_gal} shows at the top the
brightness temperature distribution at $v_{\rm LSR} = -16.6 $ \kms. USM
temperatures at the same velocity, emphasizing filamentary structures in
the CNM, are overlayed with isophotes at 1, 2.5, and 5 K. The
feature from Fig. \ref{Fig_Overlay_HO-16.6} is located at $l \sim
135\fdg8$, $b \sim$ 6\fdg7 as part of an arc with a total length of at
least 6\deg. In the bottom panel of Fig. \ref{Fig_Horo_gal} we display
the observed distribution similar to the top panel but now at $v_{\rm
  LSR} = -11.5 $ \kms. For comparison we replicate the USM structure
from the top panel at $v_{\rm LSR} = -16.6 $ \kmss with black
isophotes. It is obvious that the large scale filamentary structure has
shifted to lower latitudes and is broken in several fragments.

In Fig. \ref{Fig_Horo_cuts} we display four velocity-position cuts
through this arc-like feature at constant longitudes indicated in
Fig. \ref{Fig_Horo_gal}. In all cases we can trace
coherent features that span a range of approximately 20 \kms, shifting
in latitude by roughly 1\deg. Hence we observe a velocity gradient of 20
\kms/degree perpendicular to the mean magnetic field direction.  For
individual velocity channels the observed filamentary USM structures
like that from Fig.  \ref{Fig_Horo_gal} may be modeled as fibers. The
length of each fiber depends on the observed fragmentation. However,
continuity in velocity-position space with well defined velocity
gradients (Figs. \ref{Fig_Horo_gal} and \ref{Fig_Horo_cuts}) inhibits an
interpretation as 1D structures for most of the observed filamentary
structures. They need to be described as sheets, 2D structures, tilted
in velocity with distinct gradients. Sheets are often warped and broken.

The arc (Fig. \ref{Fig_Horo_gal}) is well defined on large scales but
was not cataloged previously. Features like this are usually classified
as expanding HI shells
\citep{Heiles1979,Heiles1984,Naomi2002,Ehlerova2005,Ehlerova2013}. An
expanding shell, originating from supernova explosions in the Galactic
plane, is affected by the planar density distribution
\citep{Ehlerova1997,Vorobyov2004,Vorobyov2005}. In case of the
filamentary feature displayed in Fig. \ref{Fig_Horo_gal}, the upper part
of the shell is expanding toward us while the lower part in latitude is
decelerated and fragmented by the stratified distribution of the ISM
close to the Galactic plane.

The shell has a dominant main feature.  Other structures, offset in
latitude, are less intense and may represent secondary shocks.
\citet{Bykov1987} developed a model for the interstellar turbulence
where shocks, produced by supernovae, are reflected by interstellar
clouds. Associated structures in the magneto-ionic medium are expected
\citep{Fletcher2007}.  \citet{Kornreich2000} argue that galactic shocks
propagating through interstellar density fluctuations may act as ``shock
pumps'', providing a mechanism for a turbulent cascade mechanism and
explaining the fractal-like structure of the cool interstellar medium.

Our data support these proposals, there are many filamentary features
that may be interpreted as secondary shocks, mostly parallel to the
Galactic plane at different latitudes. We find indications for dynamic
interactions between primary and secondary shock structures, including
position-velocity structures that might be caused by curls in the
turbulent flow, see Fig. \ref{Fig_Horo_cuts_lat}.

\end{document}